\def\draftversion{false}

\RequirePackage{ifthen}
\ifthenelse{\equal{\draftversion}{true}}{
  \documentclass[prb,galley,showpacs,preprintnumbers,citeautoscript,
      amsmath,amssymb,longbibliography]{revtex4-2}
}{
  \documentclass[citeautoscript,floatfix,aps,prb,twocolumn,
      superscriptaddress,longbibliography]{revtex4-2}
}

\usepackage{amsmath,amsfonts,latexsym,psfrag}
\usepackage[pdftex]{graphicx}

\usepackage{ragged2e} 

\usepackage{epstopdf}
\usepackage{natbib}
\usepackage{array}
\usepackage{dcolumn}
\usepackage{bm}

\usepackage{soul}  

\usepackage{float}
\usepackage[utf8]{inputenc}
\usepackage{textcomp}
\usepackage[normalem]{ulem}

\usepackage[usenames,dvipsnames]{color}

\soulregister\cite7
\usepackage[%
  colorlinks=true,
  urlcolor=blue,
  linkcolor=blue,
  citecolor=blue
]{hyperref}

\ifthenelse{\equal{\draftversion}{true}}{
  \usepackage{showlabels}
  \marginparwidth 2.7in
  \marginparsep 0.5in
  \newcounter{comm} 
  \def\commnext{\stepcounter{comm}}
  \def\commtext{{\bf\color{blue}[\arabic{comm}]}}
  \def\commmar{{\bf\color{blue}[\arabic{comm}]}}
  \def\msm#1{\commnext\marginpar{\small MS\commmar: #1}\commtext}

  \def\mlab#1{\marginpar{\small\bf #1}}
  
}{
  \def\dvm#1{}
  \def\cdm#1{}
  \def\msm#1{}
  \def\asm#1{}
  \def\miq#1{}
  \def\mlab#1{}
  
}

\newcommand\siesta{{\sc siesta}}

\DeclareMathOperator\erf{erf}
\DeclareMathOperator\erfc{erfc}

\usepackage[utf8]{inputenc}
\usepackage{newunicodechar}
\usepackage{url}
\newunicodechar{−}{-}
\begin{document}

\title{Implementation of the hybrid exchange-correlation functionals in the \siesta\ code }

\author{Yann Pouillon}%
\email[Corresponding author: ]{yann.pouillon@unican.es}%
\affiliation{Simune Atomistics, 20018 San Sebastián, Spain}

\author{Bill Clintone Oyomo}
\affiliation{ Materials Modelling Group, 
	Department of Physics, Earth and Environmental Sciences,
	The Technical University of Kenya,
	52428-00200, Nairobi, Kenya.}  

\author{ James Sifuna}
\affiliation{ Materials Modelling Group,
Department of Physics, Earth and Environmental Sciences,
The Technical University of Kenya, 52428-00200, Nairobi, Kenya.}
\affiliation{ Theoretical Condensed Matter Group, Department of Natural Science,  The Catholic University of Eastern Africa, 62157 - 00200, Nairobi, Kenya.}

\author{Mar\'{\i}a Camarasa-Gómez}
\affiliation{Centro de F\'isica de Materiales (CFM-MPC) CSIC-UPV/EHU, 20018 Donostia-San Sebastián, Spain}  

\author{Xinming~Qin}%
\affiliation{Hefei National Laboratory for Physical Sciences at the Microscale,
No.96, JinZhai Road Baohe District,
Hefei, Anhui, 230026,
China}

\author{Carlos Beltr\'{a}n}%
\affiliation{Departamento de Matem\'{a}ticas, Estad\'{i}stica y Computaci\'{o}n,
Universidad de Cantabria,
Avenida de los Castros s/n,
E-39005 Santander, Spain}

\author{Fernando G\'omez-Ortiz}
\affiliation{Theoretical Materials Physics, Q-MAT, Université de Liège, B-4000 Sart-Tilman, Belgium}

\author{Honghui~Shang}%
\email[Corresponding author: ]{shh@ustc.edu.cn}%
\affiliation{State Key Laboratory of Precision and Intelligent Chemistry, University of Science and Technology of China, Hefei, China}

\author{Javier Junquera}%
\email[Corresponding author: ]{javier.junquera@unican.es}%
\affiliation{Departamento de Ciencias de la Tierra y F\'{\i}sica de la Materia Condensada, Facultad de Ciencias,
Universidad de Cantabria,
Avenida de los Castros s/n,
E-39005 Santander,
Spain}

\begin{abstract}
We present an efficient and accurate implementation of hybrid exchange–correlation (XC) functionals in the \textsc{siesta} code, enabling large-scale simulations based on Hartree–Fock-type exact exchange combined with strictly localized numerical atomic orbitals (NAOs). Our approach exploits a fitted representation of the NAOs in terms of Gaussian-type orbitals (GTOs), which allows for the analytical evaluation of four-center electron repulsion integrals (ERIs) via the \textsc{libint} library. This framework is seamlessly integrated with \textsc{siesta}’s real-space grid and sparse-matrix infrastructure, and is combined with multiple screening techniques to control the computational complexity. We also introduce a fully analytical formulation of hybrid-functional forces and a dynamic parallel distribution scheme that ensures excellent scalability. We validate our implementation through benchmark calculations on a broad set of systems (including semiconductors, insulators, and two-dimensional materials) and demonstrate that the HSE06 functional significantly improves the prediction of band gaps compared to PBE, in close agreement with G$_{0}$W$_{0}$ and experimental data. We analyze in detail the trade-offs between accuracy and computational efficiency as a function of the number of Gaussians, basis set range, and integral screening thresholds. Our results confirm that hybrid functional calculations in \textsc{siesta} are now feasible for large extended systems, making accurate first-principles predictions of electronic and structural properties accessible at scale.
\end{abstract}

\preprint{Draft 1 - \today}


\maketitle


\section{Introduction}
\label{sec:intro}

Density-functional theory (DFT)~\cite{Hohenberg-64} has become an indispensable tool in the study of the electronic structure of materials.
Its success is largely attributed to its favorable scaling with system size compared to wavefunction-based approaches. While the latter require dealing with a many-electron wavefunction that depends on $3N$ variables for a system of $N$ electrons, resulting in an exponentially increasing computational cost, DFT reformulates the problem in terms of the electron density, a function of only three spatial coordinates. This dramatic reduction in complexity has enabled first-principles simulations of systems far beyond the reach of traditional quantum wavefunction-based methods~\cite{Martin_2004,Kohanoff_2006,Parr_book}.

The practical utility of DFT hinges on the approximations used for the exchange–correlation (XC) functional. The simplest of these, such as the local density approximation (LDA), assumes that the XC energy depends solely on the local value of the electron density~\cite{Kohn-65,Ceperley-80,Perdew-81}. Generalized gradient approximations (GGAs) improve upon LDA by incorporating the gradient of the density, capturing non-uniformities more accurately~\cite{Perdew-96.2,PBE,PBE-erratum}. These semi-local functionals have proven to be remarkably successful in predicting the structural and energetic properties of materials, typically achieving mean absolute errors of a few percent in equilibrium lattice constants, bulk moduli, and cohesive energies~\cite{Perdew-08-PBEsol, Zhang2018}.
To further improve accuracy and establish a systematic framework for functional development, the concept of ``Jacob's ladder''~\cite{Perdew2001} was introduced, a metaphorical hierarchy of XC approximations based on the inclusion of increasingly sophisticated ingredients while striving to satisfy as many known physical constraints as possible.

Despite their success, semi-local functionals suffer from inherent limitations, most notably the self-interaction error, which arises from the incomplete cancelation of an electron's interaction with itself~\cite{Cohen2008}. This deficiency manifests itself in several important observables: underestimation of band gaps~\cite{Perdew1983, Sham1983, Mori-Sanchez2008, Cohen2008}, overestimation of electron delocalization effects~\cite{Mori-Sanchez2008} and therefore failures in the treatment of many $d$ and $f$ element compounds~\cite{anisimov2000strong}, inaccurate barrier heights for chemical reactions, and significant errors in reaction and dissociation energies \cite{Cohen2008}.
%

 One promising route to mitigate these errors is the use of hybrid functionals, which incorporate a fraction of (non-local) Hartree–Fock-like exact exchange (HFX) computed from the Kohn–Sham orbitals~\cite{Becke-93,Stephens_1994}. The introduction of global and range-separated hybrid functionals~\cite{Adamo-99, HSE03-1,HSE03-2,HSE06, HSESol} has been particularly important for solid-state applications, with PBE0~\cite{Adamo-99} and HSE06~\cite{HSE03-1,HSE03-2,HSE06} becoming a significant breakthrough in the prediction of band gaps and band structures in semiconductors~\cite{Janesko_2009,Garza2016,Borlido2019,Yang2023}. These hybrid schemes have been shown to significantly improve the predictions of electronic band structures and band gaps, structural parameters~\cite{Paier2006, Paier2006err, HSESol}, charge localization~\cite{Lany2010}, defect energy levels~\cite{Chen-13,Chen-17}, as well as cohesive energies~\cite{Zhang2018} and thermochemical properties~\cite{HSE06, HSESol} across a wide variety of systems. These advances bridged the gap between (semi-)local DFT and more computationally intensive many-body approaches, providing more accurate results at a reasonable computational cost.

The main drawback of hybrid functionals lies in their significantly increased computational cost compared to conventional (semi-)local DFT calculations. This overhead is primarily due to the evaluation of electron repulsion integrals (ERIs), which dominates the construction of the non-local Hartree–Fock exchange (HFX) matrix. As a result, the development of efficient and scalable ERI algorithms is key to making hybrid functional calculations feasible for large-scale systems, as hybrid functionals have proven to obtain very accurate electronic structures in large unit cells, such as in defected-systems, surfaces, or heterostructures \cite{Guo2021, Chen2022, Ke2025, Sagredo2025}.

Existing implementations of hybrid functionals for extended systems can be broadly treated according to the underlying basis set: plane waves (PW)~\cite{Gonze2016,Lin2016,Gygi2010,DiStasio2014,DeSlippe2017,Wu2009}, real-space grids~\cite{Natan2015,Natan2016}, Gaussian-type orbitals (GTO)~\cite{Gaussian,NWChem,Quickstep,Bush2011}, Slater-type orbitals (STO)~\cite{Velde-01}, and numerical atom-centered orbitals (NAO)~\cite{Blum2009,Havu2009,Ren2012,Shang2019}. 
Although plane-wave basis sets offer systematic convergence and completeness, they lack spatial locality. Real-space implementations can overcome some of these limitations by employing efficient algorithms to compute exact exchange~\cite{Natan2015,Natan2016}. In contrast, atomic orbital basis sets (including GTOs, STOs, and NAOs) are localized by construction, leading to sparse Hamiltonian matrices and more favorable scaling with system size. These properties have made atomic basis sets particularly attractive for large-scale DFT simulations and massively parallel computations~\cite{Quickstep,Guidon_2010,Bush2011,NWChem,Ren2012}. Various efficient algorithms have been developed to accelerate the computation of exact exchange in these implementations, including linear-scaling methods~\cite{Schwegler_1996,Burant_1996,Schwegler_1997,Ochsenfeld_1998}, density fitting approaches~\cite{Polly_2004,Sodt_2008,Guidon_2010,Merlot_2014}, and advanced parallel algorithms~\cite{GTFock_1,GTFock_2,Shang2019}.

Atomic orbitals themselves may be classified into GTOs, as used in \textsc{gaussian}~\cite{Gaussian,Gaussian-web},
\textsc{cp2k}~\cite{Hutter-14,Guidon-08,kuhne2020cp2k,iannuzzi2025cp2k,CP2K-web}, or \textsc{crystal}~\cite{CRYSTAL,Dovesi-20,erba2022crystal23,Crystal-web} among others, 
STOs as in \textsc{adf}~\cite{Velde-01}, and numerical atomic orbitals, which are employed in codes such as \textsc{siesta}~\cite{SIESTA}, 
\textsc{conquest}~\cite{CONQUEST,Nakata-20}, \textsc{dmol$^3$}~\cite{DMOL}, 
\textsc{plato}~\cite{kenny2009plato},
\textsc{openmx}~\cite{OPENMX,Ozaki-05,Toyoda-11}, 
\textsc{abacus}~\cite{Chen-10,Li-16,ABACUS-hybrids},\textsc{fhi-aims}~\cite{Blum2009,Ren2012,Kokott-24,abbott2025, fhiaims-web},  
\textsc{q-chem}~\cite{Lee-22}, or \textsc{pyscf}~\cite{Sun-22}, among others. GTOs allow for the analytical evaluation of ERIs, whereas NAOs, due to their strict spatial confinement, naturally lead to lower scaling in computational effort with respect to system size.

To capitalize on the strengths of both approaches, a hybrid scheme was proposed~\cite{Shang-JCP,qin-2015-1}, later improved to enable the efficient computation of analytic forces~\cite{Qin-23} and scalable parallel execution~\cite{Shang-20}. This scheme preserves the strict locality of numerical atomic orbitals (NAOs) to enforce spatial cutoffs, while a compact auxiliary representation in terms of Gaussian-type orbitals (GTOs) enables the efficient analytical evaluation of electron-repulsion integrals (ERIs) through the {\sc libint} library~\cite{Libint}. Combined with advanced screening techniques, this approach yields highly efficient Hartree–Fock exchange (HFX) matrix construction with near-linear scaling, paving the way for large-scale hybrid-functional calculations with controlled accuracy and performance.
An alternative approach to the evaluation of ERIs for NAOs was developed by Toyoda and Ozaki~\cite{Toyoda-09,Toyoda-10}, who introduced a fully numerical formulation based on spherical Bessel transforms and successive reduction of the integral dimensionality, implemented in the {\sc liberi} library. This method enables the direct computation of ERIs and their analytic derivatives for strictly localized pseudoatomic orbitals without any fitting to analytic basis functions.

The strategy adopted in the present work is complementary in spirit. Rather than evaluating ERIs numerically, we approximate the radial part of the numerical atomic orbitals by a compact Gaussian expansion, which allows us to exploit mature and highly optimized Gaussian integral libraries, while preserving the locality and accuracy of the original NAO basis.
Here, we substantially extend and consolidate this methodology within the {\sc siesta} code. In particular, we introduce a fully internal and systematic fitting of the radial parts of native NAOs onto a Gaussian basis, avoiding the use of external fitting tools and complex workflows. The resulting Gaussian radial representation is then employed as a full replacement of the native NAO basis not only for the evaluation of exact-exchange matrix elements, but also for the remaining Hamiltonian contributions entering hybrid-functional calculations. We further propose a new truncation strategy for the auxiliary basis that guarantees continuity of the orbitals and of all their derivatives at the cutoff radius. The entire implementation is fully merged into the current {\sc siesta} code base, ensuring compatibility with recent algorithmic, numerical, and parallel developments. Finally, we provide a comprehensive validation and characterization of the new implementation, analyzing the accuracy–efficiency balance as a function of the number of Gaussians, orbital range, and screening thresholds, as well as the computational overhead with respect to semilocal DFT. We demonstrate that the parallel implementation achieves excellent load balancing and minimal communication overhead, enabling simulations of unit cells containing several hundred to a few thousand atoms.

The remainder of the paper is structured as follows. In Sec.~\ref{sec:theory}, we outline the theoretical formulation of hybrid functionals and the key expressions required for the evaluation of the Hartree–Fock exchange energy in periodic systems. Sec.~\ref{sec:nao2gto} details the transformation of strictly localized numerical atomic orbitals into GTOs to enable analytical evaluation of electron repulsion integrals. The various screening strategies employed to reduce the number of four-center integrals without compromising accuracy are described in Sec.~\ref{sec:computation-fourcenter}. Section~\ref{sec:libint} addresses the efficient computation of primitive integrals using the \textsc{libint} library. In Sec.~\ref{sec:forces}, we present the formulation and implementation of analytic forces within the hybrid framework. Section~\ref{sec:results} provides extensive benchmarks to assess the accuracy and performance of the implementation. Its parallel scalability is studied in Sec.~\ref{sec:parallel}. Finally, conclusions and future perspectives are discussed in Sec.~\ref{sec:concluding_remarks}.

\section{Theory}
\label{sec:theory}

\subsection{Exchange energy}
\label{sec:exchange-energy}

Most quantum chemistry textbooks~\cite{Szabo,Atkins,Jensen-06} introduce the Hartree-Fock approximation by first presenting the many-body Hamiltonian, typically after applying the Born-Oppenheimer approximation.
The simplest antisymmetric wave function for a system of $N$ electrons occupying $N$ spin-orbitals $\chi_i$ is then written as a single Slater determinant.
According to the variational principle, the optimal wavefunction of this form is the one which minimizes the expectation value of the Hamiltonian.
After standard manipulations, the exchange contribution to the Hartree-Fock ground-state energy is given by

\begin{align}
    E_{\mathrm{x}} &= -\frac{1}{2} \sum_{i=1}^{N} \sum_{j=1}^{N} \int \!\! \int d\mathbf{x} \, d\mathbf{x}' \,
    \chi_i^*(\mathbf{x}) \chi_j(\mathbf{x}) 
    \frac{1}{|\mathbf{r} - \mathbf{r}'|}
    \chi_j^*(\mathbf{x}') \chi_i(\mathbf{x}') \nonumber \\
    &= -\frac{1}{2} \sum_{i=1}^{N} \sum_{j=1}^{N} [ij|ji],
    \label{eq:exactexchangeener}
\end{align}

\noindent where $\mathbf{x} = (\mathbf{r}, \sigma)$ denotes both spatial and spin degrees of freedom. %
The compact notation $\int d \mathbf{x} $ in the above integrals represents $\sum_{\sigma} \int d \mathbf{r}$. The bracket $[ij|kl]$ follows the Mulliken or chemist's notation~\cite{Szabo},

\begin{equation}
    [ij|kl] = \int \!\! \int d\mathbf{x} \, d\mathbf{x}' \,
    \chi_i^*(\mathbf{x}) \chi_j(\mathbf{x}) 
    \frac{1}{|\mathbf{r} - \mathbf{r}'|}
    \chi_k^*(\mathbf{x}') \chi_l(\mathbf{x}').
\end{equation}

\noindent In this notation, spin-orbitals depending on the first electron's coordinate appear to the left; the complex conjugate is placed first.
Each spin-orbital $\chi(\mathbf{x})$ is the product of a spatial orbital $\psi(\mathbf{r})$ and a spin function. For collinear spin-polarized systems, the spin part is typically represented by orthonormal functions $\alpha(\sigma)$ and $\beta(\sigma)$ for spin-up and spin-down, respectively~\cite{Szabo}.
The spatial orbitals may differ for each spin channel, as in the unrestricted Hartree-Fock approximation. We denote this explicitly as $\psi_i^{\sigma}$, with $\sigma = \alpha, \beta$.

Integrating over the spin coordinates yields an expression involving only spatial orbitals,

\begin{widetext}

\begin{align}
    E_{\mathrm{x}} &= -\frac{1}{2} \sum_{\sigma = \{\alpha,\beta\}} \sum_{i=1}^{N_{\sigma}} \sum_{j=1}^{N_{\sigma}} 
    \int \!\! \int d\mathbf{r} \, d\mathbf{r}' \,
    \psi_i^{\sigma*}(\mathbf{r}) \psi_j^{\sigma}(\mathbf{r}) 
    \frac{1}{|\mathbf{r} - \mathbf{r}'|} 
    \psi_j^{\sigma*}(\mathbf{r}') \psi_i^{\sigma}(\mathbf{r}') \nonumber \\
    &= -\frac{1}{2} \sum_{\sigma = \{\alpha,\beta\}} \sum_{i=1}^{N_{\sigma}} \sum_{j=1}^{N_{\sigma}} 
    (ij|ji)^{\sigma},
    \label{eq:xcenerspatial}
\end{align}

\end{widetext}

\noindent where $N_{\sigma}$ is the number of electrons with spin $\sigma$. The notation $(ij|kl)^{\sigma}$ denotes spatial two-electron integrals in the same chemist's convention, but with parentheses to emphasize the absence of spin integration.

We can particularize this general expression to the case of {\sc siesta}, where the spatial orbitals are expanded on a basis of $M$ strictly localized NAOs, $\phi_{\mu} (\mathbf{r})$, as~\cite{SIESTA}

\begin{equation}
    \psi_{i}^{\sigma} (\mathbf{r}) = \sum_{\mu=1}^{M} \phi_{\mu} (\mathbf{r}) c_{\mu i}^{\sigma},
    \label{eq:expnao}
\end{equation}

\noindent where $c_{\mu i}^{\sigma}= \langle \tilde{\phi}_{\mu} \vert \psi_{i}^{\sigma} \rangle$,
and $\tilde{\phi}_{\mu}$ is the dual orbital of $\phi_{\mu}$: $\langle \tilde{\phi}_{\mu} \vert \phi_{\nu} \rangle = \delta_{\mu \nu}$.
Replacing Eq.~(\ref{eq:expnao}) into Eq.~(\ref{eq:xcenerspatial}) we arrive at 

\begin{equation}
    E_{\rm x}  = -  \frac{1}{2} \sum_{\sigma=\lbrace \alpha, \beta \rbrace}
    \sum_{\mu\eta = 1}^{M}  \sum_{\nu\lambda = 1}^{M} 
    \rho_{\eta \mu}^{\sigma \sigma} \rho_{\nu \lambda}^{\sigma \sigma}
    \left( \mu \nu \vert \lambda \eta \right),
    \label{eq:exunres}
\end{equation}

\noindent where 

\begin{equation}
    \left( \mu \nu \vert \lambda \eta \right) \equiv
     \int \int \frac{\phi_{\mu}^{\ast}(  \mathbf{r} ) \phi_{\lambda}^{\ast}(\mathbf{r}^\prime)
    \phi_{\eta} (\mathbf{r}^\prime)\phi_{\nu} (\mathbf{r}) }
     {\vert \mathbf{r}-\mathbf{r}^{\prime}\vert  } d\mathbf{r}^{\prime} 
     d\mathbf{r},
     \label{eq:four-center-int}
\end{equation}

\noindent and  we have defined a density matrix as

\begin{equation}
    \rho_{\eta \mu}^{\sigma \sigma} = \sum_{i}^{N_{\sigma}} c_{\eta i}^{\sigma} n_{i}^{\sigma} c_{i \mu}^{\sigma}.
    \label{eq:dm}
\end{equation}

\noindent In Eq.~(\ref{eq:dm}), $n_{i}^{\sigma}$ is the occupation of the \emph{spin-orbital} $\chi_{i}$ 
(i.e. takes a value between 0 and 1)
and $c_{i \mu}^{\sigma} \equiv c_{\mu i}^{\sigma \ast}$. 

In the case of an spin-unpolarized calculation, where $N$ electrons occupy $N/2$ orbitals with the same value of the occupation and the same 
spatial function, independently of the spin direction, 
(restricted Hartree-Fock approximation), the energy expression can be trivially deduced from Eq.~(\ref{eq:exunres})
\footnote{The only care that must be taken is that in the case of an spin unpolarized calculation,
{\sc siesta} takes the occupancies of the states between 0 and 2, while in Eq.~(\ref{eq:dm}) it is taken 
between 0 and 1. Therefore, a factor one half has to be considered every time the density matrix 
appears.}.

\subsection{Hybrid functionals}
\label{sec:hybrid-functionals}

Within the hybrid density functional approximation, the (semi-)local exchange of LDA, GGA or metaGGA is mixed with some fraction ($\alpha$) of non-local exact exchange given in Eq.~(\ref{eq:xcenerspatial}). 
As highlighted in the introduction, the drawback of this approach is that the computational cost for the computation of the HFX energy increases dramatically with respect to the (semi-)local case, especially in periodic systems.
One way of making the calculations more efficient consist in splitting the Coulomb operator
in a short-range (SR) and a long-range (LR) part,

\begin{equation}
   \frac{1}{r} = \underbrace{\frac{\erfc({\omega r})}{r}}_{\rm SR}  +  \underbrace{\frac{\erf({\omega r})}{r}}_{\rm LR}. 
   \label{eq:split}
\end{equation}

\noindent 
This separation provides a better description of the band gap, and keeps accuracy for molecular systems, as well as surfaces and extended systems in general. The screening parameter $\omega$ describes the separation range between the short and the long range and its optimal value depends on the local electronic structure. The long range avoids the problematic long-range Hartree-Fock exchange, while keeping the semilocal exchange in the short and the long range.
%
The larger $\omega$, the more abrupt the transition between the short- and long-range regimes, confining the short-range part to shorter distances.
A smaller $\omega$ leads to a more gradual separation, extending the range over which the transition occurs.
In this work, following the recommendation of Krukau {\it et al.}~\cite{HSE06}, a value of $\omega = 0.11$ Bohr$^{-1}$ is assumed, corresponding to a range-separation length scale of approximately 5.0 Bohr.

In the HSE06 functional, the exchange-correlation energy is computed as~\cite{HSE06}

\begin{align}
   E_{\rm xc}^{\rm HSE06} = &  \alpha E_{\rm x}^{\rm HF,SR} (\omega) + (1-\alpha) E_{\rm x}^{\rm PBE,SR} (\omega) 
   \nonumber \\ &+ E_{\rm x}^{\rm PBE,LR} (\omega) + E_{\rm c}^{\rm PBE},
\end{align}

\noindent where $E_{\rm x}^{\rm HF,SR}$
 is the short-range HF exchange, $E_{\rm x}^{\rm PBE,SR}$ and $E_{\rm x}^{\rm PBE,LR}$ are the short-range and long-range components of the PBE exchange functional obtained by integration of the model PBE exchange hole~\cite{Heyd-04,Ernzerhof-08}, and $E_{\rm c}^{\rm PBE}$ is the PBE correlation energy.
 If $\omega \rightarrow 0$, then the HSE06 functional reduces to the PBE0 hybrid functional~\cite{Adamo-99}

 \begin{align}
   E_{\rm xc}^{\rm PBE0} = &  0.25 E_{\rm x}^{\rm HF} + 0.75 E_{\rm x}^{\rm PBE} + E_{\rm c}^{\rm PBE},
\end{align}
where $E_{\rm x}^{\rm HF}$ is the HF exact exchange and $E_{\rm x}^{\rm PBE}$ is the exchange energy of the PBE functional.
\noindent On the other limit, $\omega \rightarrow \infty$, HSE06 equals the semi-local PBE functional. 
 Perturbation theory derives a value for the HF mixing parameter $\alpha = 1/4$~\cite{Perdew-96}.
 In solids, it has been proposed that the optimal $\alpha$ parameter scales as the inverse of the electronic dielectric constant~\cite{Falletta-22.1}. %
 The fraction of Fock exchange has also been determined through either the long-range screening or the generalized Koopmans' condition applied to defect charge-transition levels~\cite{Miceli-18}. 
 In the present implementation, we adopt \emph{as default} the standard parameter values for the percentage of the exact exchange, $\alpha$, and the screening parameter, $\omega$, of the HSE06 and PBE0 functionals (namely, $\alpha$ = 0.25 for PBE0, and $\alpha$ = 0.25 and $\omega$ = 0.11 Bohr$^{-1}$ for HSE06). While $\alpha$ and $\omega$ are, in principle, system- and material-dependent, their exact determination is beyond the scope of this work. However, both $\alpha$ and $\omega$ are a user-defined input parameters in the present implementation.


\subsection{Exchange potential}
\label{sec:xpotencial}

 In the previous Section, we have focused on the form of the exchange contribution to the 
 electronic energy of the multi-electronic Hamiltonian, when  a wave
 function made of a single Slater determinant is assumed.
 According to the variational principle, the best spin-orbitals that are used to 
 build this determinant are those which minimize the total electronic energy.
 Following text-book procedures~\cite{Szabo,Jensen-06}, this minimization produces a set of
 integro-differential equations whose self-consistent solution
 yields to those best spin-orbitals.
 In these equations, the non-local exchange potential term, arising from the antisymmetric nature of the
 single determinant, acting on the spin-orbital $\chi_{i}$ is defined as 
 

\begin{equation}
    \hat{V}^{\rm HFX}  \vert \chi_{i} \rangle \equiv 
    -  \sum_{j=1}^{N} \left( \int \frac{\chi_{j}^{\ast}(\mathbf{x}^\prime) \chi_{i}(\mathbf{x}^\prime) }
    {\vert \mathbf{r}-\mathbf{r}^{\prime}\vert } d\mathbf{x}^{\prime} \right) \chi_{j} (\mathbf{x}).
    \label{eq:vxopen}
\end{equation}

\noindent If we again integrate out the spin functions, the previous expression reduces to

\begin{equation}
   \hat{V}^{\rm HFX} \vert \chi_{i} \rangle = -  \sum_{j=1}^{N_{\sigma}} \left( \int \frac{\psi_{j}^{\sigma \ast}(\mathbf{r}^\prime) \psi_{i}^{\sigma}(\mathbf{r}^\prime) }
    {\vert \mathbf{r}-\mathbf{r}^{\prime}\vert } d\mathbf{r}^{\prime} \right) \psi_{j}^{\sigma} (\mathbf{r}), 
    \label{eq:vxopen2}
\end{equation}

\noindent where $N_{\sigma}$ is the number of spin-orbitals with the same spin component as the spin-orbital
$\chi_{i}(\mathbf{x})$. 
If we expand the spatial orbitals as the linear combination given in [Eq.~(\ref{eq:expnao})],
multiply by the conjugate gradient of an orbital, $\phi_{\mu}^{\ast}$, 
on the left, and integrate over $\mathbf{r}$, we arrive at

\begin{align}
   \langle \phi_{\mu} \vert \hat{V}^{\rm HFX} \vert \chi_{i} \rangle  & = - \sum_{\eta = 1}^{M} \left(\sum_{\lambda \nu = 1}^{M}
      \rho_{\nu \lambda}^{\sigma \sigma}
      \left( \mu \nu \vert \lambda \eta \right) \right) c_{\eta i}^{\sigma}
      \nonumber \\
      & = - \sum_{\eta = 1}^{M} V^{\rm HFX}_{\mu \eta}
       c_{\eta i}^{\sigma},
\end{align}

\noindent where we have defined the matrix elements of the non-local exchange potential contribution to the total Hamiltonian, expressed in the basis of NAOs, as 

\begin{equation}
    V^{\rm HFX}_{\mu \eta} = \sum_{\lambda \nu = 1}^{M}
      \rho_{\nu \lambda}^{\sigma \sigma}
      \left( \mu \nu \vert \lambda \eta \right).
      \label{eq:VHFXmueta}
\end{equation}

\subsection{Brillouin zone sampling}
\label{sec:Brillouin}

In the preceding subsections, we considered non-periodic, open systems such as atoms or molecules. In that context, the eigenfunctions of the Hamiltonian are labeled by a discrete index, and the corresponding wavefunctions are assumed to decay asymptotically.

For extended periodic solids, the generalization of the exchange operator must account for the translational invariance and the infinite nature of the system~\cite{Guidon-09}. As a consequence of Bloch's theorem, the index $j$ in Eq.~(\ref{eq:vxopen2}) is replaced by a band index $n$ and a continuous crystal momentum $\mathbf{k}$ within the first Brillouin zone.

\begin{figure}[htb]
\centering
  \includegraphics[width=\linewidth]{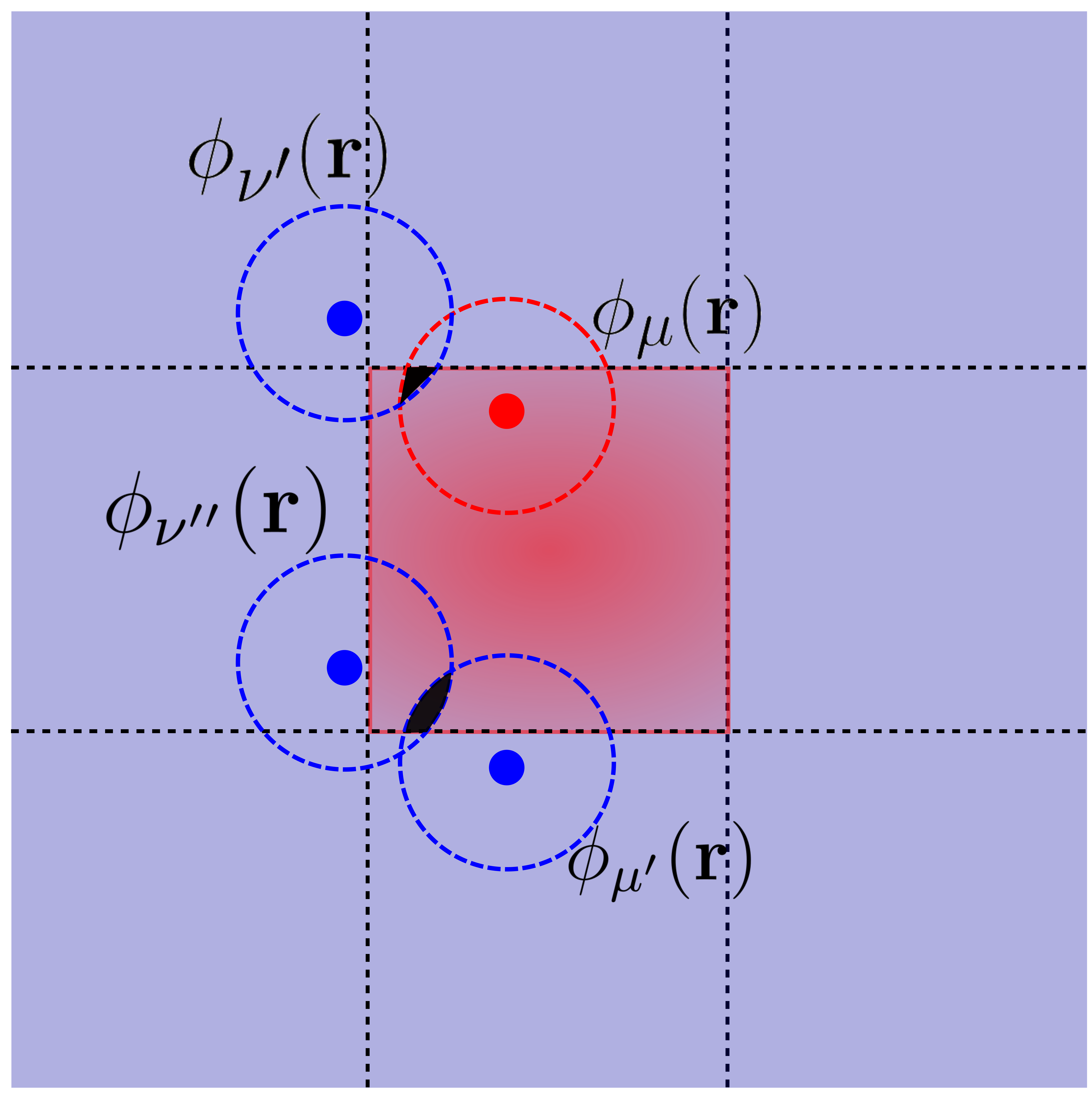} 
\caption{\label{fig:supercell} Supercell used to handle the Brillouin zone sampling in {\sc siesta}. Unit cells is marked by red. Supercell is delimited by the blue box. Notation for the different orbitals explained in the main text.}
\end{figure}

Periodic boundary conditions are incorporated via the existing \textsc{siesta} infrastructure, which manages the $\mathbf{k}$-point sampling and the Bloch phase factors. Here we replicate the algorithm explained in Sec. 8 of Ref.~\cite{SIESTA}. To enable an efficient implementation of this sampling, an auxiliary supercell (blue box in Fig.~\ref{fig:supercell}) is constructed.
This supercell, that comprises the unit cell itself (red box in Fig.~\ref{fig:supercell}) is large enough to contain all the atoms whose basis orbitals are nonzero at any grid points of the unit cell, or which overlap with any of the basis orbitals in it~\cite{SIESTA}.
The size of the supercell highly affects the efficiency of the hybrid calculation in {\sc siesta}, as will be examined in Sec.~\ref{sec:range}.
Then, the nonzero two-centre integrals between the unit-cell basis orbitals and supercell orbital are computed, without any complex phase factors (see red-dashed region in Fig.~\ref{fig:supercell}).
We also calculate the grid integrals between \emph{all} the supercell basis orbitals $\phi_{\mu^\prime}$ and $\phi_{\nu^{\prime\prime}}$ (primed indices run over all the supercell), but \emph{within the unit cell only}.
We accumulate these integrals in the corresponding matrix elements making use of the relationship

\begin{equation}
   \langle \phi_{\mu} \vert V(\mathbf{r}) \vert \phi_{\nu^\prime} \rangle = \sum_{\left( \mu^\prime \nu^{\prime \prime} \right) \equiv \left( \mu \nu^{\prime} \right) } \langle \phi_{\mu^\prime} \vert V(\mathbf{r}) f(\mathbf{r}) \vert \phi_{\nu^{\prime\prime}} \rangle,
   \label{eq:k-point-sampling}
\end{equation}

\noindent with $f(\mathbf{r})=1$ for $\mathbf{r}$ in the unit cell and is zero otherwise. $\phi_{\mu}$ is within the unit cell. The notation $\mu^\prime \equiv \mu$ indicates that $\phi_{\mu^\prime}$ and $\phi_{\mu}$ are equivalent orbitals, related by a lattice vector translation (i.e. they are periodic images). 
$\left( \mu^\prime \nu^{\prime \prime} \right) \equiv \left( \mu \nu^{\prime} \right)$ means that the sum extends over all pairs of supercell orbitals $\phi_{\mu^\prime}$ and $\phi_{\nu^{\prime\prime}}$  such that $\mu^\prime \equiv \mu$ and $\nu^{\prime\prime} \equiv \nu^{\prime}$, and the relative vector between their centers remain the same.

In particular, the exchange matrix element between two basis functions $\phi_{\mu}$ and $\phi_{\eta'}$ becomes

\begin{widetext}
\begin{equation}
    \langle \phi_{\mu} \vert \hat{V}^{\rm HFX} \vert \phi_{\eta'} \rangle = 
    -\sum_{n} \int \frac{d\mathbf{k}}{(2\pi)^{3}} n_{n \mathbf{k}}^{\sigma} 
    \left( \int \int \frac{ \phi_{\mu}^{\ast} (\mathbf{r}) \psi_{n\mathbf{k}}^{\sigma \ast}(\mathbf{r}^\prime) \phi_{\eta'}(\mathbf{r}^\prime) \psi_{n\mathbf{k}}^{\sigma} (\mathbf{r})}
    {\vert \mathbf{r}-\mathbf{r}^{\prime}\vert } \, d\mathbf{r} \, d\mathbf{r}^{\prime} \right).
    \label{eq:xsolids}
\end{equation}
\end{widetext}

Once all the real Hamiltonian matrix elements are calculated, we multiply them at every $\mathbf{k}$-point, by the corresponding phase factors and accumulate them by folding the supercell orbitals to its unit cell counterparts,

\begin{equation}
   H_{\mu \nu} (\mathbf{k}) = \sum_{\nu^{\prime} \equiv \nu} H_{\mu \nu^{\prime}} e^{i \mathbf{k} \cdot ( \mathbf{R}_{\nu^\prime} - \mathbf{R}_{\mu} )},
\end{equation}

\noindent where $\mu$ amnd $\nu$ are within the unit cell.
The resulting $N\times N$ complex eigenvalue problem, with $N$ the number of orbitals in the unit cell, is then solved at
every $\mathbf{k}$ point, finding the Bloch-state eigenfunctions and eigenvalues.

\noindent
In \textsc{siesta}, the eigenstates $\psi_{n\mathbf{k}}^{\sigma}$, whose occupations are denoted by $n_{n \mathbf{k}}^{\sigma}$, are expanded onto a basis of NAOs. Following the notation of Ref.~\cite{SIESTA}, the generalization of Eq.~(\ref{eq:expnao}) for periodic systems reads

\begin{equation}
    \psi_{n \mathbf{k}}^{\sigma} (\mathbf{r}) = 
    \sum_{\nu'} e^{i\mathbf{k} \cdot \mathbf{R}_{\nu'}} 
    \phi_{\nu'} (\mathbf{r}) \, c_{\nu' n}^{\sigma}(\mathbf{k}),
    \label{eq:expansionNAOsolid}
\end{equation}

\noindent
where the index $\nu'$ runs over all orbitals in real space, centered at $\mathbf{R}_{\nu'}$. In this case, $c_{\nu' n}^{\sigma} = c_{\nu n}^{\sigma}$, and the Bloch function $\psi_{n \mathbf{k}}^{\sigma}$ is normalized within the unit cell.

Substituting Eq.~(\ref{eq:expansionNAOsolid}) into Eq.~(\ref{eq:xsolids}), we obtain

\begin{equation}
    \langle \phi_{\mu} \vert \hat{V}^{\rm HFX} \vert \phi_{\eta'} \rangle = 
    -\sum_{\nu' \lambda'} \rho_{\nu' \lambda'}^{\sigma \sigma} 
    \left( \mu \nu' \vert \lambda' \eta' \right),
    \label{eq:xsolids-2}
\end{equation}

\noindent
where the density matrix $\rho_{\nu' \lambda'}^{\sigma \sigma}$ is the periodic counterpart of Eq.~(\ref{eq:dm}),

\begin{equation}
  \rho_{\nu' \lambda'}^{\sigma \sigma} = \sum_{n} \int \frac{d\mathbf{k}}{(2\pi)^{3}} n_{n \mathbf{k}}^{\sigma}
  e^{i \mathbf{k} \cdot ( \mathbf{R}_{\nu'} - \mathbf{R}_{\lambda'} ) }
  c_{\nu' n}^{\sigma}(\mathbf{k}) \, c_{n \lambda'}^{\sigma}(\mathbf{k}).
\end{equation}

\noindent
The four-center Coulomb integrals $\left( \mu \nu' \vert \lambda' \eta' \right)$ are evaluated in real space using the \textsc{libint} library. This approach is particularly effective for range-separated hybrid functionals such as HSE06, where the screening introduced by the error function naturally truncates long-range interactions, eliminating the need for Ewald summation or similar techniques.

The resulting exchange matrix is combined with the corresponding local or semi-local exchange-correlation contribution, weighted by the fraction specified in the chosen hybrid functional. This total exchange-correlation potential is then added to the remaining components of the Kohn–Sham Hamiltonian: the kinetic energy, the nonlocal pseudopotential, the neutral atom potential, and the Hartree potential derived from the deformation density, as defined in Eq.~(16) of Ref.~\cite{SIESTA}.

Once the matrix elements of the Hamiltonian in real space between a unit-cell orbital $\phi_{\mu}$ and a supercell orbital $\phi_{\lambda'}$ are computed, the reciprocal-space Hamiltonian is constructed for each $\mathbf{k}'$-point via Fourier summation,

\begin{equation}
    H_{\mu \lambda} (\mathbf{k}') = \sum_{\lambda' \equiv \lambda} H_{\mu \lambda'} 
    e^{i \mathbf{k}' \cdot ( \mathbf{R}_{\lambda'} - \mathbf{R}_{\mu} )},
\end{equation}

\noindent
where both $\phi_{\mu}$ and $\phi_{\lambda}$ are orbitals centered in the unit cell.

 \section{From Numerical Atomic Orbitals to Gaussian Type Orbitals}
\label{sec:nao2gto}

For the ERIs integrals, the four basis functions may be located at one, two, three, or four different atomic centers. Most modern quantum chemistry codes compute these time-consuming four-center integrals using a basis of primitive Cartesian GTOs, as introduced by Boys~\cite{Boys-50}. These are functions centered on a given position $\mathbf{R}_{I}$, whose radial dependency is given by a Gaussian function, and whose angular dependency is written in terms of Cartesian coordinates instead of polar spherical harmonics. Their general expression is

\begin{align}
  G_{k,I,l,a_{x},a_{y},a_{z}}(\mathbf{r}) = &(x-R_{Ix})^{a_x}(y-R_{Iy})^{a_y}(z-R_{Iz})^{a_z}
  \nonumber \\
  & \times\exp[-\alpha_k
  (\mathbf{r}-\mathbf{R}_{I})^2],
  \label{eq:cgaus}
\end{align}

\noindent where $(R_{Ix}, R_{Iy}, R_{Iz})$ are the cartesian coordinates of the center of the
Gaussians. $a_{x},a_{y},a_{z}$ are non-negative integers such that their sum, $l=a_{x}+a_{y}+a_{z}$, equals the angular momentum of the basis orbital.
When $a_{x}+a_{y}+a_{z} = 0$ is an $s$-type Gaussian;
when $a_{x}+a_{y}+a_{z} = 1$ is a $p$-type Gaussian;
when $a_{x}+a_{y}+a_{z} = 2$ is a $d$-type Gaussian and so on.

Indeed, this choice presents several advantages. The most important one is that the product of two Gaussians located at two different positions ($\mathbf{R}_{A}$ and $\mathbf{R}_{B}$) with different exponents ($\alpha_{A}$ and $\alpha_{B}$) can be written as a single Gaussian located at an intermediate position $\mathbf{R}_{C}$ between the two original. This allows compact formulas for all types of one- and two-electron integrals to be derived~\cite{Jensen-06}, and efficient libraries~\cite{Libint} have been implemented to compute them.

However, the GTO basis set given in Eq.~(\ref{eq:cgaus}) is not the same basis as strictly-localized NAOs used in \siesta. An intermediate step must be taken in order to adapt the two kinds of basis sets. 
To set up the notation, we briefly summarize the main characteristics of the \siesta\ basis set  
(Sec.~\ref{sec:basis-siesta}), and then devote the next two subsections to describe the adaptation of both the
radial (Sec.~\ref{sec:radialpart}) and the angular (Sec.~\ref{sec:angularpart}) parts to a basis of GTO.

At this point, we stress that the use of Gaussian-type orbitals in the present work should not be interpreted as a claim of superiority over alternative strategies based on numerical atomic orbitals or resolution-of-identity techniques. Rather, it represents a methodological choice motivated by the extensive prior success of Gaussian-based formulations for hybrid-functional and Hartree–Fock exchange calculations in several electronic-structure codes, as well as by our own previous experience with related approaches. The focus of this section is therefore on the description and implementation of one specific, well-established methodology, as realized within the {\sc siesta} framework.

\subsection{Primer on the basis of strictly localized numerical atomic orbitals}
\label{sec:basis-siesta}

In \siesta\  the atomic basis orbitals are the product of a strictly confined numerical radial function and a 
real spherical harmonic. Following the notation of Ref.~\cite{SIESTA},
for atom $I$, located at $\mathbf{R}_{I}$,

\begin{equation}
    \phi_{Ilm\zeta}(\mathbf{r}) = R_{Il\zeta}(r_{I}) Y_{lm}(\hat{\mathbf{r}}_{I})
    \label{eq:nao}
\end{equation}

\noindent where $\mathbf{r}_{I} = \mathbf{r} - \mathbf{R}_{I}$, 
$r_{I} = \vert \mathbf{r}_{I} \vert$, and
$\hat{\mathbf{r}}_{I} = \mathbf{r}_{I} / r_{I} $.
The angular momentum (labeled $l,m$) may be arbitrarily large and,
in general, there will be several orbitals (labeled by index $\zeta$)
with the same angular dependence but different radial shape,
which is conventionally called a ``multiple-$\zeta$'' basis.

Since the real spherical harmonics are not analytic at the origin, 
typically Eq.~(\ref{eq:nao}) is rewritten as

\begin{align}
    \phi_{Ilm\zeta}(\mathbf{r}) & = \left(  \frac{R_{Il\zeta}(r_{I})}{r_{I}^{l}} \right) \left[ r_{I}^{l} Y_{lm}(\hat{\mathbf{r}}_{I}) \right]
    \nonumber \\
    & = \phi_{Il\zeta}(r_{I})  \left[ r_{I}^{l} Y_{lm}(\hat{\mathbf{r}}_{I}) \right].
    \label{eq:nao-rl}
\end{align}

\noindent The radial function $\phi_{Il\zeta}(r) = R_{Il\zeta} / r^{l}$ is numerically tabulated in a fine linear radial mesh,
and interpolated afterwards by cubic splines for the required distances within the code.
Each radial function may have a different cutoff radius up to which its shape is completely free and can be introduced by
the user in an input file.
Beyond that radius, the radial part of the atomic orbital strictly vanishes.
Although the shape of the radial part is in principle totally arbitrary, the accumulated experience has proven that the numerical solution
of the Schr{\"o}dinger equation for a (confined) isolated atom with the corresponding pseudopotential is a 
very good choice in terms of accuracy versus computational cost.

The use of these finite support basis functions is combined with a discretization of space in an 
auxiliary real-space grid for the efficient representation of charge densities
and potentials, as well as the computation of the matrix elements of
the Hamiltonian that cannot be handled as two-center integrals.
These key ingredients, through the optimized handling of sparse matrices, 
are used  to compute the self-consistent Hamiltonian and overlap matrices
with a computational expense that scales linearly with system size. 

\subsection{Adapting the radial part: fitting the finite-range numerical radial functions as a linear combination of Gaussians}
\label{sec:radialpart}

%
Alongside Gaussian-based approaches, alternative strategies for the evaluation of exact exchange with numerical atomic orbitals are based on real-space formulations and Poisson solvers, and have been successfully implemented and optimized in linear-scaling electronic-structure codes. In particular, hybrid-functional calculations within the {\sc conquest} framework have been described in detail by Nakata {it et al.}~\cite{Nakata-20}. The approach adopted in the present work follows a different strategy, relying instead on a Gaussian auxiliary representation that enables the use of analytic integral engines and efficient screening within the {\sc siesta} infrastructure.
A strategy to reduce the computational cost for the computation of the four center integrals that appear in the nonlocal Hartree-Fock-type exchange hybrid functionals is to take advantage
of the properties of GTOs for the efficient analytical computation of the electron repulsion integrals,
as done in \textsc{gaussian}~\cite{Gaussian,Gaussian-web}, \textsc{crystal}~\cite{CRYSTAL,Dovesi-20,Crystal-web}, \textsc{cp2k}~\cite{Hutter-14,Guidon-08,kuhne2020cp2k,iannuzzi2025cp2k,CP2K-web},
\textsc{q-chem}~\cite{Lee-22}, or \textsc{pyscf}~\cite{Sun-22} packages, among others. 
Therefore, in order to get the best of both worlds, we approximate the radial part of the NAO
by a linear combination of Gaussians,

\begin{equation}
    \chi_{\mu}(r) = \sum_{k=1}^{N_{\rm G}}  C_{k} (\mu) \exp \left[ -\alpha_{k} (\mu) r^{2} \right],
    \label{eq:expgaus}
\end{equation}

\noindent where we use the compact notation 
$\mu \equiv \lbrace Ilm\zeta \rbrace$, and we have assumed that the
orbital is centered at the origin. 
The maximum number of Gaussians to be used in the expansion is a parameter defined by the user.
An intrinsic limitation of Gaussian representations is their inability to enforce strict spatial localization, since Gaussians extend to infinity by construction; as a result, the strictly localized behavior of NAOs can only be approximated near the cutoff, motivating the fitting and truncation strategies described below.
The parameters $C_{k}$ and $\alpha_{k}$ are determined by minimizing the goal function $G$, defined as the
residual sum over the points in the linear mesh
of the squares of the difference between the linear combination of Gaussians and the radial part of the NAO,

\begin{equation}\label{eq:errores}
    G = \sum_{i} \left[ \chi_{\mu} (r_{i}) - \phi_{\mu} (r_{i}) \right]^{2}. 
\end{equation}

\noindent In Eq.~(\ref{eq:errores}) the same weight is given to all the points in the grid.
In practical calculations, it is conceivable that this error is minimized, yet at some
particular small range of points the fitting might be poor. 
In order to avoid such behavior, some weights are imposed in the measurement of the error,
and instead of minimizing Eq.~\eqref{eq:errores}, we optimize

\begin{equation}\label{eq:errores2}
    \tilde{G} = \sum_{i} \left[ w(r_{i}) (\chi_{\mu} (r_{i}) - \phi_{\mu} (r_{i}) )\right]^{2},
\end{equation}

\noindent where $w(r_{i})$ is the weight attributed to a the $i-$th point in the linear grid. 
It takes values between $1$ and $0$, and higher $w (r_{i})$ are given to points $i$ 
where $|\chi_{\mu} (r_{i}) - \phi_{\mu} (r_{i})|$ is large. 

Several algorithms have been applied to solve this nonlinear fitting problem.
The Levenberg-Marquardt algorithm was used in Ref.~\cite{Shang-JCP}.
However, this method can only perform local minima searches and, therefore, might be trapped in one of them and cause fitting errors.
One of them is their tendency to
overfit, which typically leads to suboptimal solutions containing clusters of
nearly-identical Gaussians with coefficients several orders of magnitude
higher than the other ones. 
In order to overcome this problem, previous approaches have combined constrained genetic algorithms~\cite{global-min-1,Conn-91} which makes a global search that will be used as an initial guess for subsequent constrained local minimal search~\cite{Coleman-94,Coleman-96}. While this strategy is robust, it relies on external optimization packages (such as {\sc matlab}), increasing the complexity of the workflow for the end user and complicating long-term code maintenance.
For this reason, we propose here a new and simple fitting algorithm to determine the coefficients and exponents of the Gaussian expansion in Eq.~(\ref{eq:expgaus}), which is fully implemented within {\sc siesta} and has been proven to work reliably for the intended applications.

%


Let us assume that the values of the radial part of the atomic orbital are known at the points of a linear grid in real space,
($r_1,\dots,~r_p$), where $r_1=0$ and $\phi_{\mu} (r_{p}) = 0$. 
For the remaining part of the description, we shall omit the subscript $\mu$, 
understanding that each radial part of the NAO will have their own coefficients and exponents.
Let us assume for a moment that the exponents $\alpha_1,\dots,~\alpha_{N_{\rm G}}$ are known.
Then, the coefficients $C_1,\dots,~C_{N_{\rm G}}$ can be easily determined by solving a least-squares problem

\begin{equation}
\small
\begin{pmatrix}
      w(r_{1}) e^{-\alpha_1r_1^2} & \dots & w(r_{1}) e^{-\alpha_{N_{\rm G}}r_1^2} \\
      w(r_{2}) e^{-\alpha_1r_2^2} & \dots & w(r_{2}) e^{-\alpha_{N_{\rm G}}r_2^2} \\
      \vdots & \ddots & \vdots \\
      w(r_{p}) e^{-\alpha_1r_p^2} & \dots & w(r_{p}) e^{-\alpha_{N_{\rm G}}r_p^2}
\end{pmatrix}
\begin{pmatrix}
      C_1 \\ C_2 \\ \vdots \\ C_{N_{\rm G}}
\end{pmatrix}
=
\begin{pmatrix}
       w(r_{1}) \phi(r_1) \\ w(r_{2}) \phi(r_2) \\ \vdots \\ w(r_{p}) \phi(r_p)
\end{pmatrix}.
\label{eq:least_squares}
\end{equation}

\noindent Since the number of Gaussians in the expansion, $N_{\rm G}$, is a parameter that will strongly affect the number
of four center integrals to be computed, in a first approximation we start with a small 
(but in many cases sensible) number of $N_{\rm G} = 3$. 
This parameter will also be optimized as explained below.
We note that the number of Gaussians required for an accurate representation depends on the orbital character. In particular, previous studies have shown that a small number of Gaussians (e.g. three) may already provide an adequate description for relatively localized orbitals, such as $d$-type states, whereas more diffuse orbitals (especially $s$-type orbitals) generally require a larger number of Gaussians to properly converge their radial shape.
The solution of Eq.~(\ref{eq:least_squares}) gives, in the least-squares sense, the best set of coefficients for a given set of exponents.
Therefore, the problem reduces to determining an optimal set of Gaussian exponents. This is achieved through an alternating optimization strategy in which one parameter is optimized at a time while keeping the others fixed. The philosophy of this approach is closely related to Powell-type direct search algorithms, which reduce a multidimensional optimization problem to a sequence of one-dimensional searches~\cite{Powell-98}.

%
%

%
We start from a first guess for the exponents: $0.1,1,2,\ldots,N_{\rm G}-1$, which has shown to behave quite well in all the examples we have tested.
We note that an alternative and commonly used strategy for initializing Gaussian exponents is the use of even-tempered (geometrically spaced) sequences, which can be particularly effective when a small number of Gaussians is employed. While such an approach is certainly reasonable, the initialization strategy adopted here has been found to provide robust numerical stability across different orbital characters and chemical environments, and is therefore used throughout this work.
In the first step  we focus on $\alpha_1$ leaving the rest of exponents $\alpha_2,\dots,~\alpha_{N_{\rm G}}$ frozen, 
effectively reducing the complexity to a one-dimensional minimization problem:
find the best $\alpha_1$ that minimizes the error given by the goal function of 
Eq.~\eqref{eq:errores2}.
Once we have optimized $\alpha_1$, the procedure is repeated for $\alpha_2$, keeping constant $\alpha_1,\alpha_3,\dots,~\alpha_{N_{\rm G}}$. 
Then, we replicate this strategy running through all the exponents.

In order to minimize the one dimensional function for a given $\alpha_i$,
the interval $\left[ \alpha_{i-1}\cdot \delta \gamma,\frac{\alpha_{i+1}}{\delta \gamma}\right]$ is split into one hundred equally spaced points 
(in the case of $\alpha_1$ and $\alpha_{N_G}$, we impose global minimal and maximal bounds in the extremes).
$\delta \gamma$ is an adimensional parameter designed to avoid similar exponents for
two consecutive Gaussians, a fact that would result in numerical instabilities
in the solid-state code. 
For practical calculations, we have found that $\delta \gamma = 1.4$ is 
a sensible value.
Then, Eq.~(\ref{eq:least_squares}) is solved for all $\alpha_{i}$ within the interval,
the error defined in Eq.~\eqref{eq:errores2} is computed,
and the value that minimizes this deviation is selected.
We then repeat this procedure two more times, selecting the interval
for $\alpha_{i}$ where we evaluate the function in a neighborhood where the minimum of the previous minimization was found. 

The whole process of the $\alpha$ optimization is iterated until
the maximum change in the exponents between two consecutive steps, or the 
maximum difference between the linear combination of Gaussians and the radial part 
of the NAO at any point in the grid are below some user defined thresholds.
If one of these conditions is not fulfilled after a certain number
of self-consistent steps, the number of Gaussians in the linear fitting is 
increased by one and the full process is repeated. 

\begin{figure}[htb]
\centering
  \includegraphics[width=\linewidth]{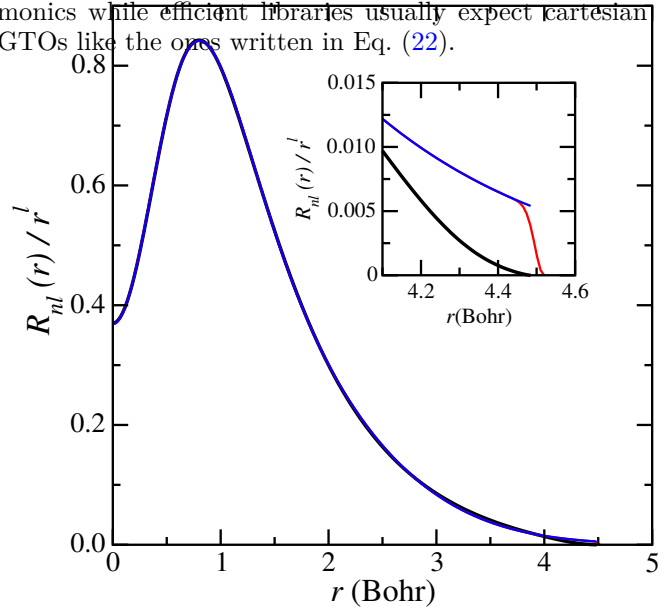} 
\caption{\label{fig:fitting-orb} Gaussian fits of the radial part of the first-$\zeta$ of the carbon 2$s$ orbital Gaussian functions. The black line represents the strictly localized native basis orbital in {\sc siesta}. The blue line shows the expansion of the NAO as a linear combination of Gaussians, given in Eq.~(\ref{eq:expgaus}) with $N_{\rm G} = 5$. These Gaussians decay exponentially but do not vanish at $r_{\rm c}$. The red line represents the product of the Gaussian combination with a cutting function, ensuring the atomic orbitals remain continuous and strictly localized within $r_{\rm c}^{\rm G}$, determined using $\epsilon_{\rm Gauss} = 5\times 10^{-3}$. The three radial parts are nearly indistinguishable to the naked eye. The inset provides a zoomed-in view around the cutoff radius.}
\end{figure}

Once the fitting procedure has finished, the native radial part of the atomic orbitals is \emph{replaced} by the optimal linear combinations of Gaussians given in Eq.~(\ref{eq:expgaus}).
Here we must remember that \textsc{siesta} requires strictly localized atomic orbitals, while the Gaussians do not vanish at any finite distance. 
To produce strictly localized orbitals, the contracted sum of Gaussians are cut at a given cutoff radius $r_{\rm c}^{\rm G}$, determined as the point where the product of the sum of Gaussians times $r^{l}$ is smaller than a user-defined threshold, $\epsilon_{\rm Gauss}$. 
A smaller threshold will result in a larger cutoff radius for each contracted sum of Gaussians. 
This produces discontinuous radial functions with a kink at the cutoff distance. 
Following ideas from Ref.~\cite{Elsasser-90}, the atomic orbital is multiplied by $1- \exp\left[-\alpha(r-r_{\rm c}^{\rm G})^{2} \right]$ for $r < r_{c}^{\rm G}$ and zero otherwise. 
This scheme does provide strict localization beyond $r_{\rm c}^{\rm G}$ and the continuous radial part of the atomic orbitals. 
However, it introduces a small bump close to $r_{\rm c}^{\rm G}$ as shown in Fig.~\ref{fig:fitting-orb}. 
This strictly localized basis set is the one chosen to compute the kinetic, non-local pseudopotential, Hartree and semi-local exchange and correlation terms of the Hamiltonian matrix elements.
Only terms related to exact exchange are computed with the uncut sum of Gaussians.
According to the experience gained in Ref.~\cite{Qin-23}, this improves the self-consistence convergence in the hybrid calculations.

Bringing together Eq.~(\ref{eq:nao-rl}) and Eq.~(\ref{eq:expgaus}), we arrive at the following expression for one
of the atomic orbitals of the \siesta\ basis sets 

\begin{align}
       \phi_{Ilm\zeta}(\mathbf{r}) & = \phi_{Il\zeta}(r_{I})  \left[ r_{I}^{l} Y_{lm}(\hat{\mathbf{r}}_{I}) \right]
        \nonumber \\
        & \approx \left( \sum_{k= 1}^{N_{\rm G}}  C_{k} (Il\zeta) \exp \left[ -\alpha_{k} (Il\zeta) r_{I}^{2} \right] \right) 
         \left[ r_{I}^{l} Y_{lm}(\hat{\mathbf{r}}_{I}) \right] 
         \nonumber \\
         & = \sum_{k = 1}^{N_{\rm G}}  C_{k} (Il\zeta) \tilde{G}_{k, I, l, m, \zeta} (\bf{r}),
    \label{eq:nao-sgaus}
\end{align}

\noindent where $\tilde{G}_{k, I, l, m, \zeta} (\bf{r})$ are real spherical harmonic Gaussians given by

\begin{equation}
    \tilde{G}_{k, I, l, m, \zeta} ({\mathbf r}) = r_{I}^{l} \exp \left[ -\alpha_{k} (Il\zeta) r_{I}^{2} \right] Y_{lm}(\hat{\mathbf{r}}_{I}),
    \label{eq:sgaus}
\end{equation}

\noindent and the $Y_{lm}(\hat{\mathbf{r}}_{I})$ correspond to real spherical harmonics. 


\subsection{Adapting the angular part: from the spherical harmonics to Cartesian Gaussian harmonics}
\label{sec:angularpart}

The second transformation required to link \siesta\ with the efficient libraries that compute the four-center integrals is related to the angular dependency of the basis set orbitals.
As already shown in Eq.~(\ref{eq:nao}) the basis set in \siesta\ is expressed in terms of real spherical harmonics while efficient libraries usually expect cartesian GTOs like the ones written in Eq.~(\ref{eq:cgaus}).

The difference between the real spherical harmonics and the Cartesian harmonic functions is the number of atomic orbitals including in the shells whose angular momentum are larger than $l=1$. 
For example, a $d$-shell has five linearly independent and orthogonal real spherical harmonic orbitals (usually labeled $xy$, $yz$, $xz$, $3z^{2}-r^{2}$, and $x^{2}-y^{2}$), but has six Cartesian orbitals ($xy$, $yz$, $xz$, $x^{2}$, $y^{2}$, and $z^{2}$), although they are not linearly independent (for instance, a linear combination like $x^{2} + y^{2} + z^{2}$ actually gives a function of $s$ symmetry).
In general, for a shell of angular momentum $l$ there will be $(2l+1)$ real spherical harmonics, but $(l+1)(l+2)/2$ Cartesian GTOs. 
A detailed transformation method can be found in Ref.~\cite{Schlegel-95}, where the spherical harmonic Gaussian functions of Eq.~(\ref{eq:sgaus}) are contracted from linear combinations of cartesian GTOs given in Eq.~(\ref{eq:cgaus}),

\begin{widetext}
\begin{equation}
    \tilde{G}_{k,I,l,m,\zeta} (\mathbf{r}) = 
    \sum_{a_{x}+a_{y}+a_{z}=l} c(l,m,a_x,a_y,a_z) G_{k, I, l, a_x,a_y,a_z,\zeta} (\mathbf{r}).
    \label{eq:fromcar2sph}
\end{equation}
\end{widetext}

\noindent Introducing this expansion in Eq.~(\ref{eq:nao-sgaus}) we arrive at the final contraction that approximates the NAOs used in \siesta\ from primitive Cartesian GTOs

\begin{widetext}
\begin{align}
       \phi_{Ilm\zeta}(\mathbf{r}) & \approx \sum_{k = 1}^{N_{\rm G}} \sum_{a_{x}+a_{y}+a_{z}=l}
       D_{k, I,l,m,a_{x},a_{y},a_{z},\zeta} \:\: G_{k, I, a_x,a_y,a_z,\zeta} (\mathbf{r}),
    \label{eq:nao-cgaus}
\end{align}
\end{widetext}

\noindent where $D_{k, I,l,m,a_{x},a_{y},a_{z},\zeta} = C_{k} (Il\zeta) \times c(l,m,a_x,a_y,a_z)$.
For example, the expansion of an orbital of the $d$-shell ($l$=2), with $N_{\rm G} = 6$ Gaussian functions
in the approximation of the radial part requires 36 primitive Cartesian GTOs.
Using the compact notation $\mu \equiv \lbrace Ilm\zeta \rbrace$, and $a \equiv \lbrace k, a_x,a_y,a_z \rbrace$, then
Eq.~(\ref{eq:nao-cgaus}) can be written as

\begin{align}
       \phi_{\mu}(\mathbf{r}) & \approx \sum_{a} D_{\mu a} G_{\mu a} (\mathbf{r}). 
    \label{eq:nao-cgaus-2}
\end{align}

Replacing the contractions given in Eq.~(\ref{eq:nao-cgaus-2}) in the four-center integrals of Eq.~(\ref{eq:four-center-int}), we arrive at

\begin{align}
    \left( \mu \nu \vert \lambda \eta \right) & \equiv
     \int \int \frac{\phi_{\mu}^{\ast}(  \mathbf{r} ) \phi_{\lambda}^{\ast}(\mathbf{r}^\prime)
    \phi_{\eta} (\mathbf{r}^\prime)\phi_{\nu} (\mathbf{r}) }
     {\vert \mathbf{r}-\mathbf{r}^{\prime}\vert  } 
     d\mathbf{r}^{\prime} d\mathbf{r} 
     \nonumber \\
     & = \sum_{a,b,c,d} 
     D_{a \mu} D_{c \lambda} D_{ \eta d} D_{\nu b} 
     \nonumber \\
     & \quad \times \int \int \frac{G_{\mu a}^{\ast}(  \mathbf{r} ) G_{\lambda c}^{\ast}(\mathbf{r}^\prime)
     G_{\eta d} (\mathbf{r}^\prime) G_{\nu b} (\mathbf{r}) }
     {\vert \mathbf{r}-\mathbf{r}^{\prime}\vert  } 
     d\mathbf{r}^{\prime} d\mathbf{r} 
     \nonumber \\
     & = \sum_{a,b,c,d} 
     D_{a \mu} D_{c \lambda} D_{\eta c} D_{\nu b} 
     \left[ \mu a, \nu b \vert \lambda c, \eta d \right],
     \label{eq:four-center-int-contract}
\end{align}

\noindent where $D_{a \mu}$ is the conjugate transpose of $D_{\mu a}$, and the primitive four-center electron-repulsion integrals between primitive Cartesian GTOs are given by $\left[ \mu a, \nu b \vert \lambda c, \eta d \right]$

After this transformation, the Cartesian GTOs are grouped into shells, 
according to the NAO's angular momentum they fit.
Thus, if $\mu \in I$, $\nu \in J$, $\lambda \in K$, and $\eta \in L$, for the 
$I, J, K, L$ shell quartet, then all the integrals $\left( \mu \nu \vert \lambda \eta \right)$ are computed together for one shell quartet at a time~\cite{Shang-20}. 

\section{Computation of the relevant four-center integrals}
\label{sec:computation-fourcenter}

As discussed in Sec.~\ref{sec:theory}, the evaluation of the exact-exchange potential matrix elements [Eqs.~(\ref{eq:VHFXmueta})--(\ref{eq:xsolids-2})] and total energy [Eq.~(\ref{eq:exunres})] requires the computation of four-center ERIs [Eq.~(\ref{eq:four-center-int})]. These involve the Coulomb interaction between two products of overlapping atomic orbitals: $(\mu \nu')$ evaluated at point $\mathbf{r}$ and $(\eta' \lambda')$ at point $\mathbf{r}'$. Owing to the long-range nature of the Coulomb kernel, $1/|\mathbf{r} - \mathbf{r}'|$, these orbital pairs can be spatially distant, making a direct computation of all such integrals computationally prohibitive.

To render the calculation tractable, a sequence of approximations is employed, designed to control the trade-off between computational efficiency and accuracy. The goal is to restrict the number of integrals evaluated without compromising the fidelity of the total energy, forces, or Hamiltonian matrix elements.

The first level of screening limits the orbital pairs $(\mu, \nu')$ considered. Beyond the basic requirement of spatial overlap, the Schwarz inequality~\cite{Haser-89} is imposed to bound the magnitude of each four-center integral,

\begin{equation}
    |(\mu \nu' | \eta' \lambda')| \leq 
    \sqrt{(\mu \nu' | \mu \nu')(\eta' \lambda' | \eta' \lambda')}.
    \label{eq:schwarz}
\end{equation}

\noindent Defining $M_{\rm e}$ as the maximum of the square roots of all two-center ERIs,

\begin{equation}
    M_{\rm e} = \max \sqrt{(\mu \nu' | \mu \nu')},
\end{equation}

\noindent the inequality becomes

\begin{equation}
    |(\mu \nu' | \eta' \lambda')| \leq 
    \sqrt{(\mu \nu' | \mu \nu') M_{\rm e}}.
    \label{eq:schwarz-2}
\end{equation}

\noindent Only pairs satisfying

\begin{equation}
    \sqrt{(\mu \nu' | \mu \nu') M_{\rm e}} \geq \epsilon_{\rm pair-list}
    \label{eq:schwarz-pairlist}
\end{equation}

\noindent are retained, where $\epsilon_{\rm pair-list}$ is a user-defined threshold~\cite{Shang-JCP}.

Atomic orbitals are grouped into shells according to their angular momentum quantum number. Two orbitals $\mu_1 \equiv (I, l, m, \zeta)$ and $\mu_2 \equiv (I, l, m', \zeta)$ that belong to the same atom $I$, share the same angular momentum $l$ and zeta index $\zeta$, but differ in magnetic quantum number $m$, are treated as part of the same shell. Four such shells define a shell quartet $(I,J,K,L)$ associated with the orbitals in Eq.~(\ref{eq:four-center-int}), where $\mu \in I$, $\nu' \in J$, $\eta' \in K$, and $\lambda' \in L$. All integrals belonging to a given shell quartet are computed collectively~\cite{Shang-20}.

This approximation leads to a compact list of interacting shell pairs $(IJ | IJ)$ satisfying $(IJ | IJ) \geq \epsilon_{\rm pair-list}$. Here, the index $I$ runs through the shells in the unit cell, while $J$ spans shells of neighboring orbitals. Shell pairs unlikely to contribute significantly are thus excluded from further consideration.

A second filtering step restricts the spatial domain of the remaining orbital pairs $(\eta', \lambda')$ to the auxiliary supercell used in {\sc siesta} for $\mathbf{k}$-point sampling, as defined in Sec.~\ref{sec:Brillouin}. The filtered list is expanded to include $(KL | KL)$, satisfying the same threshold, but with $K$ and $L$ running through the whole supercell. If $N_{IJ}$ and $N_{KL}$ are the numbers of selected shell pairs in the unit cell and supercell respectively, the maximum number of shell quartets to be considered is $N_{IJ} \times N_{KL}$.

A third approximation retains only those shell quartets where at least one of the orbitals centered at $\mathbf{r}$ overlaps directly with one at $\mathbf{r}^\prime$. This connectivity condition ensures that the sparsity pattern of the hybrid-functional Hamiltonian matches that of the semi-local functional. This approximation is particularly justified in the context of range-separated hybrids, where the Coulomb operator is split into short- and long-range parts [Eq.~(\ref{eq:split})], and only the short-range component is included in the exchange potential. Consequently, the resulting ERI matrix is considerably sparser than that of the full Coulomb operator~\cite{Izmaylov-06}.

Then, since each matrix element of the exchange potential involves a product of an ERI with a density matrix element [Eq.~(\ref{eq:xsolids-2})], a further refinement is applied. 
A large ERI can contribute in a negligible way to the energy or potential if the corresponding matrix elements between the orbitals involved in the four-center integrals are fairly small.
Following Izmaylov \textit{et al.}~\cite{Izmaylov-06}, a screening parameter $D_{\rm screening}$ is defined as the maximum value among all density matrix elements connecting the shell quartets,

\begin{equation}
    D_{\rm screening} = \max\left( |D_{\max}^{IK}|, |D_{\max}^{IL}|, |D_{\max}^{JK}|, |D_{\max}^{JL}| \right),
\end{equation}

\noindent with $\mu \in I$, $\nu^\prime \in J$, $\lambda^\prime \in K$, and $\eta^\prime \in L$. These values are obtained from the density matrix of the previous SCF cycle to ensure stability~\cite{Shang-20}. Each estimated Schwarz upper bound [Eq.~(\ref{eq:schwarz-2})] is then multiplied by $D_{\rm screening}$, and only if 

\begin{equation}
   D_{\rm screening} \times \vert \left( \mu \nu^{\prime} \vert \mu \nu^{\prime} \right) \vert^{1/2} \times \vert \left( \eta^\prime \lambda^{\prime} \vert \eta^\prime \lambda^{\prime} \right) \vert^{1/2} \le \epsilon_{\rm Schwarz}
   \label{eq:denscreen}
\end{equation}

\noindent then the four-center ERI is actually computed, where 
$\epsilon_{\rm Schwarz}$ is a second user-defined tolerance.

Through these successive screening steps, the number of four-center integrals is reduced from $\mathcal{O}(N^4)$ to approximately $\mathcal{O}(N^2)$, significantly improving computational efficiency while retaining sufficient accuracy for hybrid-functional calculations.

The final approximation takes advantage of the permutational symmetry of the ERIs,
\begin{align}
     &(\mu\nu^\prime|\eta^\prime\lambda^\prime) =(\mu\nu^\prime|\lambda^\prime \eta^\prime)=  
     (\nu^\prime \mu|\eta^\prime \lambda^\prime)=
     (\nu^\prime \mu|\lambda^\prime \eta^\prime)
     \nonumber \\
     =& 
     (\eta^\prime\lambda^\prime|\mu\nu^\prime) =
     (\lambda^\prime \eta^\prime|\mu\nu^\prime )=
     (\eta^\prime\lambda^\prime|\nu^\prime\mu) = 
     (\lambda^\prime \eta^\prime |\nu^\prime \mu),
\end{align}

\noindent saving a factor of eight in the number of integrals to be considered.

\begin{figure}[htb]
\centering
  \includegraphics[width=\columnwidth]{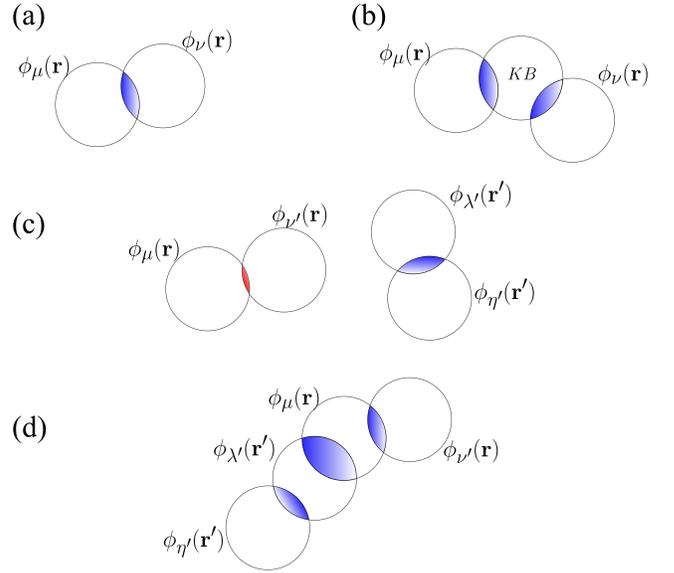} 
\caption{\label{fig:sparse} Sketch representing the sparsity of the overlap (a) and Hamiltonian (b) matrix elements in 
{\sc siesta} with only local or semi-local exchange and correlation potentials. When the exact exchange operator is considered,
then an orbital $\mu$ in the unit cell might have a non-vanishing matrix element with another orbital $\lambda^{\prime}$ 
located beyond the sum of the cutoff radii of the two orbitals (c). (d) Connectivity condition used in this work to retain the ERIs integrals that are actually computed. }
\end{figure}

%
%
%

\section{Computation of primitive ERIs}
\label{sec:libint}

%
%
%

As shown in Sec.~\ref{sec:angularpart}, the contracted ERIs appearing in Eq.~(\ref{eq:four-center-int}) can be expressed as linear combinations of primitive integrals involving only GTO [Eq.~(\ref{eq:four-center-int-contract})] 
This decomposition makes explicit the fact that each ERI in the contracted basis involves a sum over many primitive contributions, reflecting the structure of the underlying Gaussian basis expansion.
The computational challenge then reduces to the efficient evaluation of the primitive integrals $\left[ \mu a, \nu^\prime b \vert \lambda^\prime c, \eta^\prime d \right]$ for each shell quartet.

Aligning with strategies adopted in state-of-the-art hybrid DFT codes such as \textsc{cp2k}~\cite{Hutter-14,Guidon-08,kuhne2020cp2k,CP2K-web} or \textsc{honpas} with the \textsc{nao2gto} scheme~\cite{Shang-JCP,qin-2015-1, Shang-20},  we have interfaced the implementation with version~1.1.4 of the \textsc{libint} library~\cite{Libint}, a widely used engine for computing ERIs over Gaussian basis functions. \textsc{libint} provides a stable Application Programming Interface (API) tailored to many-body electronic structure calculations. 
The \textsc{libint} library is written in C and distributed under the terms of the GNU General Public License (GPLv3)~\cite{GPL}. It is actively maintained, with source code hosted on GitHub~\cite{Libint}, and is also distributed as part of the Electronic Structure Library (ESL) initiative~\cite{Oliveira-20,ESL-web}, which facilitates its integration into first-principles codes.
\textsc{libint} employs a recursive evaluation of ERIs based on the Obara–Saika algorithm~\cite{Obara-84}, complemented by the refinements introduced by Head-Gordon and Pople~\cite{Head-Gordon-88}. These methods enable the analytical computation of all required integrals and their first and second derivatives with respect to atomic positions. The use of this library ensures efficient and accurate evaluation of the primitive four-center integrals. 

%
%
%

\section{Forces}
\label{sec:forces}

Since the basis set in \textsc{siesta} depends in  the atomic positions, in the hybrid functional calculations we must calculate the HFX contribution to atomic forces.
There have been a few reports on analytical
atomic forces for periodic hybrid functional
calculations with NAOs to date~\cite{Lin-25}.
According to the recipe given in Ref.~\cite{Qin-23}, the computation of atomic forces within hybrid functional calculations follows a fully analytical approach based on the gradients of the Hartree-Fock exchange (HFX) energy with respect to atomic positions. These gradients require the evaluation of the first derivatives of the four-center electron repulsion integrals (ERIs), which are provided by the {\textsc{libint}} library in a Cartesian GTO basis.

The HFX force on an atom comprises a Pulay-type contribution due to the position dependence of NAOs, and a term involving the derivatives of the ERIs. By replacing the NAOs with their discretized fitted contracted GTO counterparts across the full hybrid functional framework, {\textsc{siesta}} avoids inconsistencies and achieves high accuracy. Furthermore, similar integral screening techniques as those used for the construction of the HFX matrix elements are employed (including Schwarz-type bounds, distance-based cutoffs, and density matrix sparsity~\cite{Guidon-08}) to minimize the number of ERI derivatives that must be computed. As a result, the HFX force calculation scales linearly with system size and can even be faster than the HFX matrix construction within the self-consistent cycle. The implementation benefits from a master-worker parallelization scheme that ensures excellent scalability and load balancing across processors.
A more detailed description on the computation of the forces can be found in Ref.~\cite{Qin-23}.
The efficient computation of the forces open the doors to perform geometry optimization and {\it ab initio} molecular dynamics simulations.
Also the polaronic behavior involving the interaction of excess of electrons with a phonon cloud becomes accessible with this implementation~\cite{Falletta-22.1,Falletta-22.2}.

\section{Results and discussion}
\label{sec:results}

\subsection{Technical details and test systems}
\label{sec:technicalities}

To assess the accuracy and performance of the hybrid functional implementation, we selected a representative set of materials with diverse bonding and structural characteristics. These include: (i) covalent semiconductors such as bulk silicon and carbon in the diamond structure; (ii) materials with mixed covalent and ionic bonding such as cubic boron nitride (c-BN); (iii) highly ionic crystals like lithium fluoride (LiF); (iv) oxides with significant ionic character, namely wurtzite ZnO and rutile TiO$_{2}$; (v) two-dimensional systems including monolayers of graphene and hexagonal boron nitride (h-BN); and (vi) black phosphorus (BP), which exhibits anisotropic covalent bonding in its orthorhombic phase. This diversity allows us to evaluate the performance of hybrid functionals under varying bonding regimes and dimensionalities.

Core electrons were replaced by {\it ab initio} norm-conserving, fully separable pseudopotentials~\cite{Kleinman-82} in the {\sc psml} format~\cite{psml}, generated using the optimized norm-conserving Vanderbilt scheme as proposed by Hamann ~\cite{norm_conserving} within the PBE flavor, and available in the Pseudo-Dojo periodic table~\cite{pseudodojo,footnotepseudo}.
Although these pseudopotentials are not strictly consistent with the hybrid exchange–correlation functional, previous systematic studies have shown that the resulting errors in structural and electronic properties are typically small (of the order of 1\%) and well controlled, making this a widely accepted and reliable approximation for hybrid-functional calculations~\cite{Yang-98}.

The one-electron Kohn–Sham eigenstates were expanded on a basis of strictly localized~\cite{Sankey-89} NAOs~\cite{Artacho-99}. These basis functions were obtained as the confined eigenfunctions of isolated pseudoatoms under the soft-confinement potential proposed in Ref.~\cite{Junquera-01}.
A double-$\zeta$ polarized basis set was used for valence states of all elements, while a single-$\zeta$ basis was applied to semicore states in Li (1$s$), Ti (3$s$, 3$p$), and Zn (3$s$, 3$p$).  The default values of all the parameters defining the shape and the range of the radial part of the native basis functions were used, ensuring a good compromise between transferability, efficiency, and softness. The only exception is h-BN, where a longer triple-$\zeta$ polarized basis set with diffuse $s$ and $p$ orbitals was employed to better describe its two-dimensional character~\cite{Garcia-Gil-09}.

The electronic density, Hartree, and exchange-correlation potentials, as well as the corresponding matrix elements between the basis orbitals, were calculated on a uniform real space grid~\cite{SIESTA}. An equivalent plane-wave cutoff of 600~Ry was used to represent the charge density. 
Reciprocal space sampling was performed using Monkhorst–Pack meshes~\cite{Monkhorst-76} commensurate with high-quality convergence: $8\times 8 \times 8$ for Si, C, LiF, and c-BN; $8\times 8 \times 12$ for TiO$_{2}$; $8\times 8 \times 6$ for ZnO; $10\times 8 \times 3$ for BP; and $18\times 18 \times 1$ for h-BN. 

The atomic positions were relaxed using the PBE functional using a conjugate-gradient algorithm until the forces and stress components fell below thresholds of 0.01 eV/$\rm{\AA}$ and 0.0002 eV/$\rm{\AA}^{3}$, respectively. The hybrid functional calculations were then performed on these PBE-relaxed geometries.

\subsection{Band structure}
\label{sec:bandstructure}

\begin{widetext}

    \begin{figure}[H]
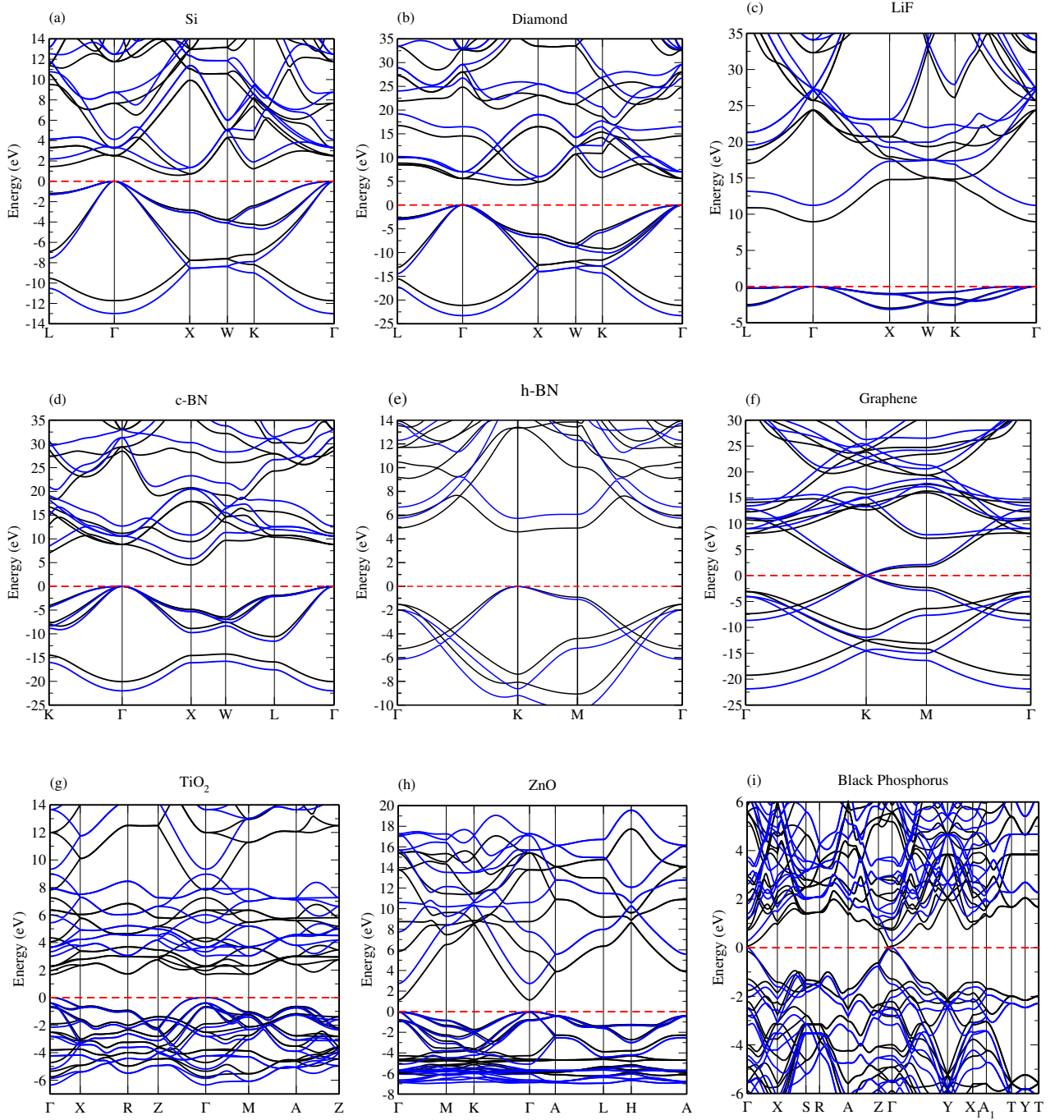

        \begin{table}[H]
            \centering
            \begin{tabular}{p{5.5cm}p{5.5cm}p{5.5cm}}
               \includegraphics[width=0.98\linewidth]{./Figures/Si.bands.eps} 
               &
               \includegraphics[width=0.98\linewidth]{./Figures/C.bands.eps} 
               &
               \includegraphics[width=0.98\linewidth]{./Figures/LiF.bands.eps}
               \vspace{0.6cm} \\ 
               \includegraphics[width=0.98\linewidth]{./Figures/c-BN.bands.eps}  
               &
               \includegraphics[width=0.98\linewidth]{./Figures/h-BN_monolayer.bands.eps} 
               &
               \includegraphics[width=0.98\linewidth]{./Figures/Graphene.bands.eps}
               \vspace{0.6cm} \\ 
               \includegraphics[width=0.98\linewidth]{./Figures/TiO2.bands.eps}  
               &
               \includegraphics[width=0.98\linewidth]{./Figures/ZnO.bands.eps} 
               &
               \includegraphics[width=0.98\linewidth]{./Figures/P.bands.eps}
               \\
            \end{tabular}
        \end{table}
        \vspace{-1cm}
        \caption{
        The electronic band structures of various materials were computed using {\sc siesta} with the PBE functional (black lines) and the HSE06 hybrid functional incorporating 25\% exact exchange (blue lines). The studied systems include:
(a) Bulk silicon (Si) in the diamond structure,
(b) Bulk carbon (C) in the diamond structure,
(c) Bulk lithium fluoride (LiF) in the rock salt structure,
(d) Bulk cubic boron nitride (c-BN) in the zinc blende structure,
(e) Monolayer hexagonal boron nitride (h-BN),
(f) Monolayer graphene,
(g) Bulk rutile titanium dioxide (TiO$_{2}$),
(h) Wurtzite zinc oxide (w-ZnO), and
(i) Black phosphorus (BP) in its orthorhombic form.
 The radial part of the native NAOs was expanded using $N_{\rm G} = 6$ Gaussian functions, with the cutoff radii determined by a tolerance of $\epsilon_{\rm Gauss} = 1 \times 10^{-5}$. To reduce computational complexity, the thresholds for four-center integral reductions were set to $1 \times 10^{-6}$.
For semiconductors and insulators, the energy zero was aligned to the top of the valence band in each case, marked with an horizontal, red-dashed line.     
        }
    \label{fig:bands-hybrids}
    \end{figure}
\end{widetext}

The electronic band structures computed with the HSE06 functional using  stringent values for all the parameters that control the accuracy of the hybrid calculations, are shown in Fig.~\ref{fig:bands-hybrids}, alongside the results obtained with the semi-local PBE functional. 

The first conclusion that can be drawn from Fig.~\ref{fig:bands-hybrids} is that, overall, the use of HSE06 leads to a systematic opening of the band gaps among all systems studied, consistent with previous results reported for this functional \cite{Paier2006, Paier2006err, Huhn2017}. This trend is summarized in Table-\ref{table:bandgap}, which compares the calculated band gaps using PBE, HSE06, and single-shot G$_{0}$W$_{0}$ \cite{Hedin1965, Hybertsen1986, Rohlfing1998, Rohlfing2000, Onida2002, Golze2019} (considered so far the \textit{gold standard} for band gap calculations in extended systems), as well as available experimental values.
The inclusion of 25\% of exact exchange significantly improves the agreement with experimental results. For instance, the band gap of bulk silicon increases from 0.59 eV (PBE) to 1.20 eV (HSE06), closely matching the experimental value of 1.12 eV.
Similar trends are found in all the materials studied, where the hybrid functional improves the well-known underestimation of the gap by semi-local functionals and yields to gaps that are close to the experimental ones. 
However, this comparison must account for the experimental conditions under which the band gap was measured.
If the experimental values were obtained at finite temperature, an extrapolation to $T$ = 0 is necessary for a proper comparison with first-principles results.
Additionally, finite-temperature experimental values may be influenced by phononic effects, which are not accounted for in the theoretical results presented here.
For this reason, alongside experimental values, we also include some results obtained at the G$_{0}$W$_{0}$ level in Table-\ref{table:bandgap}, as it provides quasi-particle energies, which are generally taken as good reference for ground state properties.
The general agreement provided by HSE06 in comparison with G$_{0}$W$_{0}$ is very good, with the hybrid results typically underestimating the G$_{0}$W$_{0}$ results.
This is consistent with previous benchmarking studies and reflects the more accurate treatment of screening in the many-body approach~\cite{Abedi2023, Camarasa-Gomez2023, kolos2025}.
In black phosphorus, the improvement is particularly notable: PBE incorrectly predicts a metallic ground state, whereas HSE06 opens a direct gap of 0.29 eV, in good agreement with G$_{0}$W$_{0}$ and experimental estimates.

The second conclusion is that not only the band gap increases, but the overall bandwidth also increases with HSE06. Such trend is expected due to the enhanced orbital localization induced by exact exchange, which reduces self-interaction and increases band dispersion~\cite{Schmidt2016, Laurien2022}.
Although exact exchange tends to localize electronic states, hybrid functionals also correct the strong delocalization error of semi-local GGAs, which increases bonding-antibonding splittings and effective hopping amplitudes. In many materials this second effect dominates, leading to wider band dispersions rather than narrowing, as reported in Refs. ~\cite{Heyd-04, Paier2006, Tran-09}. In addition, we note that band widening may also arise from several effects not related to the choice of exchange-correlation functional, including structural relaxation, basis-set incompleteness, k-point sampling, or pseudopotential differences. These factors were carefully controlled to ensure that the observed trends originate from the hybrid functional itself.

It is important to note here how the magnitude of the band gap changes with the quality of the basis set, following the expected variational trend. Minimal single-$\zeta$ and double-$\zeta$ basis sets tend to overestimate band gaps, whereas the inclusion of polarization functions substantially reduces this error. For representative covalent solids such as silicon, the difference between double-$\zeta$ plus polarization (DZP) and larger triple-$\zeta$ polarized basis sets is negligible (1.20 eV vs 1.21 eV), indicating that the band gap is already converged at the DZP level. This level of accuracy provides a suitable balance between computational efficiency and reliability for the purposes of the present work. In any case, it is always important to check this convergence for any particular problem.

\begin{table}
    \centering
    \caption{ Band gaps (in eV) of the materials represented in Fig.~\ref{fig:bands-hybrids} computed using different theoretical methods: the semi-local Perdew-Burke-Ernzerhof (PBE) functional, the hybrid HSE06 functional, and the single-shot G$_0$W$_0$ approximation. Experimental values (Expt.) are included for comparison. The HSE06 calculations were performed with stringent numerical settings: $N_{\rm G} = 6$, $\epsilon_{\rm Gauss} = 1 \times 10^{-5}$, and all remaining integral reduction thresholds set to $10^{-6}$.
    }
    \label{table:bandgap}
    \begin{tabular}{ccccc}  
    \hline
    \hline
    System                    & 
    PBE                       &
    HSE06                     &
    G$_0$W$_0$                &
    Expt.              \\
    \hline
    Si                        &
    0.59                      &
    1.20                      &
    1.17~\cite{Si_GW2015ab}   &
    1.12~\cite{Si_exp}        \\
    Diamond                   &
    4.20                      &
    5.31                      &
    5.95~\cite{G0W0_salas2022electronic}&
    5.47~\cite{C_clark1964}   \\
    LiF                       &
    8.96                      &
    11.23                    &
    13.6~\cite{LiF_sommer2012quasiparticle}&
    13.04~\cite{LiF_2004measurements}\\
    c-BN                      &
    4.51                      &
    5.84                      &
    6.1~\cite{G0W0_calcs}     &
    6.36~\cite{c-BN_evans2008}\\
    monolayer h-BN            &
    4.60                      &
    5.74                      &
    7.06~\cite{2d-BN_2016efficient}&
    5.95~\cite{h-BN_cassabois2016}\\
    TiO$_{2}$                 &
    1.67                      &
    3.01                      &
    3.14~\cite{samsonidze2014insights}&
    3.03~\cite{TiO2_amtout1995optical}\\
    ZnO                       &
    1.14                      &
    2.73                      &
    3.26~\cite{samsonidze2014insights}&
    3.24~\cite{ZnO_hashir2022experimental}\\
    Black Phosphorus          &
    0.00                      &
    0.29                      &
    0.43~\cite{rudenko2015toward}    &
    0.33~\cite{BP_1953electrical,BP_1963ElectricalAO}\\
    \hline
    \hline
    \end{tabular}
\end{table}

Despite their success in improving the description of band gaps and localized states in semiconductors and insulators, hybrid functionals often exhibit serious limitations when applied to metallic systems. The nonlocal Hartree-Fock exchange included in these functionals tends to open spurious band gaps at the Fermi level, violating the free-electron gas limit that is essential for accurately capturing metallic screening and correlation. 
Graphene, however, represents an exception for this pathological case. 
Our simulations with the HSE06 functional within the \textsc{siesta} code show that, when accurate relaxed structures are employed, HSE06 correctly preserves the gapless nature of graphene.
This agreement, however, should not be extrapolated to generic metallic systems. The accurate description of metallicity within hybrid functionals would require a proper treatment of dielectric screening, including both interband transitions and the correct evaluation of the plasma frequency, quantities that are not straightforward to define or compute consistently across different metals. For these reasons, hybrid functionals must be applied with great caution to metals, and their use in such systems remains limited in the literature~\cite{Gao2016}.
Exceptionally, undoped semimetals, such as graphene, can be well-described with hybrid functionals. The well-known reason is the weak screening of these materials due to the linear behavior of the  polarizability in the long-wavelength limit (see Ref.~\cite{DasSarma2011} and references therein), which prevents spurious opening of band gaps resulting from Hartree-Fock exchange.

In summary, the HSE06 functional offers a significant improvement over PBE in both quantitative and qualitative aspects of the electronic structure, with a performance that is comparable to that of G$_{0}$W$_{0}$ for many materials. 
Hybrid exchange–correlation functionals have a well-documented influence on structural, energetic, and vibrational properties compared to semi-local approximations. For extended solids, screened hybrid functionals such as HSE generally improve the description of lattice constants and bulk moduli relative to the PBE GGA, often yielding values closer to experiment or to higher-level methods, while maintaining comparable accuracy in bond lengths and structural parameters across a range of materials ~\cite{Hummer-09}. Regarding vibrational properties, hybrid functional calculations of phonon dispersion and lattice vibration spectra show that the inclusion of exact exchange alters the interatomic force constants and typically results in stiffer phonon modes and shifts in phonon frequencies ~\cite{Skelton-14}, which can improve agreement with experiment in many cases. In the present implementation, the formulation provides consistent total energies, forces, and stresses, enabling geometry optimization and phonon calculations within hybrid density functional theory.

\subsection{Efficiency versus accuracy}
\label{sec:effvsacc}

Figure~\ref{fig:bands-hybrids} and Table-\ref{table:bandgap} demonstrate that hybrid functionals significantly improve the predicted band gaps of a wide range of materials compared to conventional semi-local functionals such as PBE. However, this improvement in accuracy comes at the cost of a substantial increase in computational time and memory with respect to semilocal functionals. This naturally raises an important question: can we improve the efficiency of hybrid functional simulations without compromising their accuracy?

To address this question, we carry out a systematic analysis of the various approximations involved in the implementation. Specifically, we evaluate the impact of three key factors: (i) the number of Gaussian functions used in the expansion of the radial part of the NAOs (Sec.~\ref{sec:NG}), (ii) the spatial range of the orbitals (Sec.~\ref{sec:range}), and (iii) the number of four-center integrals actually computed (Sec.~\ref{sec:thresholds}), prescreened using the criteria described in Sec.~\ref{sec:computation-fourcenter}. In each case, one of these parameters is varied while keeping the others fixed, allowing us to isolate and quantify its effect on the overall computational cost and accuracy.
In every case, we compare the overload of the hybrid functionals with respect to the bare PBE calculations in {\sc siesta}.
All the calculations carried out in this Section have been run on a single core to carry out a more direct comparison without including parallelization strategies, which already improve efficiency, as we later discuss in Section \ref{sec:parallel}. These serial calculations were performed on the standard compute nodes of the Center for High Performance Computing Lengau cluster, equipped with Intel(R) Xeon(R) E5-2690 v3 CPUs running at 2.60 GHz, with 128 GB of RAM, under CentOS 7.3.

\subsubsection{Number of Gaussians}
\label{sec:NG}

The number of Gaussians used to expand the radial components of the NAOs significantly impacts the efficiency of hybrid functional simulations, as illustrated in Table-\ref{table:timing-ng}. A larger $N_{\rm G}$ reduces the goal function that measures the difference between the sum of Gaussians and the original radial part of the native NAOs [metric $\tilde{G}$, defined in  Eq.~(\ref{eq:errores2})]. But, at the same time, it increases the number of primitive four-center integrals that must be computed in Eq.~(\ref{eq:four-center-int-contract}), with the computational effort scaling approximately as $\mathcal{O}\left(N_{\rm G}^{4}\right)$. This trend is clearly reflected in the timing data: for instance, in carbon, increasing $N_{\rm G}$ from 3 to 6 leads to a 34-fold increase in the time required for a single SCF, although the prefactor of the scaling depends on several factors like the number of neighbors or the auxiliary supercell size. Therefore, this cost does not translate uniformly across materials, and a nuanced balance between accuracy and efficiency emerges.

Interestingly, the number of interacting shell pairs $N_{IJ}$ does not always increase monotonically with $N_{\rm G}$, as might be expected. 
Changing the number of Gaussians in the expansion directly affects the optimized coefficients and exponents which, in turn, alters the tails of the contracted sum of Gaussians. 
Therefore, the nontrivial dependence of $N_{IJ}$ on $N_{\rm G}$ is a consequence of the interplay between the tails of the fitted Gaussians and the screening thresholds used to determine relevant shell pairs.

The convergence of the band gap with respect to $N_{\rm G}$ is strongly material-dependent. Covalently bonded systems such as silicon, diamond, or black phosphorus exhibit a marked sensitivity to the number of Gaussians: at low $N_{\rm G}$, the band gaps are significantly overestimated, and convergence is only achieved for $N_{\rm G} \ge 4$. In contrast, ionic systems like LiF or TiO$_{2}$ show negligible variations across the entire range of $N_{\rm G}$, indicating that a minimal expansion (e.g. $N_{\rm G} = 3$) may suffice in those cases. 
The sensitivity of the band gap to the number of Gaussians in the auxiliary expansion is primarily a representation effect associated with the accuracy of the orbital shape entering the nonlocal exchange term, rather than a consequence of screening. This effect is most pronounced in covalent systems, where small changes in the exchange matrix elements can lead to sizable variations in the gap.

Taken together, these observations suggest that choosing $N_{\rm G} = 4$ provides a practical compromise between computational cost and predictive accuracy in most systems. The gaps obtained for this number of Gaussians are within some tens of meV of the most converged value shown in Table-\ref{table:bandgap}, but with a remarkable reduction in the computational time.
The exceptions are LiF and black phosphorus, where, as we show in Sec.~\ref{sec:range}, the range of the radial basis functions plays a crucial role. 
%

\begin{table*}
    \centering
    \caption{ Performance of hybrid functionals on a single processor as a function of the number of Gaussians ($N_{\rm G}$) used in the expansion of the radial part of the NAOs.
    The same number of Gaussians are used for all the atomic shells of a given atom. 
    $\tilde{G}_{\rm max}$ is the maximum value of the goal function that measures the distance between the sum of Gaussians and the original radial part of the NAOs, detailed in Eq.~(\ref{eq:errores2}) (in parenthesis, the shell where this maximum takes place).
    $N_{IJ}$ denotes the number of interacting shell pairs, as described in Sec.~\ref{sec:computation-fourcenter}.
    $t_{\rm SCF}^{\rm hybrids}$ represents the time (in seconds) required for a single self-consistent field (SCF) step with hybrid functionals, while $t_{\rm SCF}^{\rm hybrids}$ / $t_{\rm SCF}^{\rm PBE}$ indicates the relative time increase compared to a standard PBE calculation. The band gap size ($E_{\rm gap}$) is given in eV.
    All other parameters remain fixed: $\epsilon_{\rm Gauss} = 5 \times 10^{-3}$, integral reduction thresholds set to $10^{-6}$, and additional technical settings as detailed in Sec.~\ref{sec:technicalities}.
        }
    \label{table:timing-ng}
    \begin{tabular}{ccllcccc}  
    \hline
    \hline
    System                  & 
    $N_{\rm G}$             &
    \multicolumn{2}{c}{$\tilde{G}_{\rm max}$}   &
    $N_{IJ}$                &
    $t_{\rm SCF}^{\rm hybrids}$  &
    $t_{\rm SCF}^{\rm hybrids}$ / $t_{\rm SCF}^{\rm PBE}$              &
    $E_{\rm gap}$           \\
    \hline
     Si                     &
     3                      &
     \multicolumn{2}{c}{0.0089 (2-$\zeta$ 3$s$ Si)} &       
     1034                   &
     79.3                   &
     38.9                   &
     1.45                   \\
                            &
     4                      &
     \multicolumn{2}{c}{0.0087 (2-$\zeta$ 3$s$ Si)} & 
     1066                   &
     206.3                  &
     101.1                  &
     1.23                   \\
                            &
     5                      &
     \multicolumn{2}{c}{0.0087 (2-$\zeta$ 3$s$ Si)} &
     1246                   &
     964.5                  &
     472.8                  &
     1.11                   \\
                            &
     6                      &
     \multicolumn{2}{c}{0.0066 (2-$\zeta$ 3$s$ Si)} &
     1198                   &
     2004.6                 &
     982.6                  &
     1.11                   \\
     \hline
     Diamond                &
     3                      &
     \multicolumn{2}{c}{0.0507 (1-$\zeta$ 3$d$ C)} &
     1674                   &
     234.4                  &
     114.9                  &
     5.97                   \\
                            &
     4                      &
     \multicolumn{2}{c}{0.0432 (1-$\zeta$ 3$d$ C)} &
     2046                   &
     1198.3                 &
     587.6                  &
     5.28                   \\
                            &
     5                      &
     \multicolumn{2}{c}{0.0328 (1-$\zeta$ 3$d$ C)} &
     1794                   &
     1404.7                 &
     688.9                  &
     5.45                   \\
                            &
     6                      &
     \multicolumn{2}{c}{0.0283 (1-$\zeta$ 3$d$ C)} &
     2198                   &
     8073.1                 &
     3959.3                 &
     5.32                   \\
     \hline
     LiF                    &
     3                      &
     0.0104 (2-$\zeta$ 2$s$ Li)                       &
     0.0117 (2-$\zeta$ 2$p$ F)                        &
     2375                   &
     555.0                  &
     337.8                  &
     10.87                  \\
                            &
     4                      &
     0.0066 (1-$\zeta$ 1$s$ Li)                      & 
     0.0139 (2-$\zeta$ 2$p$ F)                       &
     2305                   &
     1200.9                 &
     730.9                  &
     10.84                  \\
                            &
     5                      &
     0.0066 (1-$\zeta$ 1$s$ Li)                      & 
     0.0052 (2-$\zeta$ 2$s$ F)                       &
     2305                   &
     2067.1                 &
     1258.1                 &
     10.90                  \\
                            &
     6                      &
     0.0046 (2-$\zeta$ 2$s$ Li)                      & 
     0.0049 (1-$\zeta$ 2$s$ F)                       &
     2229                   &
     3740.6                 &
     2276.7                 &
     10.86                  \\
     \hline
     c-BN                   &
     3                      &
     0.0272 (1-$\zeta$ 3$d$ B)                       &
     0.0782 (1-$\zeta$ 3$d$ N)                       &
     1832                   &
     314.5                  &
     126.6                  &
     6.03                   \\
                            &
     4                      &
     0.0192 (1-$\zeta$ 3$d$ B)                       &
     0.0757 (1-$\zeta$ 3$d$ N)                       &
     2238                   &
     1516.7                 &
     610.5                  &
     5.89                   \\
                            &
     5                      &
     0.0158 (1-$\zeta$ 3$d$ B)                       &
     0.0472 (1-$\zeta$ 3$d$ N)                       &
     1930                   &
     2282.8                 &
     919.0                  &
     5.86                   \\
                            &
     6                      &
     0.0125 (1-$\zeta$ 3$d$ B)                       &
     0.0498 (1-$\zeta$ 3$d$ N)                       &
     2384                   &
     8532.6                 &
     3435.0                 &
     5.84                   \\
     \hline
     monolayer h-BN         &
     3                      &
     0.0297 (3-$\zeta$ 2$s$ B)                       &
     0.0114 (3-$\zeta$ 2$p$ N)                       &
     6219                   &
     6522.8                 &
     206.0                  &
     5.84                   \\
                            &
     4                      &
     0.0318 (3-$\zeta$ 2$s$ B)                       &
     0.0131 (3-$\zeta$ 2$p$ N)                       &
     5772                   &
     12803.2                 &
     404.3                  &
     5.72                   \\
                            &
     5                      &
     0.0106 (3-$\zeta$ 2$s$ B)                       &
     0.0061 (3-$\zeta$ 2$s$ N)                       &
     5949                   &
     25104.3                &
     792.8                  &
     5.81                   \\
                            &
     6                      &
     0.0088 (3-$\zeta$ 2$s$ B)                       &
     0.0054 (3-$\zeta$ 2$s$ N)                       &
     5814                   &
     43592.6                &
     1376.8                 &
     5.82                   \\
     \hline
     TiO$_{2}$              &
     3                      &
     0.0135 (2-$\zeta$ 4$s$ Ti)      &
     0.1379 (1-$\zeta$ 3$d$ O)       &                 
     4392                   &
     575.5                  &
     78.8                   &
     2.96                   \\
                            &
     4                      &
     0.0093  (1-$\zeta$ 3$s$ Ti)     &
     0.1056  (1-$\zeta$ 3$d$ O)      &
     4144                   &
     929.0                  &
     127.2                  &
     2.95                   \\
                            &
     5                      &
     0.0058 (2-$\zeta$ 4$s$ Ti)       &       
     0.0850 (1-$\zeta$ 3$d$ O)        &
     4032                   &
     1884.6                 &
     258.2                  &
     2.98                   \\
                            &
     6                      &
     0.0058  (2-$\zeta$ 4$s$ Ti)     &
     0.0788  (1-$\zeta$ 3$d$ O)      &
     4020                   &
     2756.3                 &
     377.6                  &
     2.91                   \\
     \hline
     ZnO                    &
     3                      &
     0.0125 (2-$\zeta$ 4$s$ Zn)      &         
     0.1379 (1-$\zeta$ 3$d$ O)       &
     1756                   &
     71.6                   &
     41.5                   &
     2.77                   \\
                            &
     4                      &
     0.0153 (2-$\zeta$ 4$s$ Zn)      &   
     0.1056 (1-$\zeta$ 3$d$ O)       &
     1838                   &
     97.4                   &
     56.4                   &
     2.79                   \\
                            &
     5                      &
     0.0067 (1-$\zeta$ 3$s$ Zn)      & 
     0.0850 (1-$\zeta$ 3$d$ O)       &              
     1800                   &
     296.3                  &
     171.6                  &
     2.79                   \\
                            &
     6                      &
     0.0054 (1-$\zeta$ 4$s$ Zn)      &
     0.0789                 &
     1940                   &
     562.6                  &
     325.7                  &
     2.84                   \\
     \hline
     Black Phosphorus       &
     3                      &
     \multicolumn{2}{c}{0.0101 (2-$\zeta$ 3$s$ P)} &
     2688                   &
     122.6                  &
     21.1                   &
     0.52                   \\
                            &
     4                      &
     \multicolumn{2}{c}{0.0100 (2-$\zeta$ 3$s$ P)} &
     2840                   &
     263.4                  &
     45.4                   &
     0.47                   \\
                            &
     5                      &
     \multicolumn{2}{c}{0.0077 (2-$\zeta$ 3$s$ P)} &
     2872                   &
     592.1                  &
     102.2                  &
     0.45                   \\
                            &
     6                      &
    \multicolumn{2}{c}{0.0075 (2-$\zeta$ 3$s$ P)} &
     3292                   &
     2389.4                 &
     412.6                  &
     0.29                   \\
    \hline
    \hline
    \end{tabular}
\end{table*}

\subsubsection{Range of the radial parts of the atomic orbitals}
\label{sec:range}

\begin{table*}
    \centering
    \caption{Cutoff radii (in Bohr) of the atomic orbitals obtained for a given threshold $\epsilon_{\rm Gauss}$ for Si, C, Li, F, B, N, Ti, O, Zn, and P atoms using double-$\zeta$ plus polarization (DZP) basis sets. The first and second $\zeta$ functions of the $s$ and $p$ orbitals are denoted as $s_1$, $s_2$, $p_1$, and $p_2$, respectively, while $d$ refers to the polarization $d$ orbitals. The cutoff radii of the semicore orbitals are not included. For Ti and Zn, only te first-$\zeta$ of the $d$ orbital is shown. 
        }
    \label{table:cutoffrc}
    \begin{tabular}{cccccccc}  
    \hline
    \hline
    Atom                    &
    Orbital                 &
    $\epsilon_{\rm Gauss}$  &
    $s_1$                   &
    $s_{2}$                 &
    $p_{1}$                 &
    $p_{2}$                  &
    $d$                    \\
    \hline
     Si                    &
     NAO                   &
                           &
     5.392                 &
     3.847                 &
     6.844                 &
     4.952                 &
     6.844                 \\
                           &
     CGTO                  &
     $7.5 \times 10^{-3}$  &
     5.152                 &
     3.824                 &
     6.370                 &     
     4.812                 &
     5.774                 \\
                           &
                           &                          
     $5 \times 10^{-3}$    &
     5.375                 &
     3.943                 &
     6.557                 &
     4.993                 &
     5.990                 \\
                           &
                           &
     $1 \times 10^{-3}$    &
     6.182                 &
     4.383                 &
     7.543                 &
     5.646                 &
     6.790                 \\
                           &
                           &
     $1 \times 10^{-4}$    &
     7.180                 &
     4.945                 &
     8.740                 &     
     6.453                 &
     7.670                 \\
                           &
                           &
    $1 \times 10^{-5}$     &
    8.055                  &
    5.450                  &
    9.777                  &     
    7.162                  &
    8.519                  \\
    \hline
     C                     &
      NAO                  &
                      &
      4.484                &
      3.048                &
      5.436                &
      3.692                     &
      5.436                     \\
                           &
     CGTO                  &                      
     $7.5 \times 10^{-3}$  &
     4.334                 &
     3.003                 &
     5.184                 &     
     3.640                 &
     4.800                 \\
                           &
                           &
     $5 \times 10^{-3}$    &
     4.520                 &
     3.088                 &
     5.417                 &
     3.777                 &
     4.967                 \\
                           &
                           &
     $1 \times 10^{-3}$    &
     5.193                 &
     3.405                 &
     6.243                 &
     4.270                 &
     5.569                 \\
                           &
                           &
     $1 \times 10^{-4}$    &
     6.027                 &
     3.813                 &
     7.242                 &
     4.879                 &
     6.31                  \\
                           &
                           &
     $1 \times 10^{-5}$    &
     6.750                 &
     4.085                 &
     7.977                 &
     4.886                 &
     7.809                 \\
    \hline
     Li                   &
      NAO                 &
                          &
      8.808               &
      6.628               &
      8.808               &
      -                   &
      -                  \\
                          &
     CGTO                 &                      
     $7.5 \times 10^{-3}$ &
     8.149                &
     6.245                &
     8.429                &     
      -                   &
      -                   \\
                          &
                          &
     $5 \times 10^{-3}$   &
     8.524                &
     6.476                &
     8.840                &
      -                   &
      -                   \\
                          &
                          &
     $1 \times 10^{-3}$   &
     9.872                &
     7.322                &
     10.284               &
      -                   &
      -                   \\
                          &
                          &
     $1 \times 10^{-4}$   &
     11.529               &
     8.385                &
     12.013               &
       -                  &
       -                  \\
                          &
                          &
     $1 \times 10^{-5}$   &
     12.982               &
     9.313                &
     13.448               &
       -                &
       -                  \\
    \hline
     F                     &
      NAO                  &
                           &
      3.215                &
      2.050                &
      3.929                &
      2.489                &
      3.929                \\
                           &
     CGTO                  &                      
     $7.5 \times 10^{-3}$  &
     3.152                 &
     2.506                 &
     3.638                 &     
     2.726                 &
     4.891                 \\
                           &
                           &
     $5 \times 10^{-3}$    &
     3.254                 &
     2.599                 &
     3.957                 &
     2.838                 &
     5.057                 \\
                           &
                           &
     $1 \times 10^{-3}$    &
     3.629                 &
     2.939                 &
     4.973                 &
     3.239                 &
     5.657                 \\
                           &
                           &
     $1 \times 10^{-4}$    &
     4.106                 &
     3.366                 &
     6.092                 &
     3.730                 &
     6.399                 \\
                           &
                           &
     $1 \times 10^{-5}$    &
     4.465                 &
     3.515                 &
     5.779                 &
     4.127                 &
     6.023                 \\
    \hline
     B                    &
      NAO                 &
                          &
      5.254               &
      3.670               &
      6.393               &
      4.475               &
      6.393               \\
                          &
     CGTO                 &                      
     $7.5 \times 10^{-3}$ &
     5.030                &
     3.622                &
     5.935                &     
     4.406                &
     5.326                \\
                          &
                          &
     $5 \times 10^{-3}$   &
     5.254                &
     3.741                &
     6.206                &
     4.585                &
     5.519                \\
                          &
                          &
     $1 \times 10^{-3}$   &
     6.064                &
     4.183                &
     7.164                &
     5.223                &
     6.209                \\
                          &
                          &
     $1 \times 10^{-4}$   &
     7.062                &
     4.744                &
     8.321                &
     6.007                &
     7.056                \\
                          &
                          &
     $1 \times 10^{-5}$   &
     7.938                &
     4.866                &
     9.418                &
     6.372                &
     8.518                \\
    \hline
     N                    &
     NAO                  &
                          &
     3.944                &
     2.617                &
     4.781                &
     3.167                &
     4.781                \\
                          &
     CGTO                 &                      
     $7.5 \times 10^{-3}$ &
     3.841                &
     2.595                &
     4.860                &     
     3.074                &
     4.584                \\
                          &
                          &
     $5 \times 10^{-3}$   &
     4.000                &
     2.702                &
     5.101                &
     3.177                &
     4.741                \\
                          &
                          &
     $1 \times 10^{-3}$   &
     4.580                &
     3.091                &
     5.946                &
     3.555                &
     5.308                \\
                          &
                          &
     $1 \times 10^{-4}$   &
     5.301                &
     3.574                &
     6.955                &
     4.027                &
     6.008                \\
                          &
                          &
     $1 \times 10^{-5}$   &
     5.940                &
     4.351                &
     7.009                &
     4.277                &
     7.487                \\                       
    \hline
     Ti                   &
     NAO                  &
                          &
     7.610                &
     5.698                &
     7.610                &
       -                  &
     5.233               \\
                          &
     CGTO                 &                      
     $7.5 \times 10^{-3}$ &
     7.131                &
     5.480                &
     7.719                &     
       -              &
     4.955                \\
                          &
                          &
     $5 \times 10^{-3}$   &
     7.431                &
     5.668                &
     8.065                &
       -                 &
     5.169                \\
                          &
                          &
     $1 \times 10^{-3}$   &
     8.520                &
     6.360                &
     9.291                &
       -                 &
     5.917                \\
                          &
                          &
     $1 \times 10^{-4}$   &
     9.870                &
     7.236                &
    10.776                &
      -                  &
    6.812                 \\
                          &
                          &
     $1 \times 10^{-5}$   &
     11.121               &
     8.019                &
     11.616               &
       -                  &
     7.291                \\
    \hline
     O                    &
     NAO                  &
                          &
     3.532                &
     2.299                &
     4.300                &
     2.782                &
     4.300                \\
                          &
     CGTO                 &                      
     $7.5 \times 10^{-3}$ &
     3.457                &
     2.892                &
     4.472                &     
     2.898                &
     3.096                \\
                          &
                          &
     $5 \times 10^{-3}$   &
     3.572                &
     3.008                &
     4.708                &
     3.003                &
     3.191                \\
                          &
                          &
     $1 \times 10^{-3}$   &
     3.997                &
     3.431                &
     5.529                &
     3.382                &
     3.539                \\
                          &
                          &
     $1 \times 10^{-4}$   &
     4.536                &
     3.958                &
     6.501                &
     3.853                &
     3.974                \\
                          &
                          &
     $1 \times 10^{-5}$   &
     5.314                &
     3.896                &
     6.307                &
     5.055                &
     4.550                \\
    \hline
     Zn                   &
     NAO                  &
                          &
     5.524                &
     3.917                &
     5.524                &
      -                   &
     3.525                \\
                          &
     CGTO                 &                      
     $7.5 \times 10^{-3}$ &
     5.253                &
     3.960                &
     6.042                &     
       -                  &
     3.202                \\
                          &
                          &
     $5 \times 10^{-3}$   &
     5.486                &
     4.105                &
     6.281                &
       -                 &
     3.312                \\
                          &
                          &
     $1 \times 10^{-3}$   &
     6.326                &
     4.637                &
     7.140                &
       -                 &
     3.730                \\
                          &
                          &
     $1 \times 10^{-4}$   &
     7.364                &
     5.306                &
     8.196                &
       -                &
     4.238                \\
                          &
                          &
     $1 \times 10^{-5}$   &
     8.347                &
     5.281                &
     8.793                &
       -                 &
     4.852                \\
    \hline
     P                    &
     NAO                  &
                          &
     4.859                &
     3.414                &
     6.081                &
     4.321                &
     6.081                \\
                          &
     CGTO                 &                      
     $7.5 \times 10^{-3}$ &
     4.699                &
     3.429                &
     5.711                &     
     4.254                &
     4.956                \\
                          &
                          &
     $5 \times 10^{-3}$   &
     4.898                &
     3.533                &
     5.961                &
     4.408                &
     5.132                \\
                          &
                          &
     $1 \times 10^{-3}$   &
     5.619                &
     3.917                &
     6.850                &
     4.964                &
     5.763                \\
                          &
                          &
     $1 \times 10^{-4}$   &
     6.513                &
     4.408                &
     7.928                &
     5.655                &
     6.538                \\
                          &
                          &
     $1 \times 10^{-5}$   &
     7.288                &
     4.642                &
     8.869                &
     6.253                &
     8.771                \\
    \hline
    \hline
    \end{tabular}
\end{table*}     

\begin{table*}
    \centering
    \caption{ Performance of hybrid functionals on a single processor as a function of the range of the radial part of the NAOs.
    $N_{IJ}$ denotes the number of interacting shell pairs, as described in Sec.~\ref{sec:computation-fourcenter}.
    $t_{\rm SCF}^{\rm hybrids}$ represents the time (in seconds) required for a single self-consistent field (SCF) step with hybrid functionals on a single processor, while $t_{\rm SCF}^{\rm hybrids}$ / $t_{\rm SCF}^{\rm PBE}$ indicates the relative time increase compared to a standard PBE calculation. The band gap size ($E_{\rm gap}$) is given in eV.
    All other parameters remain fixed: $N_{\rm G} = 4$, integral reduction thresholds are set to $10^{-6}$, and additional technical settings are as detailed in Sec.~\ref{sec:technicalities}.
        }
    \label{table:timing-range}
    \begin{tabular}{ccccccc}  
    \hline
    \hline
    System                  & 
    $\epsilon_{\rm Gauss}$  &
    Aux. supercell          &
    $N_{IJ}$                &
    $t_{\rm SCF}^{\rm hybrids}$  &
    $t_{\rm SCF}^{\rm hybrids}$ / $t_{\rm SCF}^{\rm PBE}$              &
    $E_{\rm gap}$              \\
    \hline
     Si                        &
     $7.5 \times 10^{-3}$      &
     $5 \times 5 \times 5$     &
     982                       &
     165.2                     &
     80.9                      &
     1.19                      \\
                               &
     $5 \times 10^{-3}$        &
     $5 \times 5 \times 5$     &
     1066                      &
     206.3                     &
     101.1                     &
     1.23                      \\
                               &
     $1 \times 10^{-3}$        &
     $5 \times 5 \times 5$     &
     1582                      &
     409.9                     &
     200.9                     &
     1.26                      \\
                               &
     $1 \times 10^{-4}$        &
     $7 \times 7 \times 7$     &
     1806                      &
     486.9                     &     
     238.6                     &
     1.27                      \\
     \hline
     Diamond                   &
     $7.5 \times 10^{-3}$      &
     $5 \times 5 \times 5$     &
     1814                      &
     1087.0                    &
     533.1                     &
     5.26                      \\
                               &
     $5 \times 10^{-3}$        &
     $5 \times 5 \times 5$     &
     2046                      &
     1198.3                    &
     587.6                     &
     5.28                      \\
                               &
     $1 \times 10^{-3}$        &
     $7 \times 7 \times 7$     &
     2046                      &
     1988.3                    &
     975.1                     &
     5.27                      \\
                               &
     $1 \times 10^{-4}$        &
     $9 \times 9 \times 9$     &
     3390                      &
     3681.2                    &
     1805.4                    &
     5.30                      \\
     \hline
     LiF                       &
     $7.5 \times 10^{-3}$      &
     $9 \times 9 \times 9$     &
     2017                      &
     933.6                     &
     568.2                     &
     10.46                     \\
                               &
     $5 \times 10^{-3}$        &
     $9 \times 9 \times 9$     &
     2305                      &
     1200.9                    &
     730.9                     &
     10.84                     \\
                               &
     $1 \times 10^{-3}$        &
     $11 \times 11 \times 11$  &
     3455                      &
     2809.5                    &
     1709.9                    &
     13.37                     \\ 
                               &
     $1 \times 10^{-4}$        &
     $13 \times 13 \times 13$  &
     4551                      &
     5315.5                    &
     3235.3                    &
     13.41                     \\ 
     \hline
     c-BN                      &
     $7.5 \times 10^{-3}$      &
     $5 \times 5 \times 5$     &
     2058                      &
     1349.9                    &
     543.4                     &
     5.88                      \\
                               &
     $5 \times 10^{-3}$        &
     $7 \times 7 \times 7$     &
     2238                      &
     1516.7                    &
     610.5                     &
     5.89                      \\
                               &
     $1 \times 10^{-3}$        &
     $7 \times 7 \times 7$     &
     3304                      &
     2607.0                    &
     1049.5                    &
     5.89                      \\
                               &
     $1 \times 10^{-4}$        &
     $9 \times 9 \times 9$     &
     3844                      &
     4204.1                    &
     1692.5                    &
     5.92                      \\
     \hline
     TiO$_{2}$                 &
     $7.5 \times 10^{-3}$      &
     $5 \times 5 \times 7$     &
     3708                      &
     738.3                     &
     101.1                     &
     2.77                      \\
                               &
     $5 \times 10^{-3}$        &
     $5 \times 5 \times 7$     &
     4144                      &
     929.0                     &
     127.2                     &
     2.95                      \\
                                &
     $1 \times 10^{-3}$        &
     $5 \times 5 \times 7$     &
     6062                      &
     1597.8                    &
     218.9                     &
     3.05                      \\
                               &
     $1 \times 10^{-4}$        &
     $7 \times 7 \times 9$     &
     7530                      &
     3128.8                    &
     428.6                     &
     3.05                      \\
     \hline
     ZnO                       &
     $7.5 \times 10^{-3}$      &
     $5 \times 5 \times 5$     &
     1600                      &
     113.2                     &
     65.6                      &
     2.99                      \\
                               &
     $5 \times 10^{-3}$        &
     $5 \times 5 \times 5$     &
     1838                      &
     97.4                      &
     56.4                      &
     2.79                      \\
                               &
     $1 \times 10^{-3}$        &
     $5 \times 5 \times 5$     &
     2640                      &
     277.7                     &
     160.8                     &
     2.66                      \\
                               &
     $1 \times 10^{-4}$        &
     $7 \times 7 \times 5$     &
     3116                      &
     488.6                     &
     282.9                     &
     2.66                      \\
     \hline
     Black Phosphorous         &
     $7.5 \times 10^{-3}$      &
     $5 \times 5 \times 3$     &
     2560                      &
     242.0                     &
     41.8                      &
     0.52                      \\
                               &
     $5 \times 10^{-3}$        &
     $5 \times 5 \times 3$     &
     2840                      &
     263.4                     &
     45.4                      &
     0.47                      \\
                               &
     $1 \times 10^{-3}$        &
     $5 \times 5 \times 3$     &
     4360                      &
     486.5                     &
     84.0                      &
     0.39                      \\
                               &
     $1 \times 10^{-4}$        &
     $5 \times 5 \times 3$     &
     4776                      &
     607.4                     &
     104.9                     &
     0.38                      \\
    \hline
    \hline
    \end{tabular}
\end{table*}



The spatial extent of the NAOs plays a crucial role in determining the sparsity of the Hamiltonian matrix and, by extension, the computational efficiency of hybrid functional calculations. Table-\ref{table:timing-range} explores the effect of varying the cutoff radius $r_{\rm c}$ used to truncate the orbitals, controlled via the $\epsilon_{\rm Gauss}$ parameter defined in Sec.~\ref{sec:radialpart}. This parameter defines the threshold below which the radial function is forced to vanish. As shown in Table-\ref{table:cutoffrc}, smaller values of $\epsilon_{\rm Gauss}$ lead to longer cutoff radii and, consequently, denser Hamiltonian matrices and a greater number of relevant shell pairs $N_{IJ}$. 
When $\epsilon_{\rm Gauss}$ is set to $10^{-3}$, the radii of the Cartesian GTOs representing the occupied atomic orbitals are slightly larger than those of the original NAOs, whereas the native polarization orbitals can remain more extended at this threshold. Increasing $\epsilon_{\rm Gauss}$ to $5\times10^{-3}$ yields Cartesian GTOs with spatial ranges comparable to those of the native orbitals.

As shown in Table-\ref{table:timing-range}, increasing the range of orbitals (i.e., decreasing $\epsilon_{\rm Gauss}$) has a twofold effect. First, it increases the size of the auxiliary supercell in {\sc siesta} and the concomitant enhancement of the number of shell pairs, with a corresponding growth in the SCF time per iteration. Second, the band gap converges rather quickly with increasing orbital range. In most materials studied, the change in the band gap is less than 50 meV beyond $\epsilon_{\rm Gauss} = 10^{-3}$, and within tenths of eV if $\epsilon_{\rm Gauss} = 5 \times 10^{-3}$. This indicates that extremely extended orbitals are not necessary to obtain reliable electronic properties.
As highlighted before, the exceptions are LiF and black phosphorous, where large variations of the band gap are produced when the cutoff radii of the atomic orbitals are increased.

These findings suggest that a sensible trade-off can be achieved by setting $\epsilon_{\rm Gauss} = 5 \times 10^{-3}$, which balances computational efficiency and physical accuracy. This threshold yields sufficiently compact orbitals to retain matrix sparsity while producing converged band gaps for a wide variety of systems. Moreover, it underscores the importance of tuning the radial range of atomic orbitals in hybrid calculations to maintain favorable scaling without sacrificing predictive power.

\subsubsection{Integral screening thresholds}
\label{sec:thresholds}

To assess the impact of integral screening on the accuracy and efficiency of hybrid functional calculations, we perform a systematic benchmark by varying the screening thresholds discussed in Sec.~\ref{sec:computation-fourcenter}. In particular, we set the Schwarz pair-list threshold $\epsilon_{\rm pair-list}$ and the density screening threshold $\epsilon_{\rm Schwartz}$ [cf. Eqs.~(\ref{eq:schwarz-pairlist}) and~(\ref{eq:denscreen})] to equal values, analyzing their influence on representative materials.

In contrast to the convergence tests in previous sections, here we maintain the same basis set across all calculations, ensuring that total energy differences and derived quantities can be attributed solely to the screening parameters. This setup allows us to quantify the trade-off between computational cost and numerical precision for hybrid functional self-consistent field calculations.

Table~\ref{table:thresholds} summarizes the performance of hybrid functional calculations on a single processor for various screening thresholds ranging from $10^{-4}$ to $10^{-6}$. The number of interacting shell pairs $N_{IJ}$, the hybrid SCF time per step $t_{\rm SCF}^{\rm hybrids}$, the relative cost compared to a standard PBE SCF step, and the resulting band gap $E_{\rm gap}$ are reported.

The results indicate that the screening thresholds offer a powerful lever to control the computational workload. For all systems tested, increasing the strictness of the threshold (i.e., reducing its value) systematically increases the number of shell pairs and the SCF computational time. However, the accuracy of the band gap remains remarkably stable across thresholds, with changes typically below $\sim$0.05 eV for thresholds tighter than $10^{-4}$.

The data further suggest that thresholds tighter than $10^{-5}$ offer diminishing returns in terms of accuracy, while dramatically increasing computational effort. For instance, the number of interacting shell pairs $N_{IJ}$ saturates quickly, whereas $t_{\rm SCF}^{\rm hybrids}$ and its ratio to $t_{\rm SCF}^{\rm PBE}$ continue to increase steeply, indicating that tighter screening leads to computational overhead without significant physical benefit.

In summary, the choice of $\epsilon_{\rm pair-list} = \epsilon_{\rm Schwartz} = 10^{-5}-10^{-6}$ emerges as a balanced compromise between accuracy and performance. This value ensures nearly-converged electronic properties across diverse systems while delivering an order-of-magnitude speed-up compared to unfiltered hybrid calculations. These observations reinforce the importance of tailored screening strategies to enable a scalable and efficient use of hybrid functionals in large-scale materials simulations.

\begin{table*}
    \centering
    \caption{ Performance of hybrid functionals on a single processor as a function of the thresholds to cut the four center integrals.
    $N_{IJ}$ denotes the number of interacting shell pairs, as described in Sec.~\ref{sec:computation-fourcenter}.
    $t_{\rm SCF}^{\rm hybrids}$ represents the time (in seconds) required for a single self-consistent field (SCF) step with hybrid functionals on a single processor, while $t_{\rm SCF}^{\rm hybrids}$ / $t_{\rm SCF}^{\rm PBE}$ indicates the relative time increase compared to a standard PBE calculation. The band gap size ($E_{\rm gap}$) is given in eV.
    All other parameters remain fixed: $N_{\rm G} = 4$, and $\epsilon_{\rm Gauss} = 5 \times 10^{-3}$, and additional technical settings as detailed in Sec.~\ref{sec:technicalities}.
        }
    \label{table:thresholds}
    \begin{tabular}{ccccccc}  
    \hline
    \hline
    System                  & 
    Thresholds              &
    $N_{IJ}$                &
    $t_{\rm SCF}^{\rm hybrids}$  &
    $t_{\rm SCF}^{\rm hybrids}$ / $t_{\rm SCF}^{\rm PBE}$              &
    $E_{\rm gap}$           &
    Total Energy (eV)       \\
    \hline
     Si                     &
     $10^{-4}$              &
     970                    &
     40.9                   &
     20.0                   &
     1.31                   &
     -225.905               \\
                            &
     $10^{-5}$              &
     1066                   &
     119.2                  &
     58.4                   &
     1.25                   &
     -225.992               \\
                            &
     $10^{-6}$              &
     1066                   &
     206.3                  &
     101.1                  &
     1.23                   &
     -225.994               \\
     \hline
     Diamond                &
     $10^{-4}$              &
     1970                   &
     211.4                  &
     103.6                  &
     5.32                   &
     -322.078               \\
                            &                        
     $10^{-5}$              &
      2046                  &
      768.5                 &
      376.9                 &
      5.29                  &
      -322.121              \\
                            &                        
     $10^{-6}$              &
     2046                   &
     1198.3                 &
     587.6                  &
     5.28                   &
     -322.127               \\
     \hline
     LiF                    &
     $10^{-4}$              &
     2263                   &
     346.9                  &
     211.2                  &
      10.6                  &
     -865.863               \\
                            &
    $10^{-5}$               &
    2293                    &
    719.7                   &
    438.0                   &
    10.85                   &
    -865.881                \\
                            &
    $10^{-6}$               &
    2305                    &
    1200.9                  &
    730.9                   &
    10.84                   &
    -865.880                \\
    \hline
    c-BN                    &
     $10^{-4}$              &
     2174                   &
     309.8                  &
     124.7                  &
     5.92                   &
     -359.981               \\
                            &
    $10^{-5}$               &
    2238                    &
    809.8                   &
    326.0                   &
    5.90                    &
    -360.036                \\
                            &
    $10^{-6}$               &
    2238                    &
    1525.0                  &
    613.9                   &
    5.89                    &
    -360.040                \\
    \hline
     TiO$_2$                &
     $10^{-4}$              &
     3960                   &
     320.9                  &
      43.9                  &
       2.91                 &
     -4998.937              \\
                            &
    $10^{-5}$               &
    4136                    &
    578.7                   &
    79.3                    &
    2.96                    &
    -4999.007               \\
                            &
    $10^{-6}$               &
    4144                    &
    927.3                   &
    127.0                   &
    2.95                    &
    -4999.014               \\
    \hline
     ZnO                    &
     $10^{-4}$              &
     1722                   &
     58.0                   &
     33.6                   &
     2.77                   &
     -12213.982             \\
                            &                        
     $10^{-5}$              &
     1838                   &
     99.1                   &
     57.3                   &
     2.80                   &
     -12214.002             \\
                            &                        
     $10^{-6}$              &
     1838                   &
     145.2                  &
     84.0                   &
     2.77                   &
     -12214.004             \\
     \hline
     Black Phosphorus       &
     $10^{-4}$              &
     2648                   &
     96.8                   &
     16.7                  &
     0.46                   &
     -1484.410              \\
                            &                        
     $10^{-5}$              &
     2840                   &
     174.9                  &
     30.2                   &
     0.47                   &
     -1484.522              \\
                            &                        
     $10^{-6}$              &
     2840                   &
     266.5                  &
     46.0                   &
     0.47                   &
     -1484.527              \\
    \hline
    \hline
    \end{tabular}
\end{table*}

\subsection{Singlet--triplet splitting in the O$_2$ molecule}
\label{sec:o2-singlet-triplet}

To validate the performance of the hybrid functional implementation with respect to the computation of differences in energies, we estimated the singlet--triplet energy splitting in the oxygen molecule (O$_2$), a prototypical open-shell system with a well-established triplet ground state. Calculations were carried out using the HSE06 functional, known to provide a balanced description of exchange and correlation in molecular systems.

We first performed a spin-unrestricted geometry optimization, allowing the stabilization of the correct spin-polarized ground state. The resulting configuration corresponds to the triplet state, with a total spin moment of $2\ \mu_B$, in agreement with the $^3\Sigma_g^-$ ground state observed experimentally.

Subsequently, using the optimized triplet geometry, we carried out a spin-restricted (non-magnetic) calculation to estimate the energy of the lowest singlet configuration. The total energy differences between the singlet and the triplet were found to be 1.18~eV (considering spin contamination) and 1.86~eV, with the triplet lying lower in energy. This value is consistent with experimental estimates of the singlet–triplet splitting in O$_2$, which are 0.98 eV and 1.63 eV~\cite{Atkins2006, Schweitzer2003}, correspondingly.

This result confirms that the HSE06 functional captures the essential spin physics and multi-reference character of O$_2$, and provides a stringent test of the spin-resolved exchange implementation.

\section{Parallelization}
\label{sec:parallel}

\begin{figure}
    \centering
\includegraphics[width=0.9\linewidth]{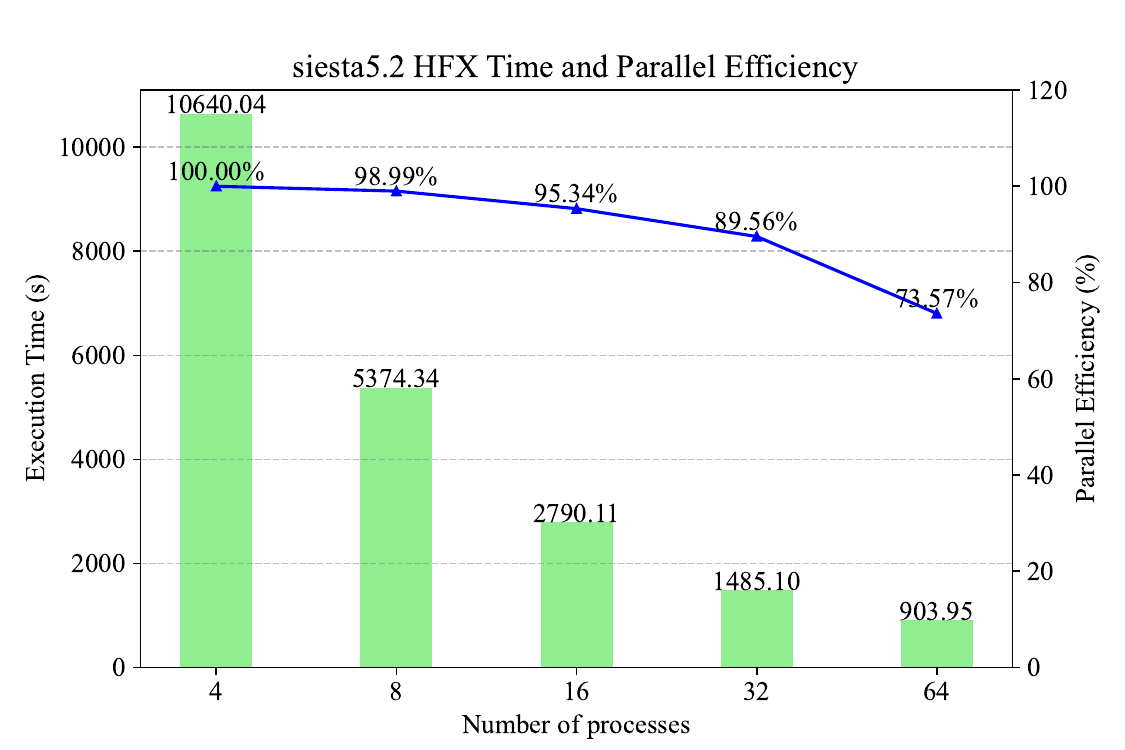}
    \caption{ Strong scaling performance of SIESTA HFX calculations for a 5$\times$5$\times$5 SrTiO$_3$ supercell with single-$\zeta$ basis set. The green bars (left $y$-axis) show the execution time in seconds, while the blue line with triangular markers (right $y$-axis) represents the parallel efficiency as a percentage. The calculation was performed on a dual-socket Intel\textregistered{} Xeon\textregistered{} Platinum 8358P system with 64 physical cores. The results demonstrate significant speedup with increasing core count, achieving an 11.8$\times$ acceleration from 4 to 64 processes, while parallel efficiency gradually decreases from 100\% to 73.57\% due to communication overhead.}
    \label{fig:parallel}
\end{figure}

To evaluate the strong scaling parallel efficiency of the \textsc{siesta} code for HFX calculations, we performed benchmark simulations on 5$\times$5$\times$5 supercell of SrTiO$_3$ with a single-$\zeta$ (SZ) basis set. We employ the parallel distribution scheme~\cite{Shang-20} for Hartree-Fock exchange calculations within the real-space \textsc{nao2gto} framework. This approach achieves optimal load balancing for ERI calculations, which is typically the most computationally intensive step. The benchmark was performed on a high-performance compute node equipped with two Intel\textregistered{} Xeon\textregistered{} Platinum 8358P processors, providing 64 physical cores. As illustrated in Fig.~\ref{fig:parallel}, the HFX execution time exhibited a substantial decrease as the number of MPI processes increased from 4 to 64, decreasing dramatically from 10,640.04~seconds (with 4 processes) to 903.95~seconds (with 64 processes), representing an 11.8$\times$ speedup and highlighting the significant acceleration achievable by parallel computing. However, the parallel efficiency showed a gradual decline with increasing core count, dropping from an ideal 100\% with 4 processes to 73.57\% with 64 processes. This declining efficiency is attributed to communication overhead. The results effectively illustrate the trade-off: increased computational resources greatly improve execution speed but simultaneously introduce inefficiencies due to heightened communication requirements among nodes.

\section{Concluding remarks}
\label{sec:concluding_remarks}

We have developed and validated an efficient and scalable implementation of hybrid exchange–correlation functionals in the \textsc{siesta} code. By expressing strictly localized NAOs as linear combinations of GTOs, we enable the analytical evaluation of four-center electron repulsion integrals using the \textsc{libint} library. This strategy, together with rigorous integral screening and an optimized treatment of real-space locality, significantly reduces the computational overhead typically associated with hybrid functionals.

Benchmark calculations across a diverse set of materials demonstrate that the HSE06 functional provides substantial improvements in the prediction of electronic band gaps, in excellent agreement with G$_{0}$W$_{0}$ results and experimental data. In particular, the implementation correctly captures the spin-triplet ground state of the O$_{2}$ molecule and the gapless nature of graphene, while improving the band structures of semiconductors and insulators.

We systematically analyzed the impact of three critical factors, namely the number of Gaussians in the radial expansion, the spatial cutoff of the atomic orbitals, and the screening thresholds for ERIs, on the computational efficiency and physical accuracy. Our findings indicate that a choice of four Gaussians per radial function and a localization threshold of $\epsilon_{\rm Gauss} = 5 \times 10^{-3}$ provide an optimal compromise. Screening thresholds in the range of $10^{-5}–10^{-6}$ further allow for linear scaling behavior without compromising predictive accuracy.

Finally, we demonstrated excellent parallel performance, with near-linear speedups and controlled memory scaling for large periodic systems. This positions \textsc{siesta} as a competitive and versatile tool for performing hybrid-functional calculations in large-scale material simulations, including geometry relaxations, molecular dynamics, and studies of charge localization or polaronic behavior.

\begin{acknowledgments}
Y.P.~acknowledges financial support from project RTC-2016-5681-7 by the Spanish Ministry of Economy, Industry and Competitiveness, co-financed by the European Structural and Investment Funds; from the Spanish State Plan for Scientific and Technical Research and Innovation 2017-2020 (Torres-Quevedo Grant PTQ-2019); from MCIN and the European Union NextGenerationEU/PRTR-C17.I1, as well as IKUR Strategy, under the collaboration agreement between Donostia International Physics Center and nanoGUNE on behalf of the Department of Education of the Basque Government.
B.C.O.~acknowledges financial support from the Erasmus+ KA-107 action, the Vice-rectorate for Internationalisation and Global Engagement of the University of Cantabria, and the Kenya Education Network Trust (KENET).
M.C.-G. acknowledges support from the Diputaci\'on Foral de Gipuzkoa through Grants 2024-FELL-000007-01 and 2025-FELL-000009-01, from the Gobierno Vasco-UPV/EHU Project No. IT1569-22, and from Grant No.~PID2024-159869NA-I00 funded by MICIU/AEI/10.13039/501100011033 and ERDF/EU.
F.G.-O. acknowledges financial support from MSCA-PF 101148906 funded by the European Union and the Fonds de la Recherche Scientifique (FNRS) through the grant FNRS-CR 1.B.227.25F.
J.J.~acknowledges financial support from Grant No.~PID2022-139776NB-C63 funded by MICIU/AEI/10.13039/501100011033 and ERDF/EU/AEI/10.13039/501100011033 and by ERDF/EU ``A way of making Europe'' by the European Union.
C.B. has been partially supported by grants MTM2017-83816-P and MTM2017-90682-REDT from the Spanish Ministry of Science (MICINN) and by
Grant 21.SI01.64658 from Universidad de Cantabria and Banco de Santander.
H.S.~acknowledges financial support from the National
 Natural Science Foundation of China (Grant No. T2222026).
The authors also gratefully acknowledge the computer resources, technical expertise, and assistance provided by the Centre for High Performance Computing (CHPC-MATS862 \& MATS1424), Cape Town, South Africa. 
\end{acknowledgments}



\bibliography{siesta-honpas,intro,bandgaps}

@article{Si_exp,
    author = {Precker, Jürgen W. and da Silva, Marcı́lio A.},
    title = {Experimental estimation of the band gap in silicon and germanium from the temperature–voltage curve of diode thermometers},
    journal = {Am. J. Phys.},
    volume = {70},
    number = {11},
    pages = {1150-1153},
    year = {2002},
    month = {11},
    issn = {0002-9505},
    doi = {10.1119/1.1512658},
    url = {https://doi.org/10.1119/1.1512658},
}

@article{C_clark1964,
  title={Intrinsic edge absorption in diamond},
  author={Clark, C. D. and Dean, P. J. and Harris, P. V.},
  journal={Proc. R. Soc. Lond. A},
  volume={277},
  number={1370},
  pages={312--329},
  year={1964},
  publisher={The Royal Society London},
  doi = {10.1098/rspa.1964.0025},
}

@article{c-BN_evans2008,
doi = {10.1088/0953-8984/20/7/075233},
url = {https://dx.doi.org/10.1088/0953-8984/20/7/075233},
year = {2008},
month = {jan},
publisher = {},
volume = {20},
number = {7},
pages = {075233},
author = {Evans, D A and McGlynn, A G and Towlson, B M and Gunn, M and Jones, D and Jenkins, T E and Winter, R and Poolton, N R J},
title = {Determination of the optical band-gap energy of cubic and hexagonal boron nitride using
luminescence excitation spectroscopy},
journal = {J. Phys.: Condens. Matter}
}

@article{TiO2_amtout1995optical,
  title = {Optical properties of rutile near its fundamental band gap},
  author = {Amtout, A. and Leonelli, R.},
  journal = {Phys. Rev. B},
  volume = {51},
  issue = {11},
  pages = {6842--6851},
  numpages = {0},
  year = {1995},
  month = {Mar},
  publisher = {American Physical Society},
  doi = {10.1103/PhysRevB.51.6842},
  url = {https://link.aps.org/doi/10.1103/PhysRevB.51.6842}
}

@article{BP_1963ElectricalAO,
  title={Electrical and Optical Properties of Crystalline Black Phosphorus},
  author={Douglas M. Warschauer},
  journal={J. Appl. Phys.},
  year={1963},
  volume={34},
  pages={1853-1860},
  url={https://api.semanticscholar.org/CorpusID:120332779}
}

@article{BP_1953electrical,
  title = {The Electrical Properties of Black Phosphorus},
  author = {Keyes, Robert W.},
  journal = {Phys. Rev.},
  volume = {92},
  issue = {3},
  pages = {580--584},
  numpages = {0},
  year = {1953},
  month = {Nov},
  publisher = {American Physical Society},
  doi = {10.1103/PhysRev.92.580},
  url = {https://link.aps.org/doi/10.1103/PhysRev.92.580}
}

@article{Si_GW2015ab,
author = {Bouhassoune, Mohammed and Schindlmayr, Arno},
title = {Ab Initio Study of Strain Effects on the Quasiparticle Bands and Effective Masses in Silicon},
journal = {Adv. Condens. Matter Phys.},
volume = {2015},
number = {1},
pages = {453125},
doi = {https://doi.org/10.1155/2015/453125},
url = {https://onlinelibrary.wiley.com/doi/abs/10.1155/2015/453125},
year = {2015}
}

@article{G0W0_calcs,
  title = {Self-consistent $\mathrm{GW}$ calculations for semiconductors and insulators},
  author = {Shishkin, M. and Kresse, G.},
  journal = {Phys. Rev. B},
  volume = {75},
  issue = {23},
  pages = {235102},
  numpages = {9},
  year = {2007},
  month = {Jun},
  publisher = {American Physical Society},
  doi = {10.1103/PhysRevB.75.235102},
  url = {https://link.aps.org/doi/10.1103/PhysRevB.75.235102}
}

@article{samsonidze2014insights,
doi = {10.1088/0953-8984/26/47/475501},
url = {https://dx.doi.org/10.1088/0953-8984/26/47/475501},
year = {2014},
month = {oct},
publisher = {IOP Publishing},
volume = {26},
number = {47},
pages = {475501},
author = {Samsonidze, Georgy and Park, Cheol-Hwan and Kozinsky, Boris},
title = {Insights and challenges of applying the $\mathrm{GW}$ method to transition metal oxides},
journal = {J. Phys.: Condens. Matter}
}

@article{ZnO_hashir2022experimental,
doi = {10.1088/1757-899X/1263/1/012025},
url = {https://dx.doi.org/10.1088/1757-899X/1263/1/012025},
year = {2022},
month = {oct},
publisher = {IOP Publishing},
volume = {1263},
number = {1},
pages = {012025},
author = {Hashir, P and Pradyumnan, P.P and Wani, Aadil Fayaz and Kaur, Kulwinder},
title = {Experimental and First-Principles Thermoelectric studies of Bulk $\mathrm{Z}$n$\mathrm{O}$},
journal = {IOP Conf. Ser.: Mater. Sci. Eng.}
}

@article{LiF_2004measurements,
title = {Measurements of insulator band parameters using combination of single-electron and two-electron spectroscopy},
journal = {Solid State Commun.},
volume = {129},
number = {6},
pages = {389-393},
year = {2004},
issn = {0038-1098},
doi = {https://doi.org/10.1016/j.ssc.2003.11.008},
url = {https://www.sciencedirect.com/science/article/pii/S003810980300975X},
author = {S. Samarin and O.M. Artamonov and A.A. Suvorova and A.D. Sergeant and J.F. Williams}
}

@article{LiF_sommer2012quasiparticle,
  title = {Quasiparticle band structure of alkali-metal fluorides, oxides, and nitrides},
  author = {Sommer, Christoph and Kr\"uger, Peter and Pollmann, Johannes},
  journal = {Phys. Rev. B},
  volume = {85},
  issue = {16},
  pages = {165119},
  numpages = {7},
  year = {2012},
  month = {Apr},
  publisher = {American Physical Society},
  doi = {10.1103/PhysRevB.85.165119},
  url = {https://link.aps.org/doi/10.1103/PhysRevB.85.165119}
}

@article{G0W0_salas2022electronic,
  title = {Electronic structure of representative band-gap materials by all-electron quasiparticle self-consistent $\mathrm{GW}$ calculations},
  author = {Salas-Illanes, Nora and Nabok, Dmitrii and Draxl, Claudia},
  journal = {Phys. Rev. B},
  volume = {106},
  issue = {4},
  pages = {045143},
  numpages = {14},
  year = {2022},
  month = {Jul},
  publisher = {American Physical Society},
  doi = {10.1103/PhysRevB.106.045143},
  url = {https://link.aps.org/doi/10.1103/PhysRevB.106.045143}
}

@Article{h-BN_cassabois2016,
author={Cassabois, G.
and Valvin, P.
and Gil, B.},
title={Hexagonal boron nitride is an indirect bandgap semiconductor},
journal={Nat. Photonics},
year={2016},
month={Apr},
day={01},
volume={10},
number={4},
pages={262-266},
issn={1749-4893},
doi={10.1038/nphoton.2015.277},
url={https://doi.org/10.1038/nphoton.2015.277}
}

@article{2d-BN_2016efficient,
  title = {Efficient many-body calculations for two-dimensional materials using exact limits for the screened potential: Band gaps of $\mathrm{MoS}_{2}$, h$-\mathrm{BN}$, and phosphorene},
  author = {Rasmussen, Filip A. and Schmidt, Per S. and Winther, Kirsten T. and Thygesen, Kristian S.},
  journal = {Phys. Rev. B},
  volume = {94},
  issue = {15},
  pages = {155406},
  numpages = {9},
  year = {2016},
  month = {Oct},
  publisher = {American Physical Society},
  doi = {10.1103/PhysRevB.94.155406},
  url = {https://link.aps.org/doi/10.1103/PhysRevB.94.155406}
}

@article{rudenko2015toward,
  title = {Toward a realistic description of multilayer black phosphorus: From $\mathrm{GW}$ approximation to large-scale tight-binding simulations},
  author = {Rudenko, A. N. and Yuan, Shengjun and Katsnelson, M. I.},
  journal = {Phys. Rev. B},
  volume = {92},
  issue = {8},
  pages = {085419},
  numpages = {9},
  year = {2015},
  month = {Aug},
  publisher = {American Physical Society},
  doi = {10.1103/PhysRevB.92.085419},
  url = {https://link.aps.org/doi/10.1103/PhysRevB.92.085419}
}

@article{Huhn2017,
  title = {One-hundred-three compound band-structure benchmark of post-self-consistent spin-orbit coupling treatments in density functional theory},
  author = {Huhn, William P. and Blum, Volker},
  journal = {Phys. Rev. Mater.},
  volume = {1},
  issue = {3},
  pages = {033803},
  numpages = {18},
  year = {2017},
  month = {Aug},
  publisher = {American Physical Society},
  doi = {10.1103/PhysRevMaterials.1.033803},
  url = {https://link.aps.org/doi/10.1103/PhysRevMaterials.1.033803}
}

@article{Paier2006err,
    author = {Paier, J. and Marsman, M. and Hummer, K. and Kresse, G. and Gerber, I. C. and Ángyán, J. G.},
    title = {Erratum: “Screened hybrid density functionals applied to solids” [J. Chem. Phys. 124, 154709 (2006)]},
    journal = {The Journal of Chemical Physics},
    volume = {125},
    number = {24},
    pages = {249901},
    year = {2006},
    month = {12},
    issn = {0021-9606},
    doi = {10.1063/1.2403866},
    url = {https://doi.org/10.1063/1.2403866}
}

@article{Hedin1965,
  title = {New Method for Calculating the One-Particle Green's Function with Application to the Electron-Gas Problem},
  author = {Hedin, Lars},
  journal = {Phys. Rev.},
  volume = {139},
  issue = {3A},
  pages = {A796--A823},
  numpages = {0},
  year = {1965},
  month = {Aug},
  publisher = {American Physical Society},
  doi = {10.1103/PhysRev.139.A796},
  url = {https://link.aps.org/doi/10.1103/PhysRev.139.A796}
}

@article{Hybertsen1986,
  title = {Electron correlation in semiconductors and insulators: Band gaps and quasiparticle energies},
  author = {Hybertsen, Mark S. and Louie, Steven G.},
  journal = {Phys. Rev. B},
  volume = {34},
  issue = {8},
  pages = {5390--5413},
  numpages = {0},
  year = {1986},
  month = {Oct},
  publisher = {American Physical Society},
  doi = {10.1103/PhysRevB.34.5390},
  url = {https://link.aps.org/doi/10.1103/PhysRevB.34.5390}
}

@article{Rohlfing1998,
  title = {Electron-Hole Excitations in Semiconductors and Insulators},
  author = {Rohlfing, Michael and Louie, Steven G.},
  journal = {Phys. Rev. Lett.},
  volume = {81},
  issue = {11},
  pages = {2312--2315},
  numpages = {0},
  year = {1998},
  month = {Sep},
  publisher = {American Physical Society},
  doi = {10.1103/PhysRevLett.81.2312},
  url = {https://link.aps.org/doi/10.1103/PhysRevLett.81.2312}
}

@article{Rohlfing2000,
  title = {Electron-hole excitations and optical spectra from first principles},
  author = {Rohlfing, Michael and Louie, Steven G.},
  journal = {Phys. Rev. B},
  volume = {62},
  issue = {8},
  pages = {4927--4944},
  numpages = {0},
  year = {2000},
  month = {Aug},
  publisher = {American Physical Society},
  doi = {10.1103/PhysRevB.62.4927},
  url = {https://link.aps.org/doi/10.1103/PhysRevB.62.4927}
}

@article{Onida2002,
  title = {Electronic excitations: density-functional versus many-body $\mathrm{G}$reen's-function approaches},
  author = {Onida, Giovanni and Reining, Lucia and Rubio, Angel},
  journal = {Rev. Mod. Phys.},
  volume = {74},
  issue = {2},
  pages = {601--659},
  numpages = {0},
  year = {2002},
  month = {Jun},
  publisher = {American Physical Society},
  doi = {10.1103/RevModPhys.74.601},
  url = {https://link.aps.org/doi/10.1103/RevModPhys.74.601}
}

@article{Golze2019,
  title={The $\mathrm{GW}$ Compendium: A Practical Guide to Theoretical Photoemission Spectroscopy},
  author={Golze, Dorothea and Dvorak, Marc and Rinke, Patrick},
  journal={Front. Chem.},
  year={2019},
  volume={7},
  url={https://api.semanticscholar.org/CorpusID:195837430}
}

@article{Abedi2023,
    author = {Abedi, Saeid and Tarighi Ahmadpour, Mahdi and Baninajarian, Samira and Kahnouji, Hamideh and Hashemifar, S. Javad and Han, Zhong-Kang and Levchenko, Sergey V.},
    title = {Statistical analysis of the performance of a variety of first-principles schemes for accurate prediction of binary semiconductor band gaps},
    journal = {J. Chem. Phys.},
    volume = {158},
    number = {18},
    pages = {184109},
    year = {2023},
    month = {05},
    issn = {0021-9606},
    doi = {10.1063/5.0138775},
    url = {https://doi.org/10.1063/5.0138775}
}

@article{Camarasa-Gomez2023,
  title = {Transferable screened range-separated hybrid functionals for electronic and optical properties of van der $\mathrm{W}$aals materials},
  author = {Camarasa-G\'omez, Mar\'{\i}a and Ramasubramaniam, Ashwin and Neaton, Jeffrey B. and Kronik, Leeor},
  journal = {Phys. Rev. Mater.},
  volume = {7},
  issue = {10},
  pages = {104001},
  numpages = {15},
  year = {2023},
  month = {Oct},
  publisher = {American Physical Society},
  doi = {10.1103/PhysRevMaterials.7.104001},
  url = {https://link.aps.org/doi/10.1103/PhysRevMaterials.7.104001}
}

@article{kolos2025,
author = {Kolos, Miroslav and Karlický, František},
title = {Predicting Fundamental Gaps of Chromium-Based $\mathrm{2D}$ Materials Using $\mathrm{GW}$ Methods},
journal = {J. Phys. Chem. C},
volume = {129},
number = {5},
pages = {2782-2787},
year = {2025},
doi = {10.1021/acs.jpcc.4c07981},
url = {https://doi.org/10.1021/acs.jpcc.4c07981}
}

@article{Gao2016,
title = {On the applicability of hybrid functionals for predicting fundamental properties of metals},
journal = {Solid State Commun.},
volume = {234-235},
pages = {10-13},
year = {2016},
issn = {0038-1098},
doi = {https://doi.org/10.1016/j.ssc.2016.02.014},
url = {https://www.sciencedirect.com/science/article/pii/S0038109816000727},
author = {Gao, Weiwei and Abtew, Tesfaye A. and Cai, Tianyi and Sun, Yi-Yang and Zhang, Shengbai and Zhang, Peihong}
}

@article{DasSarma2011,
  title = {Electronic transport in two-dimensional graphene},
  author = {Das Sarma, S. and Adam, Shaffique and Hwang, E. H. and Rossi, Enrico},
  journal = {Rev. Mod. Phys.},
  volume = {83},
  issue = {2},
  pages = {407--470},
  numpages = {0},
  year = {2011},
  month = {May},
  publisher = {American Physical Society},
  doi = {10.1103/RevModPhys.83.407},
  url = {https://link.aps.org/doi/10.1103/RevModPhys.83.407}
}

@article{Wu2009,
  title = {Order-$\mathrm{N}$ implementation of exact exchange in extended insulating systems},
  author = {Wu, Xifan and Selloni, Annabella and Car, Roberto},
  journal = {Phys. Rev. B},
  volume = {79},
  issue = {8},
  pages = {085102},
  numpages = {5},
  year = {2009},
  month = {Feb},
  publisher = {American Physical Society},
  doi = {10.1103/PhysRevB.79.085102},
  url = {https://link.aps.org/doi/10.1103/PhysRevB.79.085102}
}

@article{Gygi2010,
title = "A scalable and accurate algorithm for the computation of $\mathrm{H}$artree–$\mathrm{F}$ock exchange",
journal = "Comput. Phys. Commun.",
volume = "181",
number = "5",
pages = "855 - 860",
year = "2010",
issn = "0010-4655",
doi = "https://doi.org/10.1016/j.cpc.2009.12.021",
url = "http://www.sciencedirect.com/science/article/pii/S0010465509004135",
author = "Ivan Duchemin and François Gygi"
}

@article{Deslippe2017,
title = "Improved treatment of exact exchange in $\mathrm{Q}$uantum \textsc{espresso}",
journal = "Comput. Phys. Commun.",
volume = "214",
pages = "52 - 58",
year = "2017",
issn = "0010-4655",
doi = "https://doi.org/10.1016/j.cpc.2017.01.008",
url = "http://www.sciencedirect.com/science/article/pii/S0010465517300085",
author = "Taylor A. Barnes and Thorsten Kurth and Pierre Carrier and Nathan Wichmann and David Prendergast and Paul R.C. Kent and Jack Deslippe"
}

@article{DiStasio2014,
author = {DiStasio,Robert A.  and Santra,Biswajit  and Li,Zhaofeng  and Wu,Xifan  and Car,Roberto },
title = {The individual and collective effects of exact exchange and dispersion interactions on the ab initio structure of liquid water},
journal = {J. Chem. Phys.},
volume = {141},
number = {8},
pages = {084502},
year = {2014},
doi = {10.1063/1.4893377},

URL = { 
        https://doi.org/10.1063/1.4893377
    
},
}

@article{Natan2016,
author = {Boffi, Nicholas M and Jain, Manish and Natan, Amir},
doi = {10.1021/acs.jctc.6b00376},
issn = {1549-9618},
journal = {J. Chem. Theory Comput.},
month = {aug},
number = {8},
pages = {3614--3622},
publisher = {American Chemical Society},
title = {{Efficient Computation of the $\mathrm{H}$artree–$\mathrm{F}$ock Exchange in Real-Space with Projection Operators}},
url = {https://doi.org/10.1021/acs.jctc.6b00376},
volume = {12},
year = {2016}
}

@Article{Natan2015,
author ="Natan, Amir",
title  ="Fock-exchange for periodic structures in the real-space formalism and the $\mathrm{KLI}$ approximation",
journal  ="Phys. Chem. Chem. Phys.",
year  ="2015",
volume  ="17",
issue  ="47",
pages  ="31510-31515",
publisher  ="The Royal Society of Chemistry",
doi  ="10.1039/C5CP01093D",
url  ="http://dx.doi.org/10.1039/C5CP01093D"}

@article{Shang2019,
author = {Qin, Xinming and Shang, Honghui and Xu, Lei and Hu, Wei and Yang, Jinlong and Li, Shigang and Zhang, Yunquan},
year = {2019},
month = {05},
volume={34},
pages = {159-168},
title = {The static parallel distribution algorithms for hybrid density-functional calculations in \textsc{honpas} package},
journal = {Int. J. High Perform. Comput. Appl.},
doi = {10.1177/1094342019845046}
}

@Article{Lin2016,
author={Lin, Lin},
title={Adaptively Compressed Exchange Operator},
journal={J. Chem. Theory Comput.},
year={2016},
month={May},
day={10},
publisher={American Chemical Society},
volume={12},
number={5},
pages={2242-2249},
issn={1549-9618},
doi={10.1021/acs.jctc.6b00092},
url={https://doi.org/10.1021/acs.jctc.6b00092}
}

@article{Gonze2016,
	author = {X. Gonze and F. Jollet and F. {Abreu Araujo} and D. Adams and B. Amadon and T. Applencourt and C. Audouze and J.-M. Beuken and J. Bieder and A. Bokhanchuk and E. Bousquet and F. Bruneval and D. Caliste and M. Côté and F. Dahm and F. {Da Pieve} and M. Delaveau and M. {Di Gennaro} and B. Dorado and C. Espejo and G. Geneste and L. Genovese and A. Gerossier and M. Giantomassi and Y. Gillet and D.R. Hamann and L. He and G. Jomard and J. {Laflamme Janssen} and S. {Le Roux} and A. Levitt and A. Lherbier and F. Liu and I. Lukačević and A. Martin and C. Martins and M.J.T. Oliveira and S. Poncé and Y. Pouillon and T. Rangel and G.-M. Rignanese and A.H. Romero and B. Rousseau and O. Rubel and A.A. Shukri and M. Stankovski and M. Torrent and M.J. {Van Setten} and B. {Van Troeye} and M.J. Verstraete and D. Waroquiers and J. Wiktor and B. Xu and A. Zhou and J.W. Zwanziger},
	url={https://doi.org/10.1016/j.cpc.2016.04.003},
	Journal = { Comput. Phys. Commun.},
	Pages = {106 - 131},
	Title = {Recent developments in the {\sc ABINIT} software package},
	Volume = {205},
	Year = {2016}
	}

@article{NWChem,
	doi = {10.1016/j.cpc.2010.04.018},
	url = {https://doi.org/10.1016\%2Fj.cpc.2010.04.018},
	year = 2010,
	month = {sep},
	publisher = {Elsevier {BV}},
	volume = {181},
	number = {9},
	pages = {1477--1489},
	author = {M. Valiev and E.J. Bylaska and N. Govind and K. Kowalski and T.P. Straatsma and H.J.J. Van Dam and D. Wang and J. Nieplocha and E. Apra and T.L. Windus and W.A. de Jong},
	title = {\textsc{NWChem}: A comprehensive and scalable open-source solution for large scale molecular simulations},
	journal = {Comput. Phys. Commun.}
}

@article{Quickstep,
	doi = {10.1016/j.cpc.2004.12.014},
	url = {https://doi.org/10.1016\%2Fj.cpc.2004.12.014},
	year = 2005,
	month = {apr},
	publisher = {Elsevier {BV}},
	volume = {167},
	number = {2},
	pages = {103--128},
	author = {Joost VandeVondele and Matthias Krack and Fawzi Mohamed and Michele Parrinello and Thomas Chassaing and Jürg Hutter},
	title = {\textsc{Quickstep}: Fast and accurate density functional calculations using a mixed Gaussian and plane waves approach},
	journal = {Comput. Phys. Commun.}
}

@inproceedings{GTFock_1,
    doi = {10.1109/ipdps.2014.97},
	url = {https://doi.org/10.1109\%2Fipdps.2014.97},
	year = 2014,
	month = {may},
	publisher = {{IEEE}},
	author = {Xing Liu and Aftab Patel and Edmond Chow},
	title = {A New Scalable Parallel Algorithm for $\mathrm{F}$ock Matrix Construction},
	booktitle = {2014 {IEEE} 28th International Parallel and Distributed Processing Symposium}
}

@article{GTFock_2,
	doi = {10.1063/1.4913961},
	url = {https://doi.org/10.1063\%2F1.4913961},
	year = 2015,
	month = {mar},
	publisher = {{AIP} Publishing},
	volume = {142},
	number = {10},
	pages = {104103},
	author = {Edmond Chow and Xing Liu and Mikhail Smelyanskiy and Jeff R. Hammond},
	title = {Parallel scalability of $\mathrm{H}$artree{\textendash}$\mathrm{F}$ock calculations},
	journal = {J. Chem. Phys.}
}

@article{Schwegler_1996,
	doi = {10.1063/1.472135},
	url = {https://doi.org/10.1063\%2F1.472135},
	year = 1996,
	month = {aug},
	publisher = {{AIP} Publishing},
	volume = {105},
	number = {7},
	pages = {2726--2734},
	author = {Eric Schwegler and Matt Challacombe},
	title = {Linear scaling computation of the $\mathrm{H}$artree{\textendash}$\mathrm{F}$ock exchange matrix},
	journal = {J. Chem. Phys.}
}

@article{Burant_1996,
	doi = {10.1063/1.472627},
	url = {https://doi.org/10.1063\%2F1.472627},
	year = 1996,
	month = {nov},
	publisher = {{AIP} Publishing},
	volume = {105},
	number = {19},
	pages = {8969--8972},
	author = {John C. Burant and Gustavo E. Scuseria and Michael J. Frisch},
	title = {A linear scaling method for $\mathrm{H}$artree{\textendash}$\mathrm{F}$ock exchange calculations of large molecules},
	journal = {J. Chem. Phys.}
}

@article{Schwegler_1997,
	doi = {10.1063/1.473833},
	url = {https://doi.org/10.1063\%2F1.473833},
	year = 1997,
	month = {jun},
	publisher = {{AIP} Publishing},
	volume = {106},
	number = {23},
	pages = {9708--9717},
	author = {Eric Schwegler and Matt Challacombe and Martin Head-Gordon},
	title = {Linear scaling computation of the $\mathrm{F}$ock matrix. {II}. Rigorous bounds on exchange integrals and incremental $\mathrm{F}$ock build},
	journal = {J. Chem. Phys.}
}

@article{Ochsenfeld_1998,
	doi = {10.1063/1.476741},
	url = {https://doi.org/10.1063\%2F1.476741},
	year = 1998,
	month = {aug},
	publisher = {{AIP} Publishing},
	volume = {109},
	number = {5},
	pages = {1663--1669},
	author = {Christian Ochsenfeld and Christopher A. White and Martin Head-Gordon},
	title = {Linear and sublinear scaling formation of $\mathrm{H}$artree{\textendash}$\mathrm{F}$ock-type exchange matrices},
	journal = {J. Chem. Phys.}
}

@article{Polly_2004,
	doi = {10.1080/0026897042000274801},
	url = {https://doi.org/10.1080\%2F0026897042000274801},
	year = 2004,
	month = {nov},
	publisher = {Informa {UK} Limited},
	volume = {102},
	number = {21-22},
	pages = {2311--2321},
	author = {Robert Polly and Hans-Joachim Werner and Frederick R. Manby and Peter J. Knowles},
	title = {Fast $\mathrm{H}$artree{\textendash}$\mathrm{F}$ock theory using local density fitting approximations},
	journal = {Mol. Phys.}
}

@article{Sodt_2008,
	doi = {10.1063/1.2828533},
	url = {https://doi.org/10.1063\%2F1.2828533},
	year = 2008,
	month = {mar},
	publisher = {{AIP} Publishing},
	volume = {128},
	number = {10},
	pages = {104106},
	author = {Alex Sodt and Martin Head-Gordon},
	title = {$\mathrm{H}$artree-$\mathrm{F}$ock exchange computed using the atomic resolution of the identity approximation},
	journal = {J. Chem. Phys.}
}

@article{Guidon_2010,
	doi = {10.1021/ct1002225},
	url = {https://doi.org/10.1021\%2Fct1002225},
	year = 2010,
	month = {aug},
	publisher = {American Chemical Society ({ACS})},
	volume = {6},
	number = {8},
	pages = {2348--2364},
	author = {Manuel Guidon and Jurg Hutter and Joost VandeVondele},
	title = {Auxiliary Density Matrix Methods for $\mathrm{H}$artree-$\mathrm{F}$ock Exchange Calculations},
	journal = {J. Chem. Theory Comput.}
}

@article{Merlot_2014,
	doi = {10.1063/1.4894267},
	url = {https://doi.org/10.1063\%2F1.4894267},
	year = 2014,
	month = {sep},
	publisher = {{AIP} Publishing},
	volume = {141},
	number = {9},
	pages = {094104},
	author = {Patrick Merlot and R{\'{o}}bert Izs{\'{a}}k and Alex Borgoo and Thomas Kj{\ae}rgaard and Trygve Helgaker and Simen Reine},
	title = {Charge-constrained auxiliary-density-matrix methods for the $\mathrm{H}$artree-$\mathrm{F}$ock exchange contribution},
	journal = {J. Chem. Phys.}
}

@article{Blum2009,
author = {Blum, Volker and Gehrke, Ralf and Hanke, Felix and Havu, Paula and Havu, Ville and Ren, Xinguo and Reuter, Karsten and Scheffler, Matthias},
doi = {10.1016/j.cpc.2009.06.022},
issn = {00104655},
journal = {Comput. Phys. Commun.},
month = nov,
number = {11},
pages = {2175--2196},
publisher = {Elsevier B.V.},
title = {{Ab initio molecular simulations with numeric atom-centered orbitals}},
url = {http://linkinghub.elsevier.com/retrieve/pii/S0010465509002033},
volume = {180},
year = {2009}
}

@article{Havu2009,
author = {Havu, V. and Blum, V. and Havu, P. and Scheffler, M.},
doi = {10.1016/j.jcp.2009.08.008},
issn = {00219991},
journal = {J. Comput. Phys.},
month = dec,
number = {22},
pages = {8367--8379},
publisher = {Elsevier Inc.},
title = {{Efficient integration for all-electron electronic structure calculation using numeric basis functions}},
url = {http://linkinghub.elsevier.com/retrieve/pii/S0021999109004458},
volume = {228},
year = {2009}
}

@article{Ren2012,
author = {Ren, Xinguo and Rinke, Patrick and Blum, Volker and Wieferink, J\"{u}rgen and Tkatchenko, Alexandre and Sanfilippo, Andrea and Reuter, Karsten and Scheffler, Matthias},
doi = {10.1088/1367-2630/14/5/053020},
issn = {1367-2630},
journal = {New J. Phys.},
month = may,
number = {5},
pages = {053020},
title = {{Resolution-of-identity approach to Hartree–Fock, hybrid density functionals, RPA, MP2 and GW with numeric atom-centered orbital basis functions}},
url = {http://stacks.iop.org/1367-2630/14/i=5/a=053020?key=crossref.351b343783c2c1df1596219a941a74eb},
volume = {14},
year = {2012}
}

@article{Bush2011,
	doi = {10.1098/rspa.2010.0563},
	url = {https://doi.org/10.1098\%2Frspa.2010.0563},
	year = 2011,
	month = {apr},
	publisher = {The Royal Society},
	volume = {467},
	number = {2131},
	pages = {2112--2126},
	author = {I. J. Bush and S. Tomic and B. G. Searle and G. Mallia and C. L. Bailey and B. Montanari and L. Bernasconi and J. M. Carr and N. M. Harrison},
	title = {Parallel implementation of the \textit{ab initio} \textsc{crystal} program: electronic structure calculations for periodic systems},
	journal = {Proc. R. Soc. A}
}

@article{Stephens_1994,
    doi = {10.1021/j100096a001},
	url = {https://doi.org/10.1021\%2Fj100096a001},
	year = 1994,
	month = {nov},
	publisher = {American Chemical Society ({ACS})},
	volume = {98},
	number = {45},
	pages = {11623--11627},
	author = {P. J. Stephens and F. J. Devlin and C. F. Chabalowski and M. J. Frisch},
	title = {Ab Initio Calculation of Vibrational Absorption and Circular Dichroism Spectra Using Density Functional Force Fields},
	journal = {J. Phys. Chem.}
}

@article{Janesko_2009,
    doi = {10.1039/b812838c},
	url = {https://doi.org/10.1039\%2Fb812838c},
	year = 2009,
	publisher = {Royal Society of Chemistry ({RSC})},
	volume = {11},
	number = {3},
	pages = {443--454},
	author = {Benjamin G. Janesko and Thomas M. Henderson and Gustavo E. Scuseria},
	title = {Screened hybrid density functionals for solid-state chemistry and physics},
	journal = {Phys. Chem. Chem. Phys.}
}

@article{HSE03-1,
    doi = {10.1063/1.1564060},
	url = {https://doi.org/10.1063\%2F1.1564060},
	year = 2003,
	month = {may},
	publisher = {{AIP} Publishing},
	volume = {118},
	number = {18},
	pages = {8207--8215},
	author = {Jochen Heyd and Gustavo E. Scuseria and Matthias Ernzerhof},
	title = {Hybrid functionals based on a screened $\mathrm{C}$oulomb potential},
	journal = {J. Chem. Phys.}
}

@article{HSE03-2,
	doi = {10.1063/1.2204597},
	url = {https://doi.org/10.1063\%2F1.2204597},
	year = 2006,
	month = {jun},
	publisher = {{AIP} Publishing},
	volume = {124},
	number = {21},
	pages = {219906},
	author = {Jochen Heyd and Gustavo E. Scuseria and Matthias Ernzerhof},
	title = {Erratum: {\textquotedblleft}$\mathrm{H}$ybrid functionals based on a screened $\mathrm{C}$oulomb potential{\textquotedblright} [$\mathrm{J}$ $\mathrm{C}$hem. $\mathrm{P}$hys. 118, 8207 (2003)]},
	journal = {J. Chem. Phys.}
}

@article{HSE06,
    doi = {10.1063/1.2404663},
	url = {https://doi.org/10.1063\%2F1.2404663},
	year = 2006,
	month = {dec},
	publisher = {{AIP} Publishing},
	volume = {125},
	number = {22},
	pages = {224106},
	author = {Aliaksandr V. Krukau and Oleg A. Vydrov and Artur F. Izmaylov and Gustavo E. Scuseria},
	title = {Influence of the exchange screening parameter on the performance of screened hybrid functionals},
	journal = {J. Chem. Phys.}
}

@article{Garza2016,
  author = {Garza, Alejandro J. and Scuseria, Gustavo E.},
  title = {Predicting Band Gaps with Hybrid Density Functionals},
  journal = {J. Phys. Chem. Lett.},
  volume = {7},
  number = {20},
  pages = {4165--4170},
  year = {2016},
  doi = {10.1021/acs.jpclett.6b01807}
}

@article{Yang2023,
  author = {Yang, Jing and Falletta, Stefano and Pasquarello, Alfredo},
  title = {Range-separated hybrid functionals for accurate prediction of band gaps of extended systems},
  journal = {npj Comput. Mater.},
  volume = {9},
  pages = {108},
  year = {2023},
  doi = {10.1038/s41524-023-01064-x}
}

@article{Lany2010,
  author = {Lany, Stephan and Zunger, Alex},
  title = {Polaronic hole localization and multiple hole binding of acceptors in oxide wide-gap semiconductors},
  journal = {Phys. Rev. B},
  volume = {80},
  number = {8},
  pages = {085202},
  year = {2009},
  doi = {10.1103/PhysRevB.80.085202}
}

@article{Paier2006,
  author = {Paier, Joachim and Marsman, Martijn and Hummer, K. and Kresse, Georg and Gerber, I. C. and \'{A}ngy\'{a}n, J. G.},
  title = {Screened hybrid density functionals applied to solids},
  journal = {J. Chem. Phys.},
  volume = {124},
  number = {15},
  pages = {154709},
  year = {2006},
  doi = {10.1063/1.2187006}
}

@article{Borlido2019,
  author = {Borlido, Pedro and Aull, Thorsten and Huran, Ahmad W. and Tran, Fabien and Marques, Miguel A. L. and Botti, Silvana},
  title = {Large-Scale Benchmark of Exchange–Correlation Functionals for the Determination of Electronic Band Gaps of Solids},
  journal = {J. Chem. Theory Comput.},
  volume = {15},
  number = {9},
  pages = {5069--5079},
  year = {2019},
  doi = {10.1021/acs.jctc.9b00322}
}

@misc{Gaussian,
    author = {Frisch, M. J. and Trucks, G. W. and Schlegel, H. B. and Scuseria, G. E. and Robb, M. A. and Cheeseman, J. R. and Scalmani, G. and Barone, V. and Mennucci, B. and Petersson, G. A. and Nakatsuji, H. and Caricato, M. and Li, X. and Hratchian, H. P. and Izmaylov, A. F. and Bloino, J. and Zheng, G. and Sonnenberg, J. L. and Hada, M. and Ehara, M. and Toyota, K. and Fukuda, R. and Hasegawa, J. and Ishida, M. and Nakajima, T. and Honda, Y. and Kitao, O. and Nakai, H. and Vreven, T. and Montgomery, J. A. and Peralta, J. E. and Ogliaro, F. and Bearpark, M. and Heyd, J. J. and Brothers, E. and Kudin, K. N. and Staroverov, V. N. and Kobayashi, R. and Normand, J. and Raghavachari, K. and Rendell, A. and Burant, J. C. and Iyengar, S. S. and Tomasi, J. and Cossi, M. and Rega, N. and Millam, J. M. and Klene, M. and Knox, J. E. and Cross, J. B. and Bakken, V. and Adamo, C. and Jaramillo, J. and Gomperts, R. and Stratmann, R. E. and Yazyev, O. and Austin, A. J. and Cammi, R. and Pomelli, C. and Ochterski, J. W. and Martin, R. L. and Morokuma, K. and Zakrzewski, V. G. and Voth, G. A. and Salvador, P. and Dannenberg, J. J. and Dapprich, S. and Daniels, A. D. and Farkas and Foresman, J. B. and Ortiz, J. V. and Cioslowski, J. and Fox, D. J.},
    institution = {Gaussian, Inc.},
    journal = {Gaussian 09, Revision B.01, Gaussian, Inc., Wallingford CT},
    location = {Wallingford CT},
    title = {{Gaussian 09, Revision B.01}},
    year = {2009}
}

@article{Shang-JCP,
	doi = {10.1063/1.3610379},
	url = {https://doi.org/10.1063\%2F1.3610379},
	year = 2011,
	month = {jul},
	publisher = {{AIP} Publishing},
	volume = {135},
	number = {3},
	pages = {034110},
	author = {Honghui Shang and Zhenyu Li and Jinlong Yang},
	title = {Implementation of screened hybrid density functional for periodic systems with numerical atomic orbitals: Basis function fitting and integral screening},
	journal = {J. Chem. Phys.}
}

@article{CRYSTAL,
author = {Bush, I. J. and Tomi{\'{c}}, S. and Searle, B. G. and Mallia, G. and Bailey, C. L. and Montanari, B. and Bernasconi, L. and Carr, J. M. and Harrison, N. M.},
doi = {10.1098/rspa.2010.0563},
isbn = {1364-5021$\backslash$n1471-2946},
issn = {13645021},
journal = {Proc. R. Soc. A},
number = {2131},
pages = {2112--2126},
title = {{Parallel implementation of the \textit{ab initio} \textsc{crystal} program: Electronic structure calculations for periodic systems}},
volume = {467},
year = {2011}
}

@article{SIESTA,
	doi = {10.1088/0953-8984/14/11/302},
	url = {https://doi.org/10.1088\%2F0953-8984\%2F14\%2F11\%2F302},
	year = 2002,
	month = {mar},
	publisher = {{IOP} Publishing},
	volume = {14},
	number = {11},
	pages = {2745--2779},
	author = {Jos{\'{e}} M Soler and Emilio Artacho and Julian D Gale and Alberto Garc{\'{\i}}a and Javier Junquera and Pablo Ordej{\'{o}}n and Daniel S{\'{a}}nchez-Portal},
	title = {The \textsc{siesta} method for \textit{ab initio} order-$\mathrm{N}$ materials simulation},
	journal = {J. Phys.: Condens. Matter}
}

@article{DMOL,
	doi = {10.1063/1.458452},
	url = {https://doi.org/10.1063\%2F1.458452},
	year = 1990,
	month = {jan},
	publisher = {{AIP} Publishing},
	volume = {92},
	number = {1},
	pages = {508--517},
	author = {B. Delley},
	title = {An all-electron numerical method for solving the local density functional for polyatomic molecules},
	journal = {J. Chem. Phys.}
}

@article{OPENMX,
  title = {Variationally optimized atomic orbitals for large-scale electronic structures},
  author = {Ozaki, T.},
  journal = {Phys. Rev. B},
  volume = {67},
  issue = {15},
  pages = {155108},
  numpages = {5},
  year = {2003},
  month = {Apr},
  publisher = {American Physical Society},
  doi = {10.1103/PhysRevB.67.155108},
  url = {https://link.aps.org/doi/10.1103/PhysRevB.67.155108}
}

@article{CONQUEST,
   doi = {10.1088/0953-8984/20/29/294206},
	url = {https://doi.org/10.1088\%2F0953-8984\%2F20\%2F29\%2F294206},
	year = 2008,
	month = {jun},
	publisher = {{IOP} Publishing},
	volume = {20},
	number = {29},
	pages = {294206},
	author = {A S Torralba and M Todorovi{\'{c}} and V Br{\'{a}}zdov{\'{a}} and R Choudhury and T Miyazaki and M J Gillan and D R Bowler},
	title = {Pseudo-atomic orbitals as basis sets for the $\mathrm{O(N)}$ {DFT} code {\textsc{CONQUEST}}},
	journal = {J. Phys.: Condens. Matter}
}

@article{Elsasser-90,
doi = {10.1088/0953-8984/2/19/006},
url = {https://dx.doi.org/10.1088/0953-8984/2/19/006},
year = {1990},
month = {may},
volume = {2},
pages = {4371},
author = {C. Elsasser and N. Takeuchi and K. M. Ho and C. T. Chan and P. Braun and M. Fahnle},
title = {Relativistic effects on ground state properties of 4d and 5d transition metals},
journal = {J. Phys.: Condens. Matter}
}

@misc{abbott2025,
      title={Roadmap on Advancements of the $\mathrm{FHI}$-aims Software Package}, 
      author={Joseph W. Abbott and Carlos Mera Acosta and Alaa Akkoush and Alberto Ambrosetti and Viktor Atalla and Alexej Bagrets and Jörg Behler and Daniel Berger and Björn Bieniek and Jonas Björk and Volker Blum and Saeed Bohloul and Connor L. Box and Nicholas Boyer and Danilo Simoes Brambila and Gabriel A. Bramley and Kyle R. Bryenton and María Camarasa-Gómez and Christian Carbogno and Fabio Caruso and Sucismita Chutia and Michele Ceriotti and Gábor Csányi and William Dawson and Francisco A. Delesma and Fabio Della Sala and Bernard Delley and Robert A. DiStasio Jr. and Maria Dragoumi and Sander Driessen and Marc Dvorak and Simon Erker and Ferdinand Evers and Eduardo Fabiano and Matthew R. Farrow and Florian Fiebig and Jakob Filser and Lucas Foppa and Lukas Gallandi and Alberto Garcia and Ralf Gehrke and Simiam Ghan and Luca M. Ghiringhelli and Mark Glass and Stefan Goedecker and Dorothea Golze and Matthias Gramzow and James A. Green and Andrea Grisafi and Andreas Grüneis and Jan Günzl and Stefan Gutzeit and Samuel J. Hall and Felix Hanke and Ville Havu and Xingtao He and Joscha Hekele and Olle Hellman and Uthpala Herath and Jan Hermann and Daniel Hernangómez-Pérez and Oliver T. Hofmann and Johannes Hoja and Simon Hollweger and Lukas Hörmann and Ben Hourahine and Wei Bin How and William P. Huhn and Marcel Hülsberg and Timo Jacob and Sara Panahian Jand and Hong Jiang and Erin R. Johnson and Werner Jürgens and J. Matthias Kahk and Yosuke Kanai and Kisung Kang and Petr Karpov and Elisabeth Keller and Roman Kempt and Danish Khan and Matthias Kick and Benedikt P. Klein and Jan Kloppenburg and Alexander Knoll and Florian Knoop and Franz Knuth and Simone S. Köcher and Jannis Kockläuner and Sebastian Kokott and Thomas Körzdörfer and Hagen-Henrik Kowalski and Peter Kratzer and Pavel Kůs and Raul Laasner and Bruno Lang and Björn Lange and Marcel F. Langer and Ask Hjorth Larsen and Hermann Lederer and Susi Lehtola and Maja-Olivia Lenz-Himmer and Moritz Leucke and Sergey Levchenko and Alan Lewis and O. Anatole von Lilienfeld and Konstantin Lion and Werner Lipsunen and Johannes Lischner and Yair Litman and Chi Liu and Qing-Long Liu and Andrew J. Logsdail and Michael Lorke and Zekun Lou and Iuliia Mandzhieva and Andreas Marek and Johannes T. Margraf and Reinhard J. Maurer and Tobias Melson and Florian Merz and Jörg Meyer and Georg S. Michelitsch and Teruyasu Mizoguchi and Evgeny Moerman and Dylan Morgan and Jack Morgenstein and Jonathan Moussa and Akhil S. Nair and Lydia Nemec and Harald Oberhofer and Alberto Otero-de-la-Roza and Ramón L. Panadés-Barrueta and Thanush Patlolla and Mariia Pogodaeva and Alexander Pöppl and Alastair J. A. Price and Thomas A. R. Purcell and Jingkai Quan and Nathaniel Raimbault and Markus Rampp and Karsten Rasim and Ronald Redmer and Xinguo Ren and Karsten Reuter and Norina A. Richter and Stefan Ringe and Patrick Rinke and Simon P. Rittmeyer and Herzain I. Rivera-Arrieta and Matti Ropo and Mariana Rossi and Victor Ruiz and Nikita Rybin and Andrea Sanfilippo and Matthias Scheffler and Christoph Scheurer and Christoph Schober and Franziska Schubert and Tonghao Shen and Christopher Shepard and Honghui Shang and Kiyou Shibata and Andrei Sobolev and Ruyi Song and Aloysius Soon and Daniel T. Speckhard and Pavel V. Stishenko and Muhammad Tahir and Izumi Takahara and Jun Tang and Zechen Tang and Thomas Theis and Franziska Theiss and Alexandre Tkatchenko and Milica Todorović and George Trenins and Oliver T. Unke and Álvaro Vázquez-Mayagoitia and Oscar van Vuren and Daniel Waldschmidt and Han Wang and Yanyong Wang and Jürgen Wieferink and Jan Wilhelm and Scott Woodley and Jianhang Xu and Yong Xu and Yi Yao and Yingyu Yao and Mina Yoon and Victor Wen-zhe Yu and Zhenkun Yuan and Marios Zacharias and Igor Ying Zhang and Min-Ye Zhang and Wentao Zhang and Rundong Zhao and Shuo Zhao and Ruiyi Zhou and Yuanyuan Zhou and Tong Zhu},
      year={2025},
      eprint={2505.00125},
      archivePrefix={arXiv},
      primaryClass={cond-mat.mtrl-sci},
      url={https://arxiv.org/abs/2505.00125}, 
}

@article{HSESol,
    author = {Schimka, Laurids and Harl, Judith and Kresse, Georg},
    title = {Improved hybrid functional for solids: The $\mathrm{HSE}$sol functional},
    journal = {J. Chem. Phys.},
    volume = {134},
    number = {2},
    pages = {024116},
    year = {2011},
    month = {01},
    issn = {0021-9606},
    doi = {10.1063/1.3524336},
    url = {https://doi.org/10.1063/1.3524336},
}

@article{Zhang2018,
doi = {10.1088/1367-2630/aac7f0},
url = {https://dx.doi.org/10.1088/1367-2630/aac7f0},
year = {2018},
month = {jun},
publisher = {IOP Publishing},
volume = {20},
number = {6},
pages = {063020},
author = {Zhang, Guo-Xu and Reilly, Anthony M and Tkatchenko, Alexandre and Scheffler, Matthias},
title = {Performance of various density-functional approximations for cohesive properties of 64 bulk solids},
journal = {New J. Phys.}
}

@article{Ke2025,
  title = {Accurate point defect energy levels from non-empirical screened range-separated hybrid functionals: The case of native vacancies in $\mathrm{Z}$n$\mathrm{O}$},
  author = {Ke, Sijia and Gant, Stephen E. and Kronik, Leeor and Neaton, Jeffrey B.},
  journal = {Phys. Rev. Mater.},
  volume = {9},
  issue = {5},
  pages = {053806},
  numpages = {8},
  year = {2025},
  month = {May},
  publisher = {American Physical Society},
  doi = {10.1103/PhysRevMaterials.9.053806},
  url = {https://link.aps.org/doi/10.1103/PhysRevMaterials.9.053806}
}

@misc{fhiaims-web,
  note = {See: \url{https://fhi-aims.org}}
}

@article{Chen2022,
  title = {Nonunique fraction of $\mathrm{F}$ock exchange for defects in two-dimensional materials},
  author = {Chen, Wei and Griffin, Sin\'ead M. and Rignanese, Gian-Marco and Hautier, Geoffroy},
  journal = {Phys. Rev. B},
  volume = {106},
  issue = {16},
  pages = {L161107},
  numpages = {7},
  year = {2022},
  month = {Oct},
  publisher = {American Physical Society},
  doi = {10.1103/PhysRevB.106.L161107},
  url = {https://link.aps.org/doi/10.1103/PhysRevB.106.L161107}
}

@article{Guo2021,
author = {Guo, Hongli and Zhang, Xu and Lu, Gang},
title = {Moiré excitons in defective van der $\mathrm{W}$aals heterostructures},
journal = {Proc. Natl. Acad. Sci. U.S.A.},
volume = {118},
number = {32},
pages = {e2105468118},
year = {2021},
doi = {10.1073/pnas.2105468118},
URL = {https://www.pnas.org/doi/abs/10.1073/pnas.2105468118},
}

@article{Sagredo2025,
author = {Sagredo, Francisca and Camarasa-Gómez, María and Ricci, Francesco and Champagne, Aur{\'e}lie and Kronik, Leeor and Neaton, Jeffrey B.},
title = {The Reliability of Hybrid Functionals for Accurate Fundamental and Optical Gap Prediction of Bulk Solids and Surfaces},
journal = {J. Chem. Theory Comput.},
volume = {21},
number = {10},
pages = {5009-5015},
year = {2025},
doi = {10.1021/acs.jctc.5c00160},
URL = {     
        https://doi.org/10.1021/acs.jctc.5c00160
},
}

@article{Mori-Sanchez2008,
  title = {Localization and Delocalization Errors in Density Functional Theory and Implications for Band-Gap Prediction},
  author = {Mori-S\'anchez, Paula and Cohen, Aron J. and Yang, Weitao},
  journal = {Phys. Rev. Lett.},
  volume = {100},
  issue = {14},
  pages = {146401},
  numpages = {4},
  year = {2008},
  month = {Apr},
  publisher = {American Physical Society},
  doi = {10.1103/PhysRevLett.100.146401},
  url = {https://link.aps.org/doi/10.1103/PhysRevLett.100.146401}
}

@article{Cohen2008,
author = {Cohen, Aron J. and Mori-Sánchez, Paula and Yang, Weitao},
title = {Insights into Current Limitations of Density Functional Theory},
journal = {Science},
volume = {321},
number = {5890},
pages = {792-794},
year = {2008},
doi = {10.1126/science.1158722},
URL = {https://www.science.org/doi/abs/10.1126/science.1158722}
}

@article{Perdew2001,
    author = {Perdew, John P. and Schmidt, Karla},
    title = {Jacob’s ladder of density functional approximations for the exchange-correlation energy},
    journal = {AIP Conf. Proc.},
    volume = {577},
    number = {1},
    pages = {1-20},
    year = {2001},
    month = {07},
    issn = {0094-243X},
    doi = {10.1063/1.1390175},
    url = {https://doi.org/10.1063/1.1390175},
}

@article{Perdew-08-PBEsol,
  author    = {Perdew, John P. and Ruzsinszky, Adrienn and Csonka, G{\'a}bor I. and Vydrov, Oleg A. and Scuseria, Gustavo E. and Constantin, Lucian A. and Zhou, Xiaolan and Burke, Kieron},
  title     = {Restoring the Density-Gradient Expansion for Exchange in Solids and Surfaces},
  journal   = {Phys. Rev. Lett.},
  volume    = {100},
  number    = {13},
  pages     = {136406},
  year      = {2008},
  doi       = {10.1103/PhysRevLett.100.136406}
}

@article{kenny2009plato,
  title={\textsc{plato}: A localised orbital based density functional theory code},
  author={Kenny, Steven D and Horsfield, Andrew P},
  journal={Comput. Phys. Commun.},
  volume={180},
  number={12},
  pages={2616--2621},
  year={2009},
  publisher={Elsevier},
  url={https://doi.org/10.1016/j.cpc.2009.08.006}
}

@misc{iannuzzi2025cp2k,
      title={The \textsc{cp2k} Program Package Made Simple}, 
      author={Marcella Iannuzzi and Jan Wilhelm and Frederick Stein and Augustin Bussy and Hossam Elgabarty and Dorothea Golze and Anna Hehn and Maximilian Graml and Stepan Marek and Beliz Sertcan Gökmen and Christoph Schran and Harald Forbert and Rustam Z. Khaliullin and Anton Kozhevnikov and Mathieu Taillefumier and Rocco Meli and Vladimir Rybkin and Martin Brehm and Robert Schade and Ole Schütt and Johann V. Pototschnig and Hossein Mirhosseini and Andreas Knüpfer and Dominik Marx and Matthias Krack and Jürg Hutter and Thomas D. Kühne},
      year={2025},
      eprint={2508.15559},
      archivePrefix={arXiv},
      primaryClass={physics.comp-ph},
      url={https://arxiv.org/abs/2508.15559}, 
}

@article{Hohenberg-64,
  title = {Inhomogeneous Electron Gas},
  author = {Hohenberg, P. and Kohn, W.},
  journal = {Phys. Rev.},
  volume = {136},
  issue = {3B},
  pages = {B864--B871},
  numpages = {0},
  year = {1964},
  month = {Nov},
  publisher = {American Physical Society},
  doi = {10.1103/PhysRev.136.B864},
  url = {https://link.aps.org/doi/10.1103/PhysRev.136.B864}
}

@book{Kohanoff_2006, place={Cambridge}, title={Electronic Structure Calculations for Solids and Molecules: Theory and Computational Methods}, publisher={Cambridge University Press}, author={Kohanoff, Jorge}, year={2006},
url = {https://www.cambridge.org/core/books/electronic-structure-calculations-for-solids-and-molecules/0C0AF2B01A380912FC13816A9A0C350F#fndtn-metrics}}

@article{Miceli-18,
  title = {Nonempirical hybrid functionals for band gaps and polaronic distortions in solids},
  author = {Miceli, Giacomo and Chen, Wei and Reshetnyak, Igor and Pasquarello, Alfredo},
  journal = {Phys. Rev. B},
  volume = {97},
  issue = {12},
  pages = {121112},
  numpages = {5},
  year = {2018},
  month = {Mar},
  publisher = {American Physical Society},
  doi = {10.1103/PhysRevB.97.121112},
  url = {https://link.aps.org/doi/10.1103/PhysRevB.97.121112}
}

@article{Chen-13,
  title = {Correspondence of defect energy levels in hybrid density functional theory and many-body perturbation theory},
  author = {Chen, Wei and Pasquarello, Alfredo},
  journal = {Phys. Rev. B},
  volume = {88},
  issue = {11},
  pages = {115104},
  numpages = {8},
  year = {2013},
  month = {Sep},
  publisher = {American Physical Society},
  doi = {10.1103/PhysRevB.88.115104},
  url = {https://link.aps.org/doi/10.1103/PhysRevB.88.115104}
}

@article{Chen-17,
  title = {Accuracy of $\mathrm{GW}$ for calculating defect energy levels in solids},
  author = {Chen, Wei and Pasquarello, Alfredo},
  journal = {Phys. Rev. B},
  volume = {96},
  issue = {2},
  pages = {020101},
  numpages = {6},
  year = {2017},
  month = {Jul},
  publisher = {American Physical Society},
  doi = {10.1103/PhysRevB.96.020101},
  url = {https://link.aps.org/doi/10.1103/PhysRevB.96.020101}
}

@book{Martin_2004, place={Cambridge}, title={Electronic Structure: Basic Theory and Practical Methods}, publisher={Cambridge University Press}, author={Martin, Richard M.}, year={2004},
url = {https://www.cambridge.org/core/books/electronic-structure/DDFE838DED61D7A402FDF20D735BC63A}}

@book{Parr_book, place={Oxford}, title={Density ${F}$unctional ${T}$heory of ${A}$toms and ${M}$olecules}, publisher={Oxford University Press}, author={Parr, R. G. and Yang, W.}, year={1989},
url = {https://global.oup.com/academic/product/density-functional-theory-of-atoms-and-molecules-9780195092769?cc=es&lang=en&#}}

@article{Kohn-65,
  title = {Self-Consistent Equations Including Exchange and Correlation Effects},
  author = {Kohn, W. and Sham, L. J.},
  journal = {Phys. Rev.},
  volume = {140},
  issue = {4A},
  pages = {A1133--A1138},
  numpages = {0},
  year = {1965},
  month = {Nov},
  publisher = {American Physical Society},
  doi = {10.1103/PhysRev.140.A1133},
  url = {https://link.aps.org/doi/10.1103/PhysRev.140.A1133}
}

@article{Perdew-81,
  title = {Self-interaction correction to density-functional approximations for many-electron systems},
  author = {Perdew, J. P. and Zunger, Alex},
  journal = {Phys. Rev. B},
  volume = {23},
  issue = {10},
  pages = {5048--5079},
  numpages = {0},
  year = {1981},
  month = {May},
  publisher = {American Physical Society},
  doi = {10.1103/PhysRevB.23.5048},
  url = {https://link.aps.org/doi/10.1103/PhysRevB.23.5048}
}

@article{Ceperley-80,
  title = {Ground State of the Electron Gas by a Stochastic Method},
  author = {Ceperley, D. M. and Alder, B. J.},
  journal = {Phys. Rev. Lett.},
  volume = {45},
  issue = {7},
  pages = {566--569},
  numpages = {0},
  year = {1980},
  month = {Aug},
  publisher = {American Physical Society},
  doi = {10.1103/PhysRevLett.45.566},
  url = {https://link.aps.org/doi/10.1103/PhysRevLett.45.566}
}

@misc{ABACUS-hybrids,
      title={ABACUS: An Electronic Structure Analysis Package for the $\mathrm{AI}$ Era}, 
      author={Weiqing Zhou and Daye Zheng and Qianrui Liu and Denghui Lu and Yu Liu and Peize Lin and Yike Huang and Xingliang Peng and Jie J. Bao and Chun Cai and Zuxin Jin and Jing Wu and Haochong Zhang and Gan Jin and Yuyang Ji and Zhenxiong Shen and Xiaohui Liu and Liang Sun and Yu Cao and Menglin Sun and Jianchuan Liu and Tao Chen and Renxi Liu and Yuanbo Li and Haozhi Han and Xinyuan Liang and Taoni Bao and Nuo Chen and Hongxu Ren and Xiaoyang Zhang and Zhaoqing Liu and Yiwei Fu and Maochang Liu and Zhuoyuan Li and Tongqi Wen and Zechen Tang and Yong Xu and Wenhui Duan and Xiaoyang Wang and Qiangqiang Gu and Fu-Zhi Dai and Qijing Zheng and Jin Zhao and Yuzhi Zhang and Qi Ou and Hong Jiang and Shi Liu and Ben Xu and Shenzhen Xu and Xinguo Ren and Lixin He and Linfeng Zhang and Mohan Chen},
      year={2025},
      eprint={2501.08697},
      archivePrefix={arXiv},
      primaryClass={cond-mat.mtrl-sci},
      url={https://arxiv.org/abs/2501.08697}, 
}

@article{qin-2015-1,
  author = {
    Qin, Xinming and
    Shang, Honghui and
    Xiang, Hongjun and
    Li, Zhenyu and
    Yang, Jinlong},
  title = {\textsc{honpas}: A linear scaling open-source solution for large system
    simulations},
  journal = {Int. J. Quantum Chem.},
  volume = {115},
  number = {10},
  pages = {647-655},
  year = {2015},
  doi = {10.1002/qua.24837},
}

@article{Izmaylov-06,
author = {Izmaylov, A. F. and Scuseria, G. E. and Frisch, M. J. },
title = {Efficient evaluation of short-range Hartree-Fock exchange in large molecules and periodic systems},
journal = {J. Chem. Phys.},
volume = {125},
number = {10},
pages = {104103},
year = {2006},
doi = {10.1063/1.2347713},
URL = { https://doi.org/10.1063/1.2347713
    },
}

@article{Guidon-08,
    doi = {10.1063/1.2931945},
	url = {https://doi.org/10.1063\%2F1.2931945},
	year = 2008,
	month = {jun},
	publisher = {{AIP} Publishing},
	volume = {128},
	number = {21},
	pages = {214104},
	author = {Manuel Guidon and Florian Schiffmann and Jürg Hutter and Joost VandeVondele},
	title = {Ab initio molecular dynamics using hybrid density functionals},
	journal = {J. Chem. Phys.}
}

@Article{Guidon-09,
author={Guidon, Manuel
and Hutter, J{\"u}rg
and VandeVondele, Joost},
title={Robust Periodic $\mathrm{H}$artree−$\mathrm{F}$ock Exchange for Large-Scale Simulations Using Gaussian Basis Sets},
journal={J. Chem. Theory Comput.},
year={2009},
month={Nov},
day={10},
publisher={American Chemical Society},
volume={5},
number={11},
pages={3010-3021},
issn={1549-9618},
doi={10.1021/ct900494g},
url={https://doi.org/10.1021/ct900494g}
}

@book{global-min-1,
 author = {Goldberg, David E.},
 title = {Genetic Algorithms in Search, Optimization and Machine Learning},
 year = {1989},
 isbn = {0201157675},
 edition = {1st},
 publisher = {Addison-Wesley Longman Publishing Co., Inc.},
 address = {Boston, MA, USA},
}

@article{Conn-91,
author = {Conn, Andrew R. and Gould, Nicholas I. M. and Toint, Philippe},
title = {A Globally Convergent Augmented Lagrangian Algorithm for Optimization with General Constraints and Simple Bounds},
journal = {SIAM J. Numer. Anal.},
volume = {28},
number = {2},
pages = {545-572},
year = {1991},
doi = {10.1137/0728030},
URL = {
        https://doi.org/10.1137/0728030
},
}

@Article{Coleman-94,
author={Coleman, Thomas F.
and Li, Yuying},
title={On the convergence of interior-reflective Newton methods for nonlinear minimization subject to bounds},
journal={Math. Program.},
year={1994},
month={Oct},
day={01},
volume={67},
number={1},
pages={189-224},
issn={1436-4646},
url={https://doi.org/10.1007/BF01582221}
}

@article{Coleman-96,
author = {Coleman, Thomas F. and Li, Yuying},
title = {An Interior Trust Region Approach for Nonlinear Minimization Subject to Bounds},
journal = {SIAM J. Optim.},
volume = {6},
number = {2},
pages = {418-445},
year = {1996},
doi = {10.1137/0806023},
URL = {
        https://doi.org/10.1137/0806023
},
}

@misc{Libint,
  author = "E.~F.~Valeev and J.~T.~Fermann",
  title = "{\textsc{Libint}}: A library for the evaluation of molecular integrals of many-body operators over Gaussian functions",
  howpublished = "http://libint.valeyev.net/",
  note = "version 1.1.4",
  year =         2020
}

@misc{GPL,
  note = {See: \url{https://www.gnu.org/licenses/gpl-3.0.html}}
}

@misc{ESL-web,
  note = {See: \url{https://esl.cecam.org/Main_Page}}
}

@misc{CP2K-web,
  note = {See: \url{https://www.cp2k.org}}
}

@misc{Crystal-web,
  note = {See: \url{https://www.crystal.unito.it/index.php}}
}

@misc{Gaussian-web,
  note = {See: \url{http://gaussian.com}}
}

@article{Oliveira-20,
    author = {Oliveira, Micael J. T. and Papior, Nick and Pouillon, Yann and Blum, Volker and Artacho, Emilio and Caliste, Damien and Corsetti, Fabiano and de Gironcoli, Stefano and Elena, Alin M. and García, Alberto and García-Suárez, Víctor M. and Genovese, Luigi and Huhn, William P. and Huhs, Georg and Kokott, Sebastian and Küçükbenli, Emine and Larsen, Ask H. and Lazzaro, Alfio and Lebedeva, Irina V. and Li, Yingzhou and López-Durán, David and López-Tarifa, Pablo and Lüders, Martin and Marques, Miguel A. L. and Minar, Jan and Mohr, Stephan and Mostofi, Arash A. and O’Cais, Alan and Payne, Mike C. and Ruh, Thomas and Smith, Daniel G. A. and Soler, José M. and Strubbe, David A. and Tancogne-Dejean, Nicolas and Tildesley, Dominic and Torrent, Marc and Yu, Victor Wen-zhe},
    title = {The CECAM electronic structure library and the modular software development paradigm},
    journal = {J. Chem. Phys.},
    volume = {153},
    number = {2},
    pages = {024117},
    year = {2020},
    month = {07},
    issn = {0021-9606},
    doi = {10.1063/5.0012901},
    url = {https://doi.org/10.1063/5.0012901},
}

@article{Schlegel-95,
author = {Schlegel, H. Bernhard and Frisch, Michael J.},
title = {Transformation between Cartesian and pure spherical harmonic Gaussians},
journal = {Int. J. Quantum Chem.},
volume = {54},
number = {2},
pages = {83-87},
doi = {10.1002/qua.560540202},
url = {https://onlinelibrary.wiley.com/doi/abs/10.1002/qua.560540202},
year = {1995}
}

@article{Obara-84,
author = {Obara,S.  and Saika,A. },
title = {Efficient recursive computation of molecular integrals over Cartesian Gaussian functions},
journal = {J. Chem. Phys.},
volume = {84},
number = {7},
pages = {3963-3974},
year = {1986},
doi = {10.1063/1.450106},
}

@article{Head-Gordon-88,
author = {Head‐Gordon,Martin  and Pople,John A. },
title = {A method for two‐electron Gaussian integral and integral derivative evaluation using recurrence relations},
journal = {J. Chem. Phys.},
volume = {89},
number = {9},
pages = {5777-5786},
year = {1988},
doi = {10.1063/1.455553},
}

@article{Hutter-14,
author = {Hutter,J\"{u}rg  and Iannuzzi, Marcella and Schiffmann, Florian and VandeVondele, Joost},
title = {\textsc{cp2k}: atomistic simulations of condensed matter systems},
journal = {WIREs Comput. Mol. Sci.},
volume = {4},
number = {1},
pages = {15-25},
doi = {10.1002/wcms.1159},
url = {https://www.onlinelibrary.wiley.com/doi/abs/10.1002/wcms.1159},
year = {2014}
}

@article{Boys-50,
author = {Boys, S. F.  and Egerton, Alfred Charles },
title = {Electronic wave functions - I. A general method of calculation for the stationary states of any molecular system},
journal = {Proc. R. Soc. Lond. A Math. Phys. Sci.},
volume = {200},
number = {1063},
pages = {542-554},
year = {1950},
doi = {10.1098/rspa.1950.0036},
URL = {https://royalsocietypublishing.org/doi/abs/10.1098/rspa.1950.0036}
}

@book{Atkins,
  title={Molecular Quantum Mechanics},
  author={Atkins, P.W. and Friedman, R.S.},
  isbn={9780199541423},
  lccn={2011289349},
  url={https://books.google.es/books?id=9k-cAQAAQBAJ},
  year={2011},
  publisher={Oxford University Press Oxford}
}

@book{Szabo,
  address = {Mineola},
  author = {Szabo, A. and Ostlund, N. S.},
  url = {https://chemistlibrary.wordpress.com/wp-content/uploads/2015/02/modern-quantum-chemistry.pdf},
  edition = {},
  publisher = {Dover Publications, Inc.},
  title = {Modern Quantum Chemistry: Introduction to Advanced Electronic
                  Structure Theory},
  year = 1996
}

@book{ Jensen-06, 
       author = {Jensen, F.}, 
       title = {Introduction to Computational Chemistry}, 
       year = {2006}, 
       isbn = {0470011874}, 
       publisher = {John Wiley \& Sons, Inc.}, 
       address = {Hoboken, NJ, USA},
       url = {https://www.wiley.com/en-us/Introduction+to+Computational+Chemistry%2C+3rd+Edition-p-9781118825990}
       }

@article{Shang-20,
title = {The dynamic parallel distribution algorithm for hybrid density-functional calculations in \textsc{honpas} package},
journal = {Comp. Phys. Commun.},
volume = {254},
pages = {107204},
year = {2020},
issn = {0010-4655},
doi = {https://doi.org/10.1016/j.cpc.2020.107204},
url = {https://www.sciencedirect.com/science/article/pii/S0010465520300448},
author = {H. Shang and L. Xu and B. Wu and X. Qin and Y. Zhang and J. Yang}
}

@article{Haser-89,
author = {Häser, M. and Ahlrichs, R.},
title = {Improvements on the direct SCF method},
journal = {J. Comput. Chem.},
volume = {10},
pages = {104-111},
doi = {https://doi.org/10.1002/jcc.540100111},
url = {https://onlinelibrary.wiley.com/doi/abs/10.1002/jcc.540100111},
year = {1989}
}

@article{norm_conserving,
  title={Optimized norm-conserving \textsc{V}anderbilt pseudopotentials},
  author={Hamann, DR},
  journal={Phys. Rev. B},
  volume={88},
  number={8},
  pages={085117},
  year={2013},
  publisher={APS},
  url={https://doi.org/10.1103/PhysRevB.88.085117}
}

@article{psml,
  title={The $\mathrm{PSML}$ format and library for norm-conserving pseudopotential data curation and interoperability},
  author={Garc{\'\i}a, Alberto and Verstraete, Matthieu J and Pouillon, Yann and Junquera, Javier},
  journal={Comput. Phys. Commun.},
  volume={227},
  pages={51},
  year={2018},
  publisher={Elsevier},
  url={https://doi.org/10.1016/j.cpc.2018.02.011}
}

@article{pseudodojo,
  title={The \textsc{PseudoDojo}: Training and grading a 85 element optimized norm-conserving pseudopotential table},
  author={Van Setten, MJ and Giantomassi, Matteo and Bousquet, Eric and Verstraete, Matthieu J and Hamann, Don R and Gonze, Xavier and Rignanese, G-M},
  journal={Comput. Phys. Commun.},
  volume={226},
  pages={39},
  year={2018},
  publisher={Elsevier},
  url={https://doi.org/10.1016/j.cpc.2018.01.012}
}

@misc{footnotepseudo,
 note = {The scalar relativistic \textsc{oncvpsp} v0.4.1 pseudopotentials with stringent accuracy were used.}
}

@article{Kleinman-82,
  title = {Efficacious Form for Model Pseudopotentials},
  author = {Kleinman, Leonard and Bylander, D. M.},
  journal = {Phys. Rev. Lett.},
  volume = {48},
  issue = {20},
  pages = {1425--1428},
  numpages = {0},
  year = {1982},
  month = {May},
  publisher = {American Physical Society},
  doi = {10.1103/PhysRevLett.48.1425},
  url = {https://link.aps.org/doi/10.1103/PhysRevLett.48.1425}
}

@article{Monkhorst-76,
  title = {Special points for  $\mathrm{B}$rillouin-zone integrations},
  author = {Monkhorst, H. J. and Pack, J. D.},
  journal = {Phys. Rev. B},
  volume = {13},
  issue = {12},
  pages = {5188--5192},
  numpages = {0},
  year = {1976},
  month = {Jun},
  publisher = {American Physical Society},
  doi = {10.1103/PhysRevB.13.5188},
  url = {https://link.aps.org/doi/10.1103/PhysRevB.13.5188}
}

@article{Sankey-89,
  title = {Ab initio multicenter tight-binding model for molecular-dynamics simulations and other applications in covalent systems},
  author = {Sankey, O. F. and Niklewski, D. J.},
  journal = {Phys. Rev. B},
  volume = {40},
  issue = {6},
  pages = {3979--3995},
  numpages = {0},
  year = {1989},
  month = {Aug},
  publisher = {American Physical Society},
  doi = {10.1103/PhysRevB.40.3979},
  url = {https://link.aps.org/doi/10.1103/PhysRevB.40.3979}
}

@article{Artacho-99,
author = {Artacho, E. and Sánchez-Portal, D. and Ordejón, P. and García, A. and Soler, J. M.},
title = {Linear-Scaling ab-initio Calculations for Large and Complex Systems},
journal = {Phys. Stat. Sol. (b)},
volume = {215},
number = {1},
pages = {809-817},
doi = {https://doi.org/10.1002/(SICI)1521-3951(199909)215:1<809::AID-PSSB809>3.0.CO;2-0},
url = {https://onlinelibrary.wiley.com/doi/abs/10.1002/%28SICI%291521-3951%28199909%29215%3A1%3C809%3A%3AAID-PSSB809%3E3.0.CO%3B2-0},
year = {1999}
}

@article{Junquera-01,
  title = {Numerical atomic orbitals for linear-scaling calculations},
  author = {Junquera, Javier and Paz, \'Oscar and S\'anchez-Portal, Daniel and Artacho, Emilio},
  journal = {Phys. Rev. B},
  volume = {64},
  issue = {23},
  pages = {235111},
  numpages = {9},
  year = {2001},
  month = {Nov},
  publisher = {American Physical Society},
  doi = {10.1103/PhysRevB.64.235111},
  url = {https://link.aps.org/doi/10.1103/PhysRevB.64.235111}
}

@article{Garcia-Gil-09,
  title = {Optimal strictly localized basis sets for noble metal surfaces},
  author = {Garc\'{\i}a-Gil, Sandra and Garc\'{\i}a, Alberto and Lorente, Nicol\'as and Ordej\'on, Pablo},
  journal = {Phys. Rev. B},
  volume = {79},
  issue = {7},
  pages = {075441},
  numpages = {9},
  year = {2009},
  month = {Feb},
  publisher = {American Physical Society},
  doi = {10.1103/PhysRevB.79.075441},
  url = {https://link.aps.org/doi/10.1103/PhysRevB.79.075441}
}

@article{Toyoda-11,
  title = {Exchange functional by a range-separated exchange hole},
  author = {Toyoda, Masayuki and Ozaki, Taisuke},
  journal = {Phys. Rev. A},
  volume = {83},
  issue = {3},
  pages = {032515},
  numpages = {7},
  year = {2011},
  month = {Mar},
  publisher = {American Physical Society},
  doi = {10.1103/PhysRevA.83.032515},
  url = {https://link.aps.org/doi/10.1103/PhysRevA.83.032515}
}

@article{Becke-93,
    author = {Becke, Axel D.},
    title = {A new mixing of $\mathrm{H}$artree-$\mathrm{F}$ock and local density‐functional theories},
    journal = {J. Chem. Phys.},
    volume = {98},
    number = {2},
    pages = {1372-1377},
    year = {1993},
    month = {01},
    issn = {0021-9606},
    doi = {10.1063/1.464304},
    url = {https://doi.org/10.1063/1.464304},
}

@article{Nakata-20,
    author = {Nakata, Ayako and Baker, Jack S. and Mujahed, Shereif Y. and Poulton, Jack T. L. and Arapan, Sergiu and Lin, Jianbo and Raza, Zamaan and Yadav, Sushma and Truflandier, Lionel and Miyazaki, Tsuyoshi and Bowler, David R.},
    title = {Large scale and linear scaling $\mathrm{DFT}$ with the {\textsc{CONQUEST}} code},
    journal = {J. Chem. Phys.},
    volume = {152},
    number = {16},
    pages = {164112},
    year = {2020},
    month = {04},
    issn = {0021-9606},
    doi = {10.1063/5.0005074},
    url = {https://doi.org/10.1063/5.0005074},
}

@article{Chen-10,
  title={Systematically improvable optimized atomic basis sets for ab initio calculations},
  author={Chen, Mohan and Guo, GC and He, Lixin},
  journal={J. Phys.: Condens. Matter},
  volume={22},
  number={44},
  pages={445501},
  year={2010},
  publisher={IOP Publishing},
  doi={10.1088/0953-8984/22/44/445501}
}

@article{Li-16,
  title={Large-scale ab initio simulations based on systematically improvable atomic basis},
  author={Li, Pengfei and Liu, Xiaohui and Chen, Mohan and Lin, Peize and Ren, Xinguo and Lin, Lin and Yang, Chao and He, Lixin},
  journal={Comput. Mater. Sci.},
  volume={112},
  pages={503--517},
  year={2016},
  publisher={Elsevier},
  url={https://doi.org/10.1016/j.commatsci.2015.07.004}
}

@article{Velde-01,
  title={Chemistry with \textsc{ADF}},
  author={Te Velde, G t and Bickelhaupt, Friedrich Matthias and Baerends, Evert Jan and Fonseca Guerra, C and van Gisbergen, Stan JA and Snijders, Jaap G and Ziegler, Tom},
  journal={J. Comput. Chem.},
  volume={22},
  number={9},
  pages={931--967},
  year={2001},
  publisher={Wiley Online Library},
  url={https://doi.org/10.1002/jcc.1056}
}

@article{erba2022crystal23,
  title={\textsc{crystal23}: A program for computational solid state physics and chemistry},
  author={Erba, Alessandro and Desmarais, Jacques K and Casassa, Silvia and Civalleri, Bartolomeo and Don{\`a}, Lorenzo and Bush, Ian J and Searle, Barry and Maschio, Lorenzo and Edith-Daga, Loredana and Cossard, Alessandro and others},
  journal={J. Chem. Theory Comput.},
  volume={19},
  number={20},
  pages={6891--6932},
  year={2022},
  publisher={ACS Publications},
  url={https://doi.org/10.1021/acs.jctc.2c00958}
}

@article{kuhne2020cp2k,
  title={\textsc{cp2k}: An electronic structure and molecular dynamics software package-Quickstep: Efficient and accurate electronic structure calculations},
  author={K{\"u}hne, Thomas D and Iannuzzi, Marcella and Del Ben, Mauro and Rybkin, Vladimir V and Seewald, Patrick and Stein, Frederick and Laino, Teodoro and Khaliullin, Rustam Z and Sch{\"u}tt, Ole and Schiffmann, Florian and others},
  journal={J. Chem. Phys.},
  volume={152},
  number={19},
  pages={194103},
  year={2020},
  publisher={AIP Publishing},
  url={https://doi.org/10.1063/5.0007045}
}

@article{Kokott-24,
    author = {Kokott, Sebastian and Merz, Florian and Yao, Yi and Carbogno, Christian and Rossi, Mariana and Havu, Ville and Rampp, Markus and Scheffler, Matthias and Blum, Volker},
    title = {Efficient all-electron hybrid density functionals for atomistic simulations beyond 10000 atoms},
    journal = {J. Chem. Phys.},
    volume = {161},
    number = {2},
    pages = {024112},
    year = {2024},
    month = {07},
    issn = {0021-9606},
    doi = {10.1063/5.0208103},
    url = {https://doi.org/10.1063/5.0208103},
}

@book{anisimov2000strong,
  title={Strong $\mathrm{C}$oulomb Correlations in Electronic Structure Calculations},
  author={Anisimov, V.I.},
  isbn={9789056991319},
  lccn={2002421700},
  series={Advances in Condensed Matter Science},
  url={https://books.google.es/books?id=3sjJkC8yN8QC},
  year={2000},
  publisher={Taylor \& Francis}
}

@article{Heyd-04,
    author = {Heyd, Jochen and Scuseria, Gustavo E.},
    title = {Assessment and validation of a screened $\mathrm{C}$oulomb hybrid density functional},
    journal = {J. Chem. Phys.},
    volume = {120},
    number = {16},
    pages = {7274-7280},
    year = {2004},
    month = {04},
    issn = {0021-9606},
    doi = {10.1063/1.1668634},
    url = {https://doi.org/10.1063/1.1668634},
}

@article{Ernzerhof-08,
    author = {Ernzerhof, Matthias and Perdew, John P.},
    title = {Generalized gradient approximation to the angle- and system-averaged exchange hole},
    journal = {J. Chem. Phys.},
    volume = {109},
    number = {9},
    pages = {3313-3320},
    year = {1998},
    month = {09},
    issn = {0021-9606},
    doi = {10.1063/1.476928},
    url = {https://doi.org/10.1063/1.476928},
}

@article{Perdew-96,
    author = {Perdew, John P. and Ernzerhof, Matthias and Burke, Kieron},
    title = {Rationale for mixing exact exchange with density functional approximations},
    journal = {J. Chem. Phys.},
    volume = {105},
    number = {22},
    pages = {9982-9985},
    year = {1996},
    month = {12},
    issn = {0021-9606},
    doi = {10.1063/1.472933},
    url = {https://doi.org/10.1063/1.472933},
}

@article{Adamo-99,
    author = {Adamo, Carlo and Barone, Vincenzo},
    title = {Toward reliable density functional methods without adjustable parameters: The $\mathrm{PBE0}$ model},
    journal = {J. Chem. Phys.},
    volume = {110},
    number = {13},
    pages = {6158-6170},
    year = {1999},
    month = {04},
    issn = {0021-9606},
    doi = {10.1063/1.478522},
    url = {https://doi.org/10.1063/1.478522},
}

@article{Perdew-96.2,
author = {Perdew, John P. and Burke, Kieron},
title = {Comparison shopping for a gradient-corrected density functional},
journal = {Int. J. Quant. Chem.},
volume = {57},
number = {3},
pages = {309-319},
doi = {https://doi.org/10.1002/(SICI)1097-461X(1996)57:3<309::AID-QUA4>3.0.CO;2-1},
url = {https://onlinelibrary.wiley.com/doi/abs/10.1002/%28SICI%291097-461X%281996%2957%3A3%3C309%3A%3AAID-QUA4%3E3.0.CO%3B2-1},
year = {1996}
}

@article{PBE,
  title = {Generalized Gradient Approximation Made Simple},
  author = {Perdew, John P. and Burke, Kieron and Ernzerhof, Matthias},
  journal = {Phys. Rev. Lett.},
  volume = {77},
  issue = {18},
  pages = {3865--3868},
  numpages = {0},
  year = {1996},
  month = {Oct},
  publisher = {American Physical Society},
  doi = {10.1103/PhysRevLett.77.3865},
  url = {https://link.aps.org/doi/10.1103/PhysRevLett.77.3865}
}

@article{PBE-erratum,
  title = {Erratum: Generalized Gradient Approximation Made Simple [Phys. Rev. Lett. 77, 3865 (1996)]},
  author = {Perdew, John P. and Burke, Kieron and Ernzerhof, Matthias},
  journal = {Phys. Rev. Lett.},
  volume = {78},
  issue = {7},
  pages = {1396--1396},
  numpages = {0},
  year = {1997},
  month = {Feb},
  publisher = {American Physical Society},
  doi = {10.1103/PhysRevLett.78.1396},
  url = {https://link.aps.org/doi/10.1103/PhysRevLett.78.1396}
}

@Article{Lin-25,
author={Lin, Peize
and Ji, Yuyang
and He, Lixin
and Ren, Xinguo},
title={Efficient Hybrid-Functional-Based Force and Stress Calculations for Periodic Systems with Thousands of Atoms},
journal={J. Chem. Theory Comput.},
year={2025},
month={Apr},
day={08},
publisher={American Chemical Society},
volume={21},
number={7},
pages={3394-3409},
issn={1549-9618},
doi={10.1021/acs.jctc.4c01635},
url={https://doi.org/10.1021/acs.jctc.4c01635}
}

@ARTICLE{Qin-23,
  AUTHOR={Qin, Xinming  and Shang, Honghui  and Yang, Jinlong },
TITLE={Efficient implementation of analytical gradients for periodic hybrid functional calculations within fitted numerical atomic orbitals from {\textsc{NAO2GTO}}},
JOURNAL={Front. Chem.},
VOLUME={11},
PAGES={27 July 2023},
YEAR={2023},
URL={https://www.frontiersin.org/journals/chemistry/articles/10.3389/fchem.2023.1232425},
ISSN={2296-2646}}

@article{Dovesi-20,
    author = {Dovesi, Roberto and Pascale, Fabien and Civalleri, Bartolomeo and Doll, Klaus and Harrison, Nicholas M. and Bush, Ian and D’Arco, Philippe and Noël, Yves and Rérat, Michel and Carbonnière, Philippe and Causà, Mauro and Salustro, Simone and Lacivita, Valentina and Kirtman, Bernard and Ferrari, Anna Maria and Gentile, Francesco Silvio and Baima, Jacopo and Ferrero, Mauro and Demichelis, Raffaella and De La Pierre, Marco},
    title = {The {\textsc{CRYSTAL}} code, 1976–2020 and beyond, a long story},
    journal = {J. Chem. Phys.},
    volume = {152},
    number = {20},
    pages = {204111},
    year = {2020},
    month = {05},
    issn = {0021-9606},
    doi = {10.1063/5.0004892},
    url = {https://doi.org/10.1063/5.0004892},
}

@Article{Lee-22,
author={Lee, Joonho
and Rettig, Adam
and Feng, Xintian
and Epifanovsky, Evgeny
and Head-Gordon, Martin},
title={Faster Exact Exchange for Solids via occ-$\mathrm{RI-K}$: Application to Combinatorially Optimized Range-Separated Hybrid Functionals for Simple Solids with Pseudopotentials Near the Basis Set Limit},
journal={J. Chem. Theory Comput.},
year={2022},
month={Dec},
day={13},
publisher={American Chemical Society},
volume={18},
number={12},
pages={7336-7349},
issn={1549-9618},
doi={10.1021/acs.jctc.2c00742},
url={https://doi.org/10.1021/acs.jctc.2c00742}
}

@article{Sun-22,
    author = {Sun, Qiming and Zhang, Xing and Banerjee, Samragni and Bao, Peng and Barbry, Marc and Blunt, Nick S. and Bogdanov, Nikolay A. and Booth, George H. and Chen, Jia and Cui, Zhi-Hao and Eriksen, Janus J. and Gao, Yang and Guo, Sheng and Hermann, Jan and Hermes, Matthew R. and Koh, Kevin and Koval, Peter and Lehtola, Susi and Li, Zhendong and Liu, Junzi and Mardirossian, Narbe and McClain, James D. and Motta, Mario and Mussard, Bastien and Pham, Hung Q. and Pulkin, Artem and Purwanto, Wirawan and Robinson, Paul J. and Ronca, Enrico and Sayfutyarova, Elvira R. and Scheurer, Maximilian and Schurkus, Henry F. and Smith, James E. T. and Sun, Chong and Sun, Shi-Ning and Upadhyay, Shiv and Wagner, Lucas K. and Wang, Xiao and White, Alec and Whitfield, James Daniel and Williamson, Mark J. and Wouters, Sebastian and Yang, Jun and Yu, Jason M. and Zhu, Tianyu and Berkelbach, Timothy C. and Sharma, Sandeep and Sokolov, Alexander Yu. and Chan, Garnet Kin-Lic},
    title = {Recent developments in the {\textsc{PySCF}} program package},
    journal = {J. Chem. Phys.},
    volume = {153},
    number = {2},
    pages = {024109},
    year = {2020},
    month = {07},
    issn = {0021-9606},
    doi = {10.1063/5.0006074},
    url = {https://doi.org/10.1063/5.0006074},
}

@article{Ozaki-05,
  title = {Efficient projector expansion for the ab initio $\mathrm{LCAO}$ method},
  author = {Ozaki, T. and Kino, H.},
  journal = {Phys. Rev. B},
  volume = {72},
  issue = {4},
  pages = {045121},
  numpages = {8},
  year = {2005},
  month = {Jul},
  publisher = {American Physical Society},
  doi = {10.1103/PhysRevB.72.045121},
  url = {https://link.aps.org/doi/10.1103/PhysRevB.72.045121}
}

@article{Perdew1983,
  title = {Physical Content of the Exact $\mathrm{K}$ohn-$\mathrm{S}$ham Orbital Energies: Band Gaps and Derivative Discontinuities},
  author = {Perdew, John P. and Levy, Mel},
  journal = {Phys. Rev. Lett.},
  volume = {51},
  issue = {20},
  pages = {1884--1887},
  numpages = {0},
  year = {1983},
  month = {Nov},
  publisher = {American Physical Society},
  doi = {10.1103/PhysRevLett.51.1884},
  url = {https://link.aps.org/doi/10.1103/PhysRevLett.51.1884}
}

@article{Sham1983,
  title = {Density-Functional Theory of the Energy Gap},
  author = {Sham, L. J. and Schl\"uter, M.},
  journal = {Phys. Rev. Lett.},
  volume = {51},
  issue = {20},
  pages = {1888--1891},
  numpages = {0},
  year = {1983},
  month = {Nov},
  publisher = {American Physical Society},
  doi = {10.1103/PhysRevLett.51.1888},
  url = {https://link.aps.org/doi/10.1103/PhysRevLett.51.1888}
}

@article{Falletta-22.1,
  title = {Polarons free from many-body self-interaction in density functional theory},
  author = {Falletta, Stefano and Pasquarello, Alfredo},
  journal = {Phys. Rev. B},
  volume = {106},
  issue = {12},
  pages = {125119},
  numpages = {17},
  year = {2022},
  month = {Sep},
  publisher = {American Physical Society},
  doi = {10.1103/PhysRevB.106.125119},
  url = {https://link.aps.org/doi/10.1103/PhysRevB.106.125119}
}

@article{Falletta-22.2,
  title = {Many-Body Self-Interaction and Polarons},
  author = {Falletta, Stefano and Pasquarello, Alfredo},
  journal = {Phys. Rev. Lett.},
  volume = {129},
  issue = {12},
  pages = {126401},
  numpages = {6},
  year = {2022},
  month = {Sep},
  publisher = {American Physical Society},
  doi = {10.1103/PhysRevLett.129.126401},
  url = {https://link.aps.org/doi/10.1103/PhysRevLett.129.126401}
}

@article{Schmidt2016,
  title = {One- and many-electron self-interaction error in local and global hybrid functionals},
  author = {Schmidt, Tobias and K\"ummel, Stephan},
  journal = {Phys. Rev. B},
  volume = {93},
  issue = {16},
  pages = {165120},
  numpages = {15},
  year = {2016},
  month = {Apr},
  publisher = {American Physical Society},
  doi = {10.1103/PhysRevB.93.165120},
  url = {https://link.aps.org/doi/10.1103/PhysRevB.93.165120}
}

@article{Laurien2022,
  title = {Benchmarking exchange-correlation potentials with the mstar60 dataset: Importance of the nonlocal exchange potential for effective mass calculations in semiconductors},
  author = {Laurien, Magdalena and Rubel, Oleg},
  journal = {Phys. Rev. B},
  volume = {106},
  issue = {4},
  pages = {045204},
  numpages = {13},
  year = {2022},
  month = {Jul},
  publisher = {American Physical Society},
  doi = {10.1103/PhysRevB.106.045204},
  url = {https://link.aps.org/doi/10.1103/PhysRevB.106.045204}
}

@article{Powell-98, title={Direct search algorithms for optimization calculations}, volume={7}, DOI={10.1017/S0962492900002841}, journal={Acta Numer.}, author={Powell, M. J. D.}, year={1998}, pages={287}}

@article{Yang-98,
  title = {Hybrid functional pseudopotentials},
  author = {Yang, Jing and Tan, Liang Z. and Rappe, Andrew M.},
  journal = {Phys. Rev. B},
  volume = {97},
  issue = {8},
  pages = {085130},
  numpages = {9},
  year = {2018},
  month = {Feb},
  publisher = {American Physical Society},
  doi = {10.1103/PhysRevB.97.085130},
  url = {https://link.aps.org/doi/10.1103/PhysRevB.97.085130}
}

@article{Toyoda-09,
    author = {Toyoda, Masayuki and Ozaki, Taisuke},
    title = {Numerical evaluation of electron repulsion integrals for pseudoatomic orbitals and their derivatives},
    journal = {J. Chem. Phys.},
    volume = {130},
    number = {12},
    pages = {124114},
    year = {2009},
    month = {03},
    issn = {0021-9606},
    doi = {10.1063/1.3082269},
    url = {https://doi.org/10.1063/1.3082269},
}

@article{Toyoda-10,
title = {$\mathrm{LIBERI}$: Library for numerical evaluation of electron-repulsion integrals},
journal = {Comp. Phys. Commun.},
volume = {181},
number = {8},
pages = {1455-1463},
year = {2010},
issn = {0010-4655},
doi = {https://doi.org/10.1016/j.cpc.2010.03.019},
url = {https://www.sciencedirect.com/science/article/pii/S0010465510001141},
author = {Masayuki Toyoda and Taisuke Ozaki}
}

@article{Hummer-09,
  title = {Heyd-$\mathrm{S}$cuseria-$\mathrm{E}$rnzerhof hybrid functional for calculating the lattice dynamics of semiconductors},
  author = {Hummer, Kerstin and Harl, Judith and Kresse, Georg},
  journal = {Phys. Rev. B},
  volume = {80},
  issue = {11},
  pages = {115205},
  numpages = {12},
  year = {2009},
  month = {Sep},
  publisher = {American Physical Society},
  doi = {10.1103/PhysRevB.80.115205},
  url = {https://link.aps.org/doi/10.1103/PhysRevB.80.115205}
}

@article{Skelton-14,
  title = {Thermal physics of the lead chalcogenides $\mathrm{P}$b$\mathrm{S}$, $\mathrm{P}$b$\mathrm{S}$e, and $\mathrm{P}$b$\mathrm{T}$e from first principles},
  author = {Skelton, Jonathan M. and Parker, Stephen C. and Togo, Atsushi and Tanaka, Isao and Walsh, Aron},
  journal = {Phys. Rev. B},
  volume = {89},
  issue = {20},
  pages = {205203},
  numpages = {10},
  year = {2014},
  month = {May},
  publisher = {American Physical Society},
  doi = {10.1103/PhysRevB.89.205203},
  url = {https://link.aps.org/doi/10.1103/PhysRevB.89.205203}
}

@article{Tran-09,
  title = {Accurate Band Gaps of Semiconductors and Insulators with a Semilocal Exchange-Correlation Potential},
  author = {Tran, Fabien and Blaha, Peter},
  journal = {Phys. Rev. Lett.},
  volume = {102},
  issue = {22},
  pages = {226401},
  numpages = {4},
  year = {2009},
  month = {Jun},
  publisher = {American Physical Society},
  doi = {10.1103/PhysRevLett.102.226401},
  url = {https://link.aps.org/doi/10.1103/PhysRevLett.102.226401}
}

@book{Atkins2006,
  author    = {Atkins, Peter and de Paula, Julio},
  title     = {Atkins{'} Physical Chemistry},
  edition   = {8},
  year      = {2006},
  publisher = {W. H. Freeman},
  address   = {New York},
  pages     = {482--483},
  isbn      = {978-0-7167-8759-4}
}

@article{Schweitzer2003,
  author  = {Schweitzer, Christian and Schmidt, Rolf},
  title   = {Physical Mechanisms of Generation and Deactivation of Singlet Oxygen},
  journal = {Chem. Rev.},
  volume  = {103},
  number  = {5},
  pages   = {1685--1757},
  year    = {2003},
  month   = {May},
  doi     = {10.1021/cr010371d}
}
\clearpage


%
%


\appendix
\renewcommand\thefigure{\thesection.\arabic{figure}}



\end{document}